\documentclass[twocolumn,fleqn,usenatbib]{mnras}

\usepackage{newtxtext,newtxmath}

\usepackage[T1]{fontenc}

\DeclareRobustCommand{\VAN}[3]{#2}
\let\VANthebibliography\thebibliography
\def\thebibliography{\DeclareRobustCommand{\VAN}[3]{##3}\VANthebibliography}

\newcommand{\ms}[1]{{\color{black}#1}}

\usepackage{graphicx}
\usepackage{color}
\usepackage[nointegrals]{wasysym}
\usepackage{mathtools}
\usepackage{subcaption}
\usepackage{booktabs}
\usepackage{newtxtext,newtxmath}
\usepackage[T1]{fontenc}
\usepackage{bm}
\usepackage{amsmath}
\title[Temperature inversions on satellites]{Efficient and inefficient hydrodynamic escape of exo-satellite atmospheres driven by irradiation from their young giant planets}

   \author[M. Schulik et al.]{Matth\"aus Schulik $^{1,3}$\thanks{E-mail: mschulik@ic.ac.uk (Imperial College),}
          James E. Owen$^{1,3}$,
          Richard A. Booth $^{2}$,
          Shun Fai Ling $^{1}$, and 
          Shun Ping Wong $^{1}$ 
          \\  
$^{1}$Imperial Astrophysics, Imperial College London, Blackett Laboratory, Prince Consort Rd, London, SW7 2AZ, UK\\
$^{2}$School of Physics and Astronomy, University of Leeds, Leeds, LS2 9JT, UK \\
$^{3}$Department of Earth, Planetary, and Space Sciences, The University of California, Los Angeles, 595 Charles E. Young Drive East, Los Angeles, CA 90095, USA}

\date{Accepted XXX. Received YYY; in original form ZZZ}

\pubyear{\the\year{}}

\begin{document}
\label{firstpage}
\pagerange{\pageref{firstpage}--\pageref{lastpage}}
\maketitle

\begin{abstract}
The bolometric radiation from a central body is potentially a powerful driver of atmospheric escape from planets or satellites. When heated above their equilibrium temperatures those satellites, due to their low surface gravity, are be prone to significant atmospheric erosion. Such high temperatures can be reached through a known mechanism: a large ratio of the irradiation to re-radiation opacities of the atmospheric species. We investigate this mechanism for irradiating black-bodies of sub-stellar temperatures and find that specific molecules exist, such as $\rm NH_3$ and $\rm CH_4$, which develop temperature inversions under the irradiation of young post-formation giant planets.
These non-isothermal temperature profiles lead to escape rates that can significantly exceed isothermal Parker-model escape rates evaluated at the satellite's equilibrium temperature.
Our results indicate that exo-satellites can lose most of their atmospheric mass through this mechanism
if the cooling of the exo-satellite's interior is not too rapid. In all scenarios, we find a hierarchical ordering of escape rates of atmospheric species due to thermal decoupling in the upper atmosphere. This thermal decoupling leads to a natural depletion of $\rm CH_4$ and retention of $\rm NH_3$ in our models.
We find that giant planets with masses above 2$m_{\rm Jup}$, for cold starts and above 1$m_{\rm Jup}$ in hot start scenarios are able to remove the majority of a Titan analogue's atmosphere. 
Hence, finding and characterizing exomoon atmospheres in hypothetical future surveys can constrain the post-formation cooling behaviour of giant planets.
\end{abstract}




\section{Introduction}

\ms{Low-mass} bodies such as planets and satellites are born with gaseous atmospheres, acquired either through accretion from their parent discs, or from outgassing events after these discs disperse. Once the disc has dispersed and most gas-driven formation processes have ceased, the body will exist in a hot post-formation state. In this state, its atmosphere is susceptible to hydrodynamic removal processes driven either by tidal forces, a process akin to binary Roche-lobe overflow \citep{erkaev2007}, or by an irradiation-driven outflow \citep{gross1972, sekiya1980, zahnlekasting1986, lammer2008}. 
These outflows occur when the constituent particles in the atmosphere are sufficiently collisional \ms{to be treated as a fluid} \citep{volkov2013}. The resulting atmospheric temperatures, which are directly linked to the escape rates, depend critically on the part of the stellar spectrum which is driving the outflow, \ms{due to different penetration depths and heating mechanisms}. The prototypical energy-limited escape process \citep{watson1981} \ms{previously applied to the solar system satellites \citep[e.g.][]{erkaev2021}}, requires inefficient \ms{radiative} cooling by the escaping gas species and is driven by the high-energy flux $\rm F_{XUV}$ of the star \citep{Lammer2003, baraffe2004,  MurrayClay2009, Owen2012}. 

Alternatively, the bolometric stellar flux $\rm F_{bol}$ is typically $10^4$-$10^6$ times greater than $\rm F_{XUV}$, depending on the exact stellar age and type \citep[e.g.][]{shkolnik2014, france2018, mcdonald2019}, but this large energy input is usually cooled away very efficiently in the molecular parts of atmospheres, hence typically driving weaker mass-loss rates than their XUV counterparts \citep{owen2024}. 
These bolometrically driven hydrodynamic escape rates have been studied for close-in planets, and are typically estimated in the form of an isothermal Parker-wind \citep{Ginzburg2018, gupta2020}. The \ms{key free parameter driving this type of outflow is} the isothermal temperature, $T(r)={\rm constant}$. The value of this temperature is often further approximated as the equilibrium temperature of the planet, i.e. $T(r)=T_{\rm eq}$ and the mass-loss rates $\dot{M}$ are then approximated as $\dot{M}=\dot{M}_{\rm Parker}(T=T_{\rm eq})$. 

However, both theoretical work \citep[e.g.][]{guillot2010} and studies of the solar system atmospheres  \citep[e.g.][]{haynes2015, wyttenbach2015, yan2020} have shown how the equilibrium temperature can be significantly exceeded. Further, recent simulation work by \citet{misener2025} has demonstrated that including more realistic bolometric radiative transfer results in non-isothermal outflows with the resultant mass-loss rates being very sensitive to the form of the atmospheric temperature profile. 

While the above theoretical studies have shown that close-in, low-mass planets can be significantly impacted \ms{by stellar-driven bolometric mass-loss, and for objects of even lower masses such as satellites, the bolometric irradiation from their close giant planets could form a short-lived but important contribution to their mass-loss histories.}

In this work, we investigate the mass-loss rates from atmospheres \ms{with bolometrically driven temperature inversions, i.e.} with \rm $T (r)>T_{eq}$, for young and rocky exo-satellites. This is similar to the already known mechanism to form inversions via specific optically active molecules such as TiO and VO in exoplanets \citep[][]{haynes2015}. \ms{We will show that} the formation of atmospheric temperature inversions is particularly favourable for cold satellites irradiated by young, luminous gas giants \ms{via a new intermediate mass-loss mechanism, connecting isothermal winds and energy-limited escape}, besides already well-known mechanism to form inversions in exoplanet atmospheres.
Studying the retention or loss of exo-satellite atmospheres is important beyond considerations of radiation transport. In cases where the satellite-planet atmospheric spectrum cannot be separated, an exo-satellite's atmosphere can introduce false signals into inferences about the host planet's atmosphere \citep{Rein2014}. Furthermore, with upcoming instrumentation such as the ELTs, studying exo-satellite atmospheres could become technically feasible \citep{Kleisioti2024,vanWoerkom2024}.

\begin{figure}
\begin{subfigure}{0.43\textwidth} 
   \centering
   \includegraphics[width=1.0\textwidth]{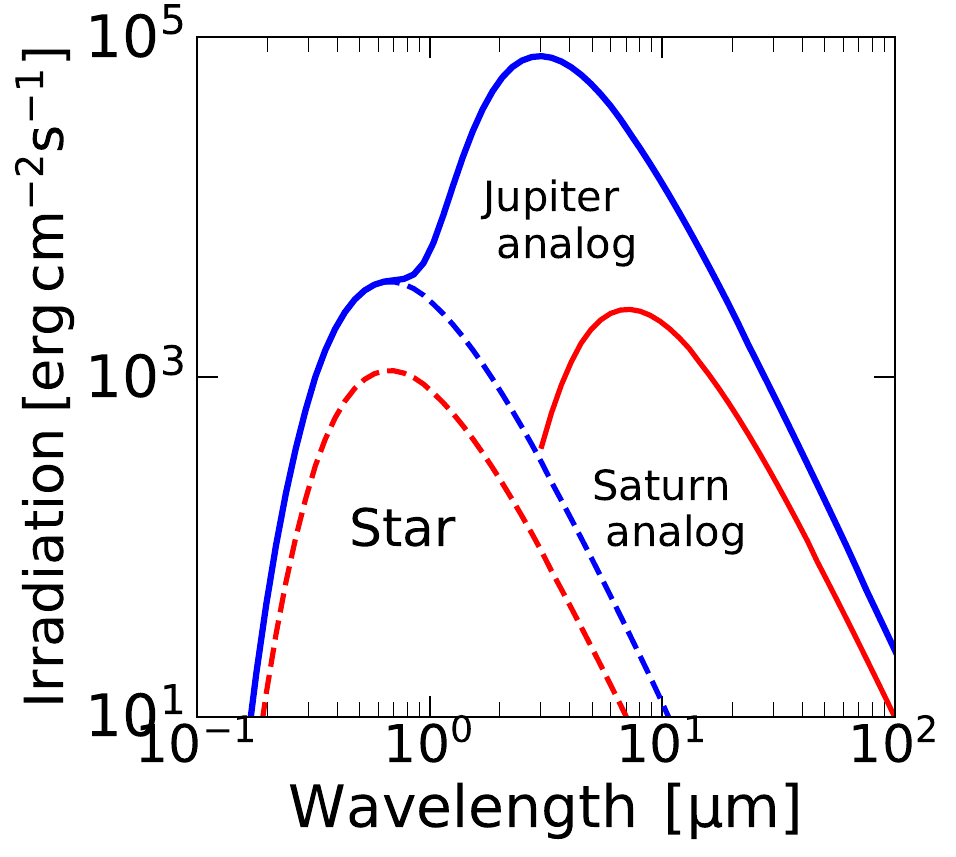}
\end{subfigure} 

\caption{ A comparison of the different irradiation levels ($\lambda F_{\lambda}$) at an exosatellite's orbit, just after disc dispersal, when planet formation is complete. The total irradiation flux can be dominated by the planetary hosts' luminosity, which is mainly determined by the giant planet's mass and its accretion history. A Jupiter-mass planet (blue) at $5$au semimajor-axis distance will dominate over the stellar radiation (dashed), while for a Saturn-mass planet (red) at $10$au semimajor-axis distance, the contributions will be of similar magnitude. }
\label{fig:irradiation_spectra}
\end{figure}

\ms{This paper is structured as follows. In Section \ref{sec:methods}, we list the equations solved in our hydrodynamic outflow problem and introduce important radiative quantities to understand the physics. We then discuss a general, parameterized mechanism to reach bolometrically-driven temperature inversions for favourable radiative opacities in Section \ref{sec:mechanism}. 
In Section \ref{sec:proofofconcept}, we then link this mechanism to the opacity structure of molecules that might be present in the atmospheres of exo-satellites. We show that specific molecules exist which feature strong temperature inversions and short atmospheric lifetimes due to favourable opacity ratios in the giant planet irradiation regime.

In order to go beyond simple lifetime estimations, we perform evolutionary 1-D radiative hydrodynamics simulations in Section \ref{sec:fullmodels_section}, with a particular emphasis on multi-species and multi-band effects. } In Section \ref{sec:exosatellites}, we follow the evolution of exo-satellites atmospheres concurrently with the cooling curve of a population of exo-giant planet hosts.
 
We then follow this modelling with a brief application of our scenario to the solar system satellites Ganymede and Titan in Section \ref{sec:ganymedetitan}.
Finally, we give a discussion and outlook on possible more complex models, including the effects of other dominant atmospheric components such as water steam, and speculative applications of our model to giant planet formation theory in Section \ref{sec:discussion}.

\section{Equations solved}
\label{sec:methods}

\ms{Our goal is to compute atmospheric mass-loss rates from exo-satellites. Thus, we must solve the equations of hydrodynamics coupled with the radiation transport of photons. Given the well-known challenge of solving directly for steady-state solutions \citep[e.g.][]{MurrayClay2009}, we choose to solve instead a time-dependent problem, which we evolve to steady-state.

This is a challenging, non-linear multi-physics problem, for which we employ the numerical code  \textsc{Aiolos }\citep{schulik2023}. }
The equations per species, $s$, for the mass density, momentum and total energy $\rho_s, \rho_s u_s, E_s$ and radiative mean field density $J$, are solved in spherically symmetric geometry for the independent coordinate, $r$, under the action of mutual collisional drag and collisional thermalization are:

\begin{align}
    \frac{\partial \rho_s}{\partial t} + \frac{1}{r^2} \partial_r (r^2\rho u ) &= 0\label{eq:fullequations_mass} \\
    \frac{\partial \rho_s u_s}{\partial t} + \frac{1}{r^2}\partial_r (r^2\rho_s u^2_s) + \partial_r p_s &= \nonumber \\
    -
    \rho_s \frac{\partial \Phi}{\partial r} - &\rho_s \sum_{\rm species\; s,s'} \alpha_{ss'}(v_s-v_{s'}) \label{eq:fullequations_momentum}
\end{align}

\begin{align}
    \frac{4\pi}{c}\frac{\partial J}{\partial t} +\frac{1}{r^2} \partial_r (r^2 F) &=\; -\sum_{\rm species\; s} \rho_s {\kappa}^{s}_{\rm out} \;\,(4\pi J - 4 \sigma T^4_s) \label{eq:fullequations_photondensity}
\end{align}
where we use the total energy  $E_s = \frac{1}{2} \rho_s u_s^2 + \rho_s e_s$, with the ideal gas closure relations $e_s = c_{V,s} T_s$, with $c_{V,s}$ the heat capacity at constant volume. Further, adopting the ideal gas law yield $p_s = e_s \rho_s(\gamma_{\rm ad,s}-1)$, where $\gamma_{\rm ad,s}$ is the adiabatic index of the species, not to be confused with the radiative quantity $\gamma$, defined below. 
\ms{We single out the energy equation per species to stress the role of the opacities $\rm \kappa_{in}$ and $\rm \kappa_{out}$ for the in- and outgoing radiation, respectively.}

\begin{align}
    \frac{\partial E_s}{\partial t} + \frac{1}{r^2} \partial_r (r^2 u_s (E_s+p_s)) &=  \nonumber \\
-\rho_s u_s \frac{\partial \Phi}{\partial r} +& \rho_s \sum_{\rm species\; s,s'} \alpha'_{ss'}(T_s-T_s')   \nonumber \\
  + \sum_{\rm bands\; b}\; \rho_s \kappa^{s,b}_{\rm in}\;\; \frac{1}{4}\,S_b \exp(-\tau_b) &  + \;\;\rho_s {\kappa}^{s}_{\rm out}\;\, (4\pi J -  4 \sigma  T_s^4)
    \label{eq:fullequations_energy}
\end{align}
\ms{where the last and second-to-last terms are the radiative heating and radiative cooling terms, proportional to $\rm \kappa_{in}$ and $\rm \kappa_{out}$ in potentially multiple bands, $b$, with irradiation fluxes $S_b$. In our approach, the outgoing radiation is only computed in a single band}.
The strength of gravitational field is $\Phi(r) = Gm_{\rm satellite}(r)/r + T$ with a tidal component $T=\frac{3}{2} G m_{\rm giant}r^2/a^3_{\rm giant}$ due to the satellites' semimajor-axis distance $a_{\rm giant}$ to the giant planet\ms{, under the assumption $m_{\rm satellite} \ll m_{\rm giant}$ \citep{murraydermott2000}}. We generally denote masses $m$ non-capitalized and mass-loss rates $\dot{M}$ in order to easier distinguish them. Finally, the collision rates $\alpha_{ss'}$ we use to compute the momentum and the thermal couplings are taken to be neutral-neutral collision rates from \cite{schunk1980}.

The radiative equation is closed via the flux-limited diffusion approximation \citep{levermore1981}
\begin{align}
F = - \frac{\lambda c}{\rho \kappa_{R}} \nabla J , 
\label{eq:flux_limited_diffusion}
\end{align}
with the flux-limiter $\lambda$ given by \cite{kley1989}, and more details on its use in atmospheric outflows are discussed in \citet{schulik2023}. The quantity $\kappa_{\rm R}$ is the total Rosseland mean opacity, a function of all species' Rosseland means $\kappa^{s}_{R}$. In general, Rosseland opacities are not additive; but the atmospheric composition we consider is variable; hence, we restrict ourselves to a simple maximum rule in computing this number from its constituents, i.e. $\kappa_R=\rm max(\kappa_{R,s1}, \kappa_{R,s2}, \kappa_{R,s3},..)$.
The outgoing radiative quantities $J$ and $F$ are computed in a single band throughout this work, \ms{their band index is therefore dropped}.

\subsection{Behaviour of temperature solutions and opacities}
\label{sec:analytic_temperature_solutions}

The solutions to the above equations yield mass-loss rates and temperature profiles. The temperature profiles, which set the mass-loss rates, are determined by the opacities to the incoming radiation per band $b$ and species $s$, i.e. $\kappa^{b,s}_{\rm in}$, appearing in the last term on the r.h.s of Eqn. \ref{eq:fullequations_energy}, and the opacity of the gas to its own radiation $\kappa^{s}_{\rm out}$, appearing in the cooling term, the second-to-last term on the r.h.s of  Eqn. \ref{eq:fullequations_energy}. The number of bands of incoming radiation is denoted as $N_b$. For $N_b=1$, this system of equations, when in local radiative equilibrium, can be solved to give the atmosphere's thermal structure. These solutions reproduce the results in the analytical work by \citep{guillot2010}, hereafter G10. Where G10 define the ratio of these opacities as: 
 \begin{align}
    \gamma &\equiv \frac{ \kappa_{\rm in}}{ \kappa_{\rm out}},
    \label{eq:gamma_definition}
\end{align}
which is a key parameter determining the temperature structure. We drop the indices $b$, $s$ when referencing single-species, single-band models.
As both opacities control the magnitude of the source and sink terms for the energy in Eqn. \ref{eq:fullequations_energy}, then $\gamma$ can be understood as a number correlating with the ratio of the relative importance of heating and cooling, with the slight complication that it is to be understood as a radiative inefficiency parameter, not an efficiency parameter.

The mean opacities $\kappa_{\rm in}$ and $\kappa_{\rm out}$ can be calculated as the appropriate mean, to be discussed later, over a wavelength-dependent opacity function $\kappa(\lambda, P,T)$, where $\lambda$ is the spectral wavelength. The outcome of the opacity averaging, which matters greatly to our model, is further discussed in Section \ref{sec:gammavaluesreal}, whereas in this subsection, we only discuss the provenance of $\kappa(\lambda, P,T)$.
The function $\kappa(\lambda, P,T)$ is \ms{computed using the correlated-k method} \citep{goody1989, lacis1991} using the correlated-k databases in the Petitradtrans package \citep[][and refs. therein]{molliere2019} where available, and the DACE database \citep[][and refs. therein]{grimm2015}\footnote{dace.unige.ch} otherwise. \ms{The mean opacity calculations, averaging over the corr-k tables, were done} using the Exo-k toolsuite \citep{leconte2021}.
These tools allow us to compute the mean opacities as pressure and temperature-dependent quantities ${\kappa}^{s}_{\rm R}$, ${\kappa}^{s}_{\rm out}$, which are computed via Exo-k as a single-band average over the entire spectrum. We emphasize that the latter two quantities remain single-banded throughout this work, whereas $\kappa^{s,b}_{\rm in}$ later becomes multi-banded.

The opacity function yielding the multi-band mean irradiation opacities $\kappa^{s,b}_{\rm in}(T_{\rm irr},T_{\rm gas})$ are taken at constant $T_{\rm gas}=200 K$ and $P=1 \rm bar$, appropriate for our exo-satellites, and preclude significant pressure-broadening. Taking a single $T_{\rm gas}$ and $P$ for the irradiation opacities is a simplification in our models, but given computational constraints, we are forced to employ it and leave relaxing this approximation to future, more accurate modelling. Since our goal is the first attempt to scan the parameter space relevant for exo-satellites, along with the fact that $T_{\rm gas}$ {does not vary by more than $\pm$ 100K, giving rise to variations in $\kappa_{\rm in}^{s,b,}$ at the $\lesssim10\%$ level in the upper atmosphere, see also Appendix \ref{sec:appendix_corrk},} this is an appropriate approximation at this stage. \ms{Instead, our study here is focused on the exponential sensitivity of mass-loss rates with temperature.}

The number of incoming bands $N_b$ \ms{is in this initial discussion is set to} $N_b=1$, to showcase comparisons of temperature profiles to the work of G10 \ms{and to present a general physical understanding of} our mechanism for the enhancement of the bolometrically driven mass-loss rates. Only for $N_b=1$ does a single value of $\gamma$ form a meaningful parameterization of the temperature profile.
Simulations with $N_b>1$ (section~\ref{sec:fullmodels}) are later used to approach more realistic temperature profiles and mass-loss rates. Those runs represent the heating function more faithfully, i.e. the last term on the r.h.s in Eqn. \ref{eq:fullequations_energy}, which impacts the mass-loss rates.
We now study the case of $N_b=1$ for some important species in cold, molecular atmospheres.

\section{The Mechanism}
\label{sec:mechanism}

We will first discuss how hydrodynamic mass loss rates react to non-isothermal temperature profiles, particularly bolometrically driven temperature inversions. Subsequently we use an analytic method to calculate the structure of non-isothermal temperature profiles from first principles by linking it to \ms{the opacity ratio} $\gamma$. Finally, we show that certain molecules, under the right irradiation conditions, possess values of $\gamma$ adequate for our enhanced atmospheric mass-loss scenario.
\ms{ As we introduce a number of different temperatures, where some are measures of the radiation field and do not actually appear in the atmospheric temperature profiles, we summarize the most important temperature concepts and vocabulary in Tab.  \ref{tab:temperature_table}.}

\subsection{The transition around $\gamma = 1$ and a new escape regime  }

In general, the atmospheric mass loss rate in a spherically symmetric geometry in 1-D steady-state outflows at any radius $r$ is the mass flux $\dot{M}(r) = 4\pi r^2 u(r) \rho(r) = const$. When evaluated at the sonic point radius $r_s$, this form specializes to 
\begin{align}
    \dot{M} = 4\pi r^2_{s} c_{s} \rho_s,
\end{align}
where $c_s$ is the isothermal speed of sound profile $c(r)$ evaluated at the sonic point, $u(r)$ the velocity profile, $\rho(r)$ the density profile and $\rho_s=\rho(r_s)$ is the density of the escaping gas at the sonic point. We can \ms{approximate} this expression, using the sonic point relation $r_s\approx G\,m/(2c_s^2)$ \citep{parker1958}, to find:

\begin{align}
    \dot{M} = \pi \frac{(G\,m)^2}{c_s^3} \rho(r_s)
    \label{eq:parkermdott}
\end{align}
where $m$ can be the mass of any atmosphere-holding body, this is not yet limited to satellites. At this point the remaining challenge is to compute the \textit{local} sound speed at $r_s$ and the \textit{global} solution to the density profile. Fortunately, the density profile is near-hydrostatic \ms{even up to the sonic point} \citep{catlingbook2017}, and it is generally an acceptable approximation to take the density profile as being hydrostatic, i.e. $\rho(r_s)\approx \rho_{\rm stat}(r_s)$, even in non-isothermal atmospheres.
After integration of the hydrostatic law one then obtains $\rho_{\rm stat}(r)=\rho_0 \exp{\left[- \frac{G\;m}{c_s^2}(\frac{1}{r_0}-\frac{1}{r})\right]}$, where $\rho_0$ is the density at the lower integration boundary. Thus one finds:

\begin{align}
    \dot{M}(T_{\rm skin}) = \pi \frac{G\;m}{c_s^3}\rho_0 \exp{\left[- \frac{G\;m}{c_s^2 r_0}\right]} \times \exp(+2) \times \exp(-1/2)
    \label{eq:parkermdott2}
\end{align}
where, 
\begin{align}
 c_s^2=\frac{k_B\;T_{\rm skin}}{\mu}
 \label{eq:soundspeed_isothermal}
\end{align}
is the sound speed at large radii due to the isothermal ``skin'' temperature $ T_{\rm skin}$ (i.e. the temperature in the outer regions of the atmosphere), $k_B$ the Boltzmann constant and $\mu$ the mean particle mass. The factors $\exp(2)$ originate in inserting $r=r_s$, and $\exp(-1/2)$ is an empirical correction due to the non-hydrostatic nature of the flow \ms{at the sonic point} \citep{lamerscassinelli1999}.

In order to move beyond the isothermal approximation 
we use the analytic model by G10. Their model gives a solution for the hydrostatic temperature structure in the radiative parts of the atmosphere, depending mainly on $\gamma$.\footnote{A few other dependencies exist in the G10 model, such as various unity factors encoding the ratios of geometric moments of the radiation field, but different choices of those do not change our model results drastically.} Thus, for the argument we are presenting, we will approximate their solutions here as a doubly-isothermal temperature structure.

\begin{figure}
\begin{subfigure}{0.44\textwidth} 
   \centering
   \includegraphics[width=1.0\textwidth]{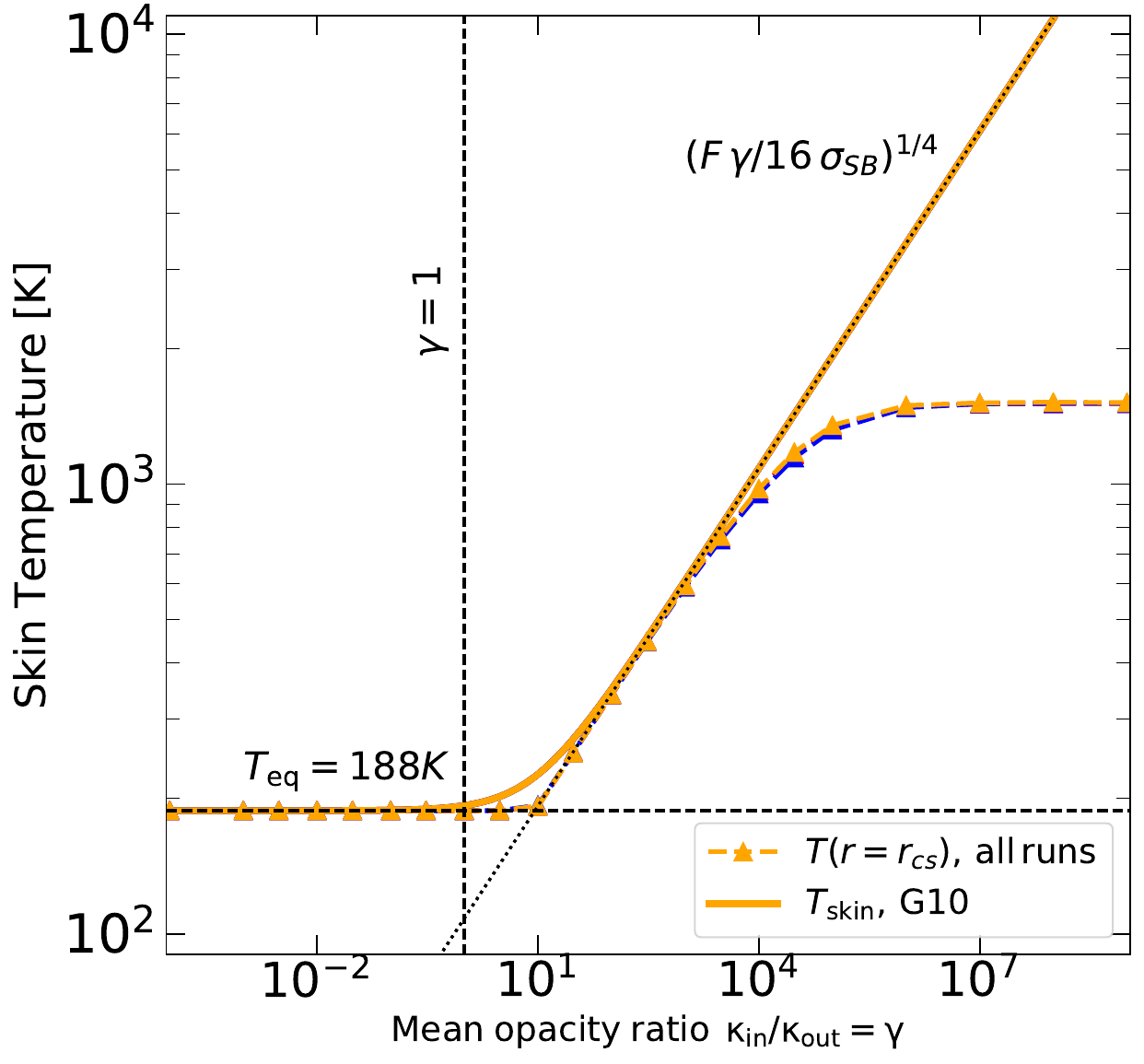}
\end{subfigure}%

 \begin{subfigure}{0.44\textwidth} 
   \centering
   \includegraphics[width=1.0\textwidth]{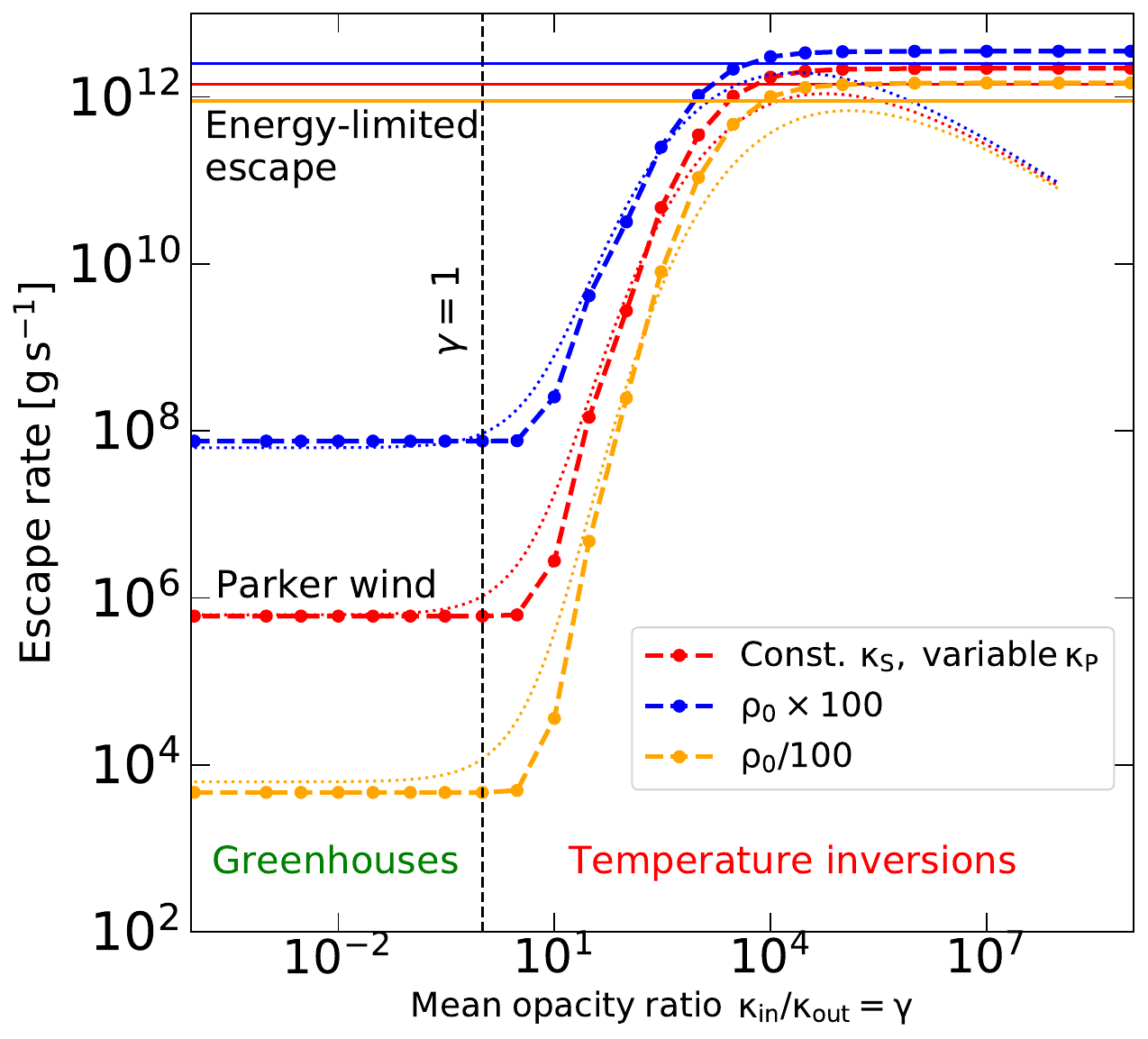}
\end{subfigure}%

\caption{Comparison of analytic and numerical results over a wide range of $\gamma$. \textbf{Top}: Shown are the numerical temperatures at the sonic point as dotted line and the solid lines show Eqn. \ref{eq:tskin_guillot}. The deviation at large $\gamma$ indicates the transition to energy-limited escape. \textbf{Bottom}: Mass-loss rates are shown as a function of the cooling inefficiency parameter, where the points show our simulation results.  Simple model escape rates using Eqn. \ref{eq:parkermdott1.5} are shown as the thin, dotted lines. Anti-greenhouses ($\gamma>1$) possess temperature inversions, driving stronger escape rates than greenhouse atmospheres ($\gamma<1$). In the extreme limit of $\gamma\ll1$ we find Parker winds at $T$=$T_{\rm eq}$, and the mass-loss rates depend on the amount of gas in the wind-launching region. At $\gamma\gg1$, the escape rates transition into an energy-limited regime, i.e. Eqn. \ref{eq:energylimit}. In this case, the mass-loss rate becomes density-independent, and the skin temperatures become saturated by efficient PdV-cooling. }
\label{fig:varying_gamma_temperatureplot}
\end{figure}

The G10 approach allows the temperature at large radii ($T_{\rm skin}$), to be computed from first principles. 
Formally $T_{\rm skin}$ is only the temperature at $r\rightarrow \infty$. Still, in practice this temperature is a good approximation in a large volume of the \ms{outermost} heated layer, i.e. above the optically thick radius $\rm r(\tau_{bol}=1)\equiv r_{1}$ \citep{schulik2023}. Hence, we take $T(r) = T_{\rm skin}$ for $r \in [r_1,\infty)$ in this region of the atmosphere. 
The temperature {at lower altitude} is set to be $T(r)=T_{\rm eq}/2^{1/4}$, in the radius interval $r \in [r_0,r_1)$ (for the $2^{1/4}$ factor see also the classic works by \citealt{schwarzschild1906, Milne1921, sagan1969}, \ms{that accounts for re-radiation of thermal photons into two half-hemispheres}). Note that the latter is a further simplification over the G10 models, in which we neglect downward diffusion into the lower atmosphere for $\gamma<1$ \ms{in order to highlight they key physics}.
Hence, for our doubly-isothermal atmosphere, only $T_{\rm skin}=T_{\rm skin}(\gamma)$ is of a nontrivial value and needs further consideration.
With G10, $\rm T_{skin}$ can be expressed as:

\begin{align}
    T_{\rm skin}^4 = 
    \frac{1}{2} T_{\rm eq}^{4} \left( 1 + \frac{\gamma}{2} \right) \rightarrow T_{\rm skin} =
        \begin{cases}
      \approx T_{\rm eq}/2^{1/4} & \rm if \;\gamma \ll 1\\
      >T_{\rm eq}/2^{1/4} & \rm if \; \gamma \gtrsim 1\\
    \end{cases}       
    \label{eq:tskin_guillot}
\end{align}
where $T_{\rm eq}$ is given by the emitted luminosity of the giant planet at black-body temperature $T_{\rm irr}$, such that $T^4_{\rm eq} = \frac{1}{4}T^4_{\rm irr} \left(\frac{R_{\rm Giant}}{a_{\rm Giant}} \right)^2$,  with $R_{\rm giant}$ is the host's radius. At a young system age and far from the host star,  $T^4_{\rm eq}$, is dominated by contributions from the giant planet host over those of the system's host star, as illustrated in Fig. \ref{fig:irradiation_spectra}, motivating our neglection of the stellar contribution in Eqn. \ref{eq:tskin_guillot}. 

\begin{table*}
    \centering
    \begin{tabular}{c|c|}
    \hline
        Symbol & Description \\
    \hline
        $T_s(r)$ & The radial temperature profile for species $s$\\ 
        $T_{\rm surf}$ & The temperature of the isothermal satellite body. Sets the lower atmospheric temperature, $T_s(r_{\rm satellite})=T_{\rm surf}$. \\
        $T_{\rm skin}$ & $T_s(r\rightarrow \infty)$ \\ 
        $T_{\rm irr}$ & Black-body temperature of the irradiating giant planet. \\ 
        $T_{\rm eq}$ & Equilibrium temperature, a measure for the total energy flux at the top of the atmosphere. \\ & \ms{In this work, directly proportional to $( T^4_{\rm irr} \times R^2_{\rm giant})^{1/4}$.} \\ 
        $\tau_{\rm R}$ & Rosseland-mean optical depth, measured from large r inwards \\
    \hline
    \end{tabular}
    \caption{A temperature notation encyclopedia.}
    \label{tab:temperature_table}
\end{table*}

Then, it follows from Eqn. \ref{eq:parkermdott}  that the mass-loss rates obey

\begin{align}
    \dot{M}(T_{\rm eq}, T_{\rm skin}) = \pi \frac{G\,m}{c_{skin}^3}\rho_0 &\exp{\left[- G\,m/c_{\rm eq}^2 \left(\frac{1}{r_0} - \frac{1}{r_1} \right)\right]} \times \frac{T_{\rm eq}}{ T_{\rm skin}} \times \nonumber \\
 \times\exp(+1.5)\; \times\;  &\exp{\left[- G\,m/c_{\rm skin}^2 \left( \frac{1}{r_1} - \frac{1}{r_{\rm B, skin}} \right) \right]} 
    \label{eq:parkermdott1.5}
\end{align}
where the bolometric absorption radius $r_1$ is \textit{a priori} unknown, $r_{\rm B, skin}=G\,m/(2\,c_{\rm skin}^2)$, and the two sound speeds $c_{\rm eq}$ and $c_{\rm skin}$ correspond to the temperatures $T_{\rm eq}$ and $T_{\rm skin}$, respectively. \ms{The factor $T_{\rm eq}/T_{\rm skin}$ accounts for pressure continuity at the temperature discontinuity}. For the sake of simplicity, in comparing analytical to simulated escape rates, we \ms{use} the value of $r_1$ \ms{which we find in} our simulation, but in general, one would have to iterate over the temperature-dependent optical depth \ms{ $\tau =1= \rho(r_1) \kappa H$ } to find $r_1$.

\ms{ In Eqn. \ref{eq:parkermdott1.5}, when considering $\gamma>1$, temperatures  $T_{\rm skin} > T_{\rm eq}$ are achieved,} it then becomes evident that a high mass-loss limit regime for Parker-winds exists.
Because $\gamma$ can be interpreted as a radiative inefficiency number, the existence of the high mass-loss regime can be straightforwardly understood: for large $\gamma$, cooling is inefficient compared to heating. Thus, heating must be increasingly balanced by adiabatic cooling, driving stronger mass-loss.

In Greenhouse atmospheres with $\gamma<1$\ms{, the hot and cold temperatures swap place, and hence those atmospheres}
would feature some expansion in their lower atmospheres, albeit over a small radial extent. Their upper atmospheres are cold at $T=T_{\rm eq}$, and the net effect is that a $\gamma<1$ atmosphere has much lower escape rates than a $\gamma>1$ atmosphere. {We note that in a naive application of our Eqn. \ref{eq:parkermdott1.5}, atmospheres with $\gamma<1$ could also yield $\dot{M}$>$\dot{M}_{\rm Parker}$. However, this would require the ingoing and outgoing radiation to be well separated in wavelength, such as is the case for stellar irradiation on temperate planets, where even then the formation of a convective layer prevents high temperatures. In such a temperature case, the instellation is deposited deep in the diffusive part of the atmosphere, elevating the temperatures above $T_{\rm eq}$. Atmospheres with large scale heights could then be prone to high mass-loss rates. This situation does not arise for the satellite atmospheres considered here, as the ingoing and outgoing radiation passes through the same troughs of the opacity function; therefore, the ingoing thermal radiation does not become diffusive.} 
The $\gamma$-dependent atmospheric prototypes are further explored in double-gray simulations in the \ms{context of the evolution of small exoplanets by} \citet{misener2025}.

We show numerical simulation results for the value of $T_{\rm skin}$ while varying $\gamma$ as a free parameter in Fig. \ref{fig:varying_gamma_temperatureplot} and contrast those with the expectations from G10. Further, the simulations confirm the analytic mass-loss rates from Eqn. \ref{eq:parkermdott1.5} against our numerical radiation hydrodynamic double-grey model, executed using \textsc{Aiolos}. In the simulations, we do not directly vary the quantity $\gamma$, but instead its constituent opacities $\kappa_{\rm in}$ and $\kappa_{\rm out}$ 
from Eqn. \ref{eq:gamma_definition}.
We choose to keep $\kappa_{\rm in}$ constant in order not to change the absorption altitude and vary $\kappa_{\rm out}$ instead, which translates to changing the effective strength of the radiative loss term in the gas energy equation.
Higher values of $\kappa_{\rm out}$ then amount to stronger cooling and lower $\gamma$.

The adopted simulation parameters for the atmospheric gas are that of $\rm NH_3$, i.e. a mean molecular weight $\mu=17 a.m.u$, an adiabatic heat ratio of $\gamma_{\rm ad}=6/4$, a base density of $\rho_0=10^{-3}\rm g\,cm^{-3}$ and a {surface} temperature of $T(r=r_{\rm surf})$ $=T_{\rm eq} = 188K$, which is also the isothermal initial condition of the simulation.
The exo-satellite parameters are a mass of $\rm m=0.025 m_{\oplus} \approx\rm 1 m_{Titan}$, radius $\rm r=2.63\times 10^{8}cm$ and a bolometric irradiation flux with $F_0=3\times 10^5 \rm erg\,cm^{-2}\,s^{-1}$.
The base density $\rho_0$ is varied by two orders of magnitude up and down from the nominal value to show that the regime around $\gamma=1$ depends on density, whereas, for very large $\gamma$, as discussed in section~\ref{sec:highgamma}, this dependency vanishes.

It becomes evident that around the transition value of $\gamma=1$, an increase in $\rm T_{skin}$ is observed at the magnitude predicted by the G10 model. Hence, the mass-loss rates increase exponentially, corresponding to the increasing density at the sonic point. This dramatic increase in mass-loss rates \ms{will have} important consequences for the (non-)retention of atmospheres, should $\gamma>1$ be reached under realistic conditions. We will now briefly discuss the physics in the limit of $\gamma \rightarrow \infty$ for the sake of completeness and then proceed to investigate realistic conditions under which $\gamma>1$.

\begin{figure*}
\hspace*{-0.5cm}

\begin{subfigure}{0.45\textwidth} 
   \centering
   \includegraphics[width=1.0\textwidth]{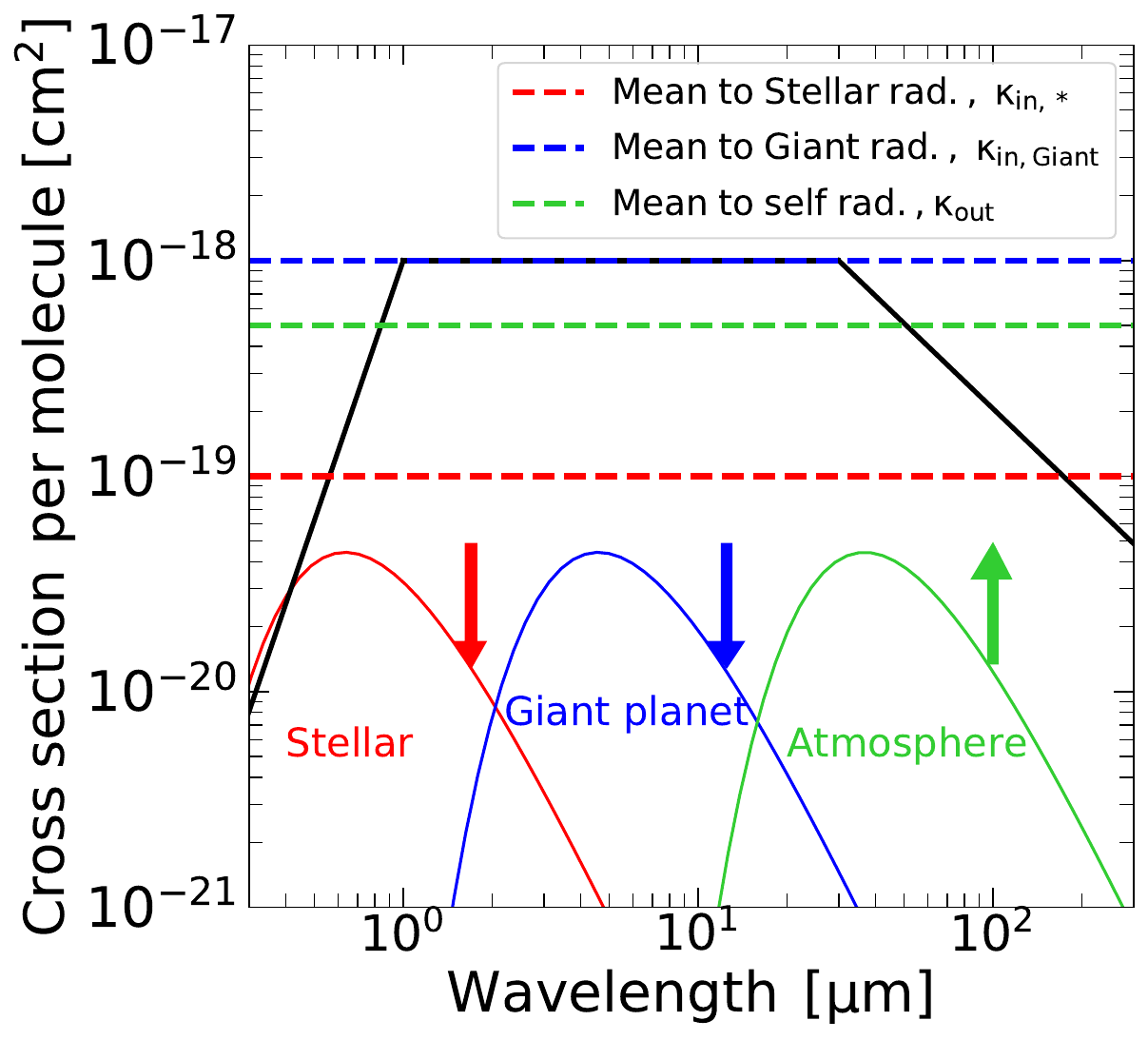}
\end{subfigure}%
\begin{subfigure}{0.45\textwidth} 
   \centering
   \includegraphics[width=1.0\textwidth]{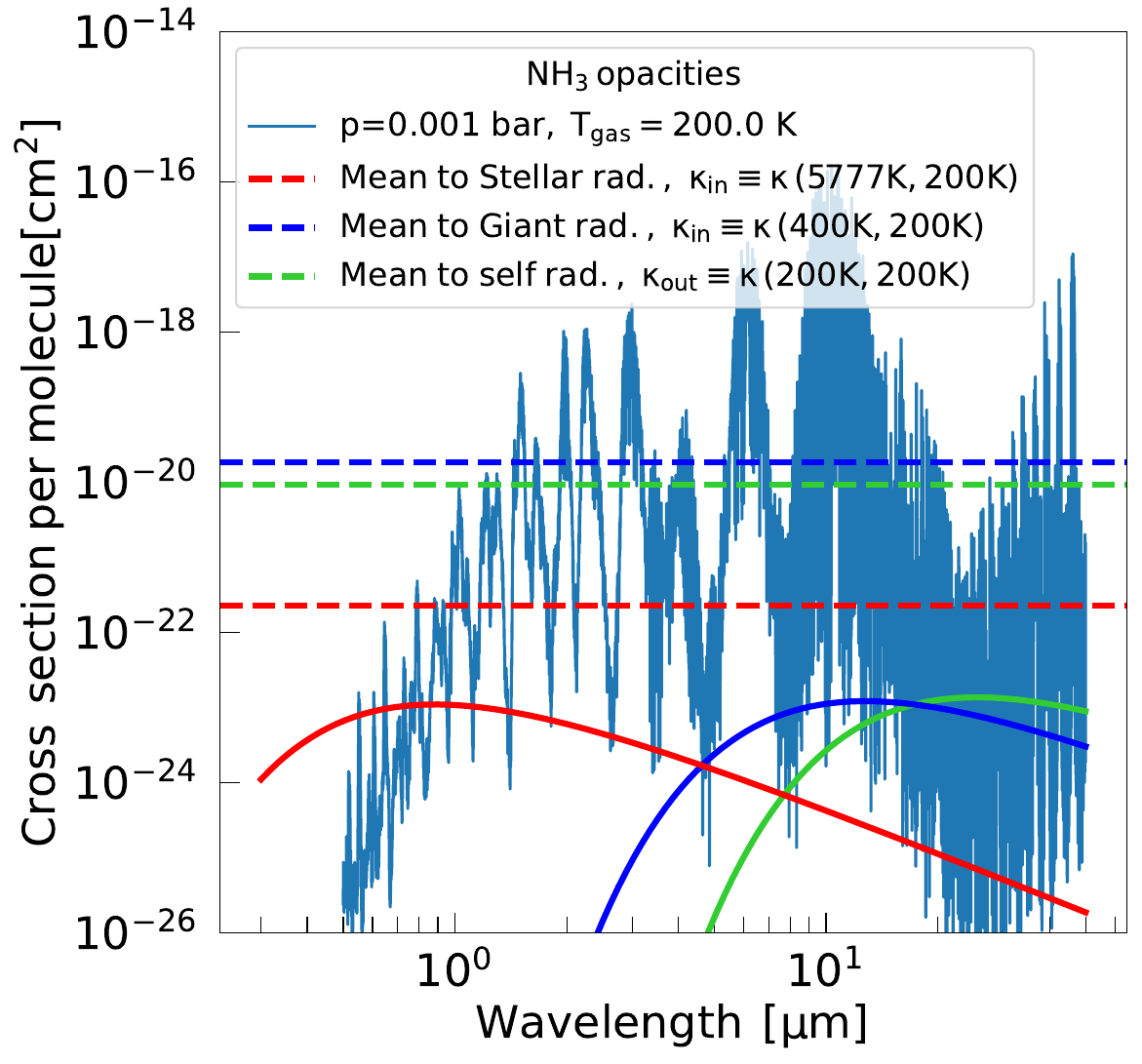}
\end{subfigure}%

\caption{
\ms{A comparison between the values of the various mean opacities defined by Eqns. \ref{eq:opacity_mean_twotemp}-\ref{eq:opacity_mean_singletemp}, the wavelength dependence of the different radiation fields and actual cross-sections.}
The left panel sketches the conditions under which $\gamma>1$ occurs when averaging opacity functions $\kappa(\lambda)$ (solid black line). When irradiated by a stellar black-body, the opacity ratio of incoming to outgoing mean opacities is $  \gamma < 1$ (the constant red value is below the constant green value). When the irradiating black-body cools, it moves up to the high opacity range in the infrared, but the outgoing radiation remains in the far-infrared, resulting in $\gamma  > 1$ (the constant blue line is above the constant green one) and vigorous hydrodynamic escape occurs.
The right panel shows that $\kappa_{\rm out}$ readily remains low for realistic and cold, molecules such as $\rm NH_3$, as the natural spacing between the molecular band peaks outpaces the width of a black-body.
}
\label{fig:mean_opacity}
\end{figure*}

\subsection{The high $\gamma$ regime}
\label{sec:highgamma}

While not the main focus of this paper, we note the existence of a distinct limit for very high $\gamma$, in which the mass-loss rates become independent of the base density. This limit is found in our bolometrically-driven simulations and is physically identical to energy-limited flows previously studied in the case of photoevaporation\footnote{This should not be surprising since radiative cooling is negligible in a photoionized gas unless it's heated beyond $10^4$~K \citep{owen2019}, thus, energetically at least a photoevaporative flow is similar to a $\gamma\gg 1$ bolometric outflow. }.

This is no coincidence and can be understood as follows:
any gas, independent \ms{of whether it is being heated by} bolometric or high-energy irradiation, which is sufficiently slow at cooling radiatively 
allows for \cite{watson1981}'s energy limited theory to be applicable. \ms{Because slow cooling can be given by a small $\rm \kappa_{out}$,  equivalent to large $\gamma$ it is expected to find energy-limited flows for $\gamma\rightarrow \infty$.  In this limit, adiabatic cooling is the dominant cooling mechanism, which at the same time gives rise to atmospheric escape.}
The solutions to Eqn. \ref{eq:fullequations_energy} then take the form of \cite{sekiya1980}'s and \cite{watson1981}'s solution of a simple Bernoulli integral, and, neglecting a few minor terms, this yields the familiar: 
\begin{align}
    \dot{M} = \pi F_0 \frac{r^3_{\tau=1}}{Gm}
    \label{eq:energylimit}
\end{align}
We illustrate the transition between the ``Parker-wind'' and ``energy-limited'' regimes in Fig. \ref{fig:varying_gamma_temperatureplot}, \ms{ where it becomes obvious in the numerical simulation data. The analytic model, however, Eqn. \ref{eq:parkermdott1.5}, is invalid in this limit and undershoots the energy limited values. This non-monotonic solution behaviour is due to the saturation of the exponential term and the takeover by the $1/c_s^3$-prefactors, which would predict unrealistically low densities at the sonic point.}  As expected from energy-limited theory, the mass-loss rates for all three densities are very close to each other, reflecting only minor differences in $r_{\rm \tau=1}$, but not a direct dependency on the base density \citep[similar to the discussion in][for planetary photoevaporation]{MurrayClay2009}.

\subsection{ Values of $\gamma$ for real molecules - the case for hot giant planets and cold satellites}
\label{sec:gammavaluesreal}

To move beyond $\kappa_{\rm in }$, $\kappa_{\rm out}$ as constant, free parameters, and to connect them to real opacity data, we compute their values as two-temperature and one-temperature Planck-means \citep{mihalasmihalas}.

The mean opacities then encapsulate the complications of large numbers of molecular lines at individual excitation levels at a local gas temperature $T$ and local gas density $\rho$. One can define several Planck-mean mean opacities $\kappa_{\rm mean}$, which describes how strongly the gas opacities absorb and emit a given radiation field.
Generally, the Planck-mean is a function depending on two temperatures. The temperature of the gas ($T_{\rm gas}$) provides the complex shape of the opacity function, and the irradiating photon field, which we parameterize by the black-body function, $B$, of temperature $T_{\rm irr}$, such that:

\begin{align}
    \kappa_{\rm mean}(T_{\rm irr}, T_{\rm gas})&\equiv\frac{\int_0^{\infty} d\lambda \; B(T_{\rm irr}) \;\kappa(\lambda, P, T_{\rm gas})}{\int_0^{\infty} d\lambda \; B(T_{\rm irr})} \label{eq:opacity_mean_twotemp}.
\end{align}
where $P$ is the total pressure of all species.
This function can now be evaluated in two different ways to obtain the mean opacities making up $\gamma$, via:
\begin{align}
    \kappa_{\rm in}&\equiv\kappa_{\rm mean}(T_{\rm irr}, T_{\rm gas})\\
    \kappa_{\rm out}&\equiv\kappa_{\rm mean}(T_{\rm gas},T_{\rm gas})
\label{eq:opacity_mean_singletemp}
\end{align}
\ms{which}, in $\rm \kappa_{out}$ \ms{encodes} information about the absorption strength of the self-emitted photons, while $\rm \kappa_{in}$, \ms{contains} information about the strength of the interaction of the gas with the external irradiation, parameterized with $T_{\rm irr}$. Therefore, $\gamma$ inherits dependencies on $T_{\rm irr}, T_{\rm gas}$ and $P$. 
Setting $T_{\rm gas}, P = \rm const.$, it is crucial to study how $\gamma(T_{\rm irr})$ evolves. For any band $b$, i.e. when one computes $\kappa^b_{\rm in}$, the integral boundaries are adjusted between the band wavelengths $\lambda^b_{\rm lower}$ to $\lambda^b_{\rm upper}$.

With the typical structure of $\kappa(\lambda, T_{\rm gas}, P)$ for molecular gas, as sketched in Fig. \ref{fig:mean_opacity} (Left), two conditions for obtaining $\gamma > 1$, or equivalently $\kappa_{\rm in} > \kappa_{\rm out}$ become clear:
(i) the maximum of $B(T_{\rm irr})$ must be located in the near-infrared, and (ii) $B(T_{\rm gas})$ must be located on the decreasing branch of the opacity function, i.e. in the far-infrared.

Certain real molecular gases, as shown for $\rm NH_3$ at $10^{-4}$ bar, $T_{\rm gas}=200K$ in Fig. \ref{fig:mean_opacity} (Right), fulfil those conditions. Their $\kappa(\lambda)$ possess naturally strong \ms{bands} in the near-infrared due to the existence of many dense, excited rovibrational states, keeping $\kappa_{\rm in}$ at a high value. Additionally, the natural band spacing increases with respect to the width of a black body; therefore, the far-infrared minimum indicated in Fig. \ref{fig:mean_opacity} (Left) is bound to exist at cold gas temperatures. Low pressures also help keep the voids between the bands due to weak pressure broadening, ultimately guaranteeing low $\kappa_{\rm out}$.

Hence these two effects conspire to yield $\gamma>1$ when cold molecular gas is irradiated by a moderately warm 
black-body. This is the exact situation expected for the upper atmosphere of a newly formed satellite orbiting a newly formed gas giant. Hence, our focus on the scenario involving exo-satellites orbiting gas giant planets in this work.
We also stress that not all molecular opacity functions fulfil those conditions, as shall be demonstrated in the next sub-section.

For stellar irradiating black bodies, as illustrated in Fig., \ref{fig:mean_opacity} (Left), irradiation is absorbed less efficiently in the visible, as known from, e.g. Earth's atmosphere, resulting in $\gamma<1$.

\begin{figure*}
\hspace*{-0.5cm}

\begin{subfigure}{0.33\textwidth} 
   \centering
   \includegraphics[width=1.0\textwidth]{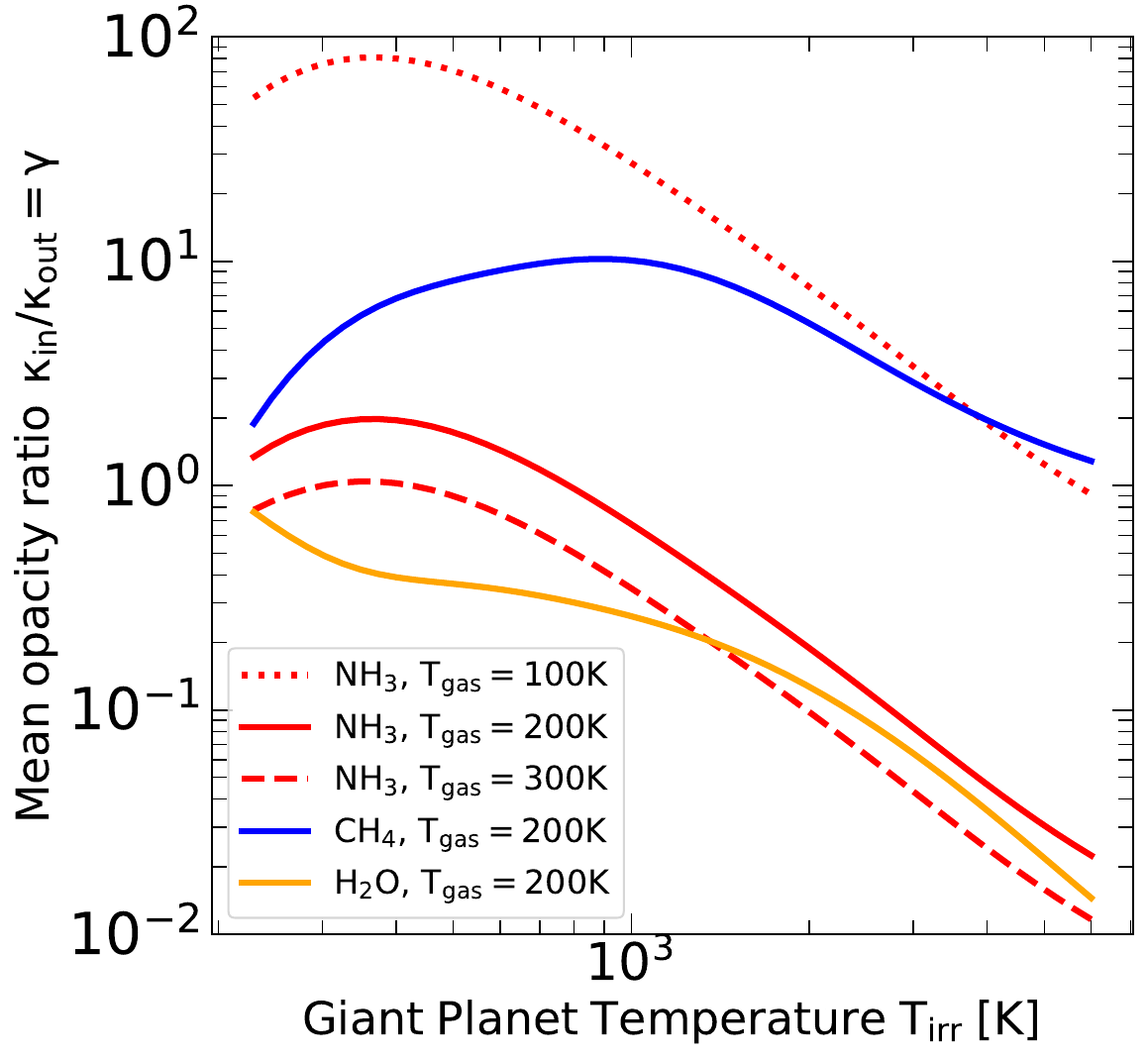}
\end{subfigure}%
\begin{subfigure}{0.33\textwidth} 
   \centering
   \includegraphics[width=1.0\textwidth]{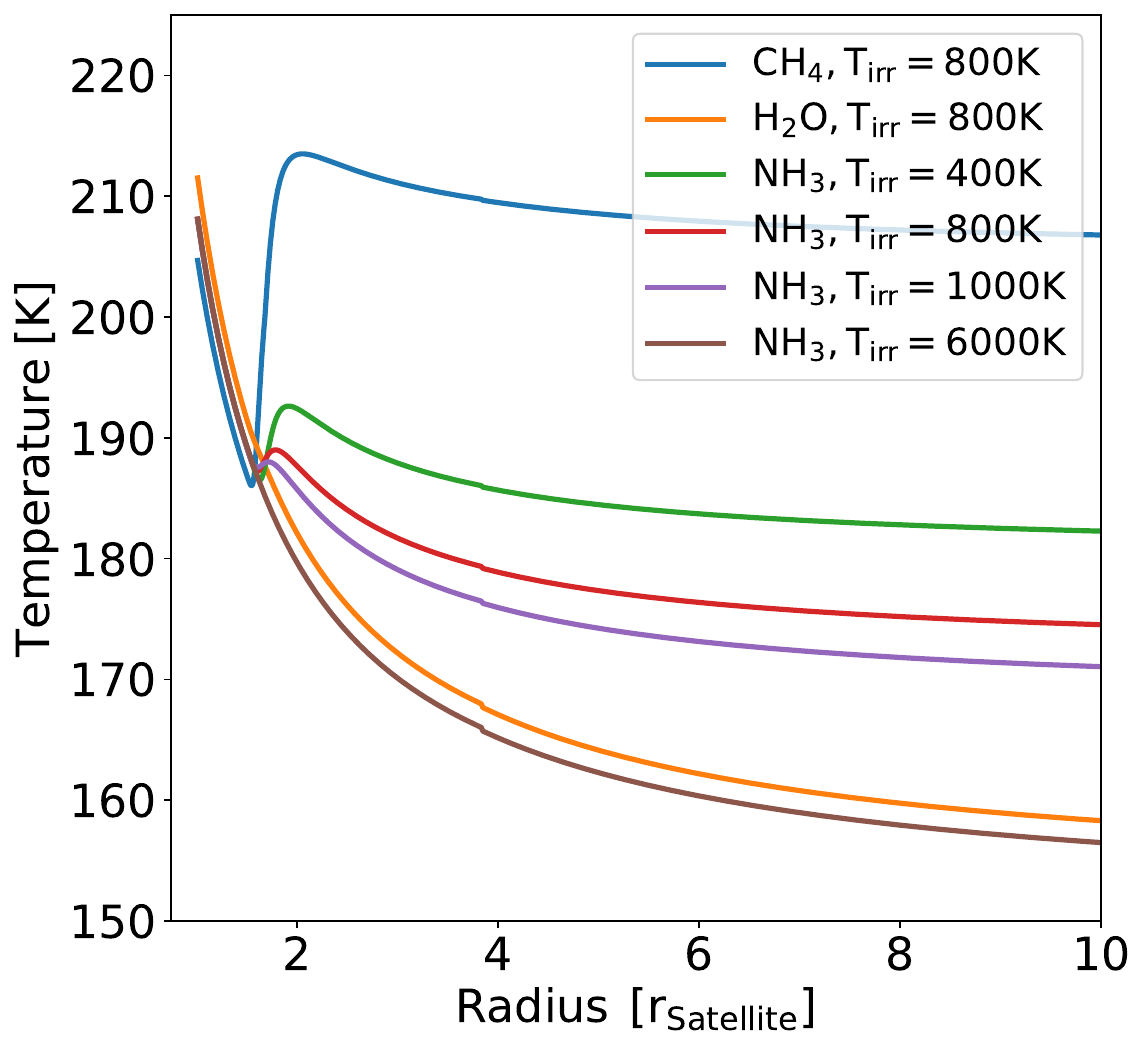}
\end{subfigure}%
\begin{subfigure}{0.315\textwidth} 
   \centering
   \includegraphics[width=1.0\textwidth]{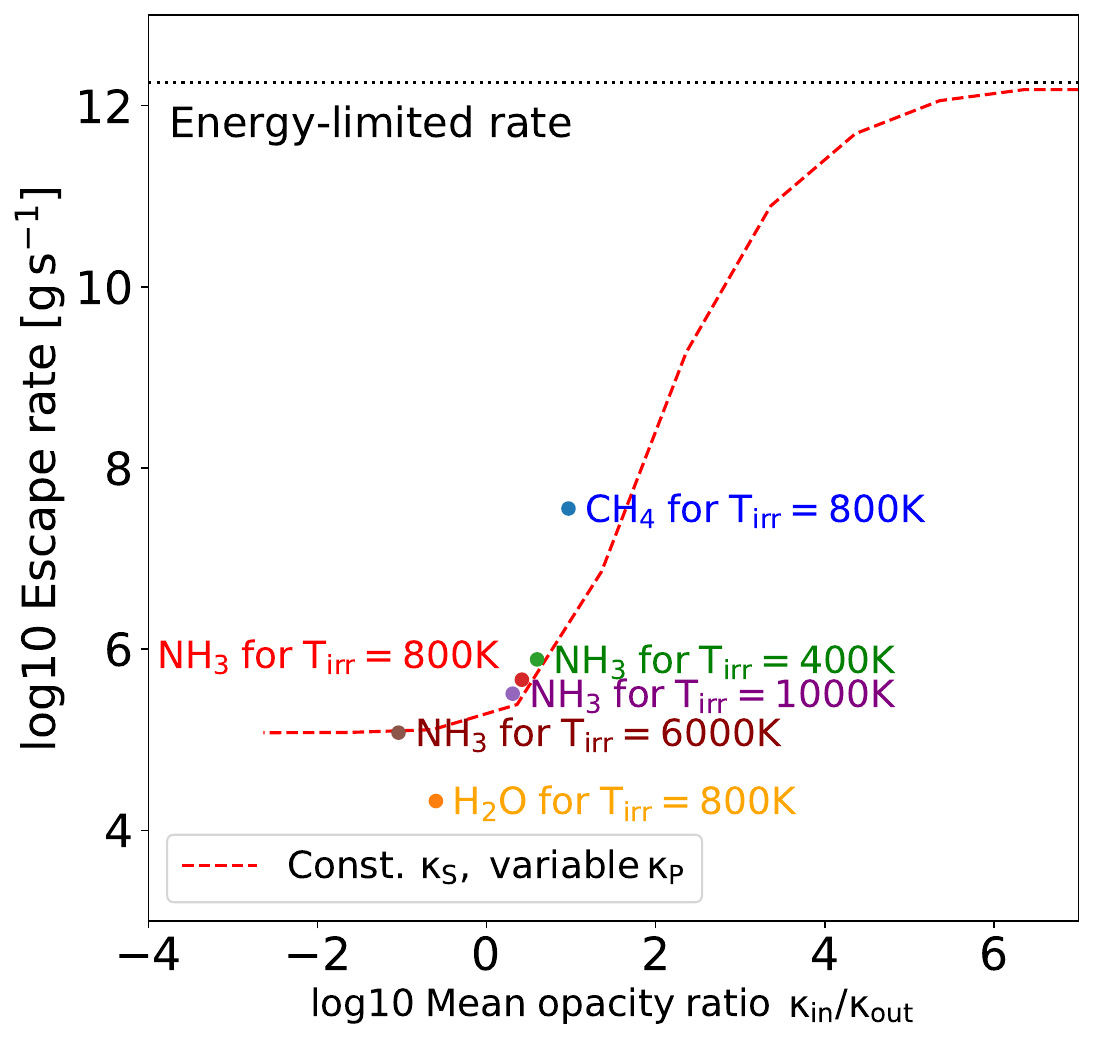}
\end{subfigure} 
\caption{ Results from double-grey simulations. \textbf{Left}: Values of $\gamma$ shown as a function of different temperatures of irradiating black-bodies, \ms{for different gas temperatures $T_{\rm gas}$, spanning reasonable post-formation scenarios}. Species with local maxima, such as $\rm NH_3$ and $\rm CH_4$ are the focus of this work. \textbf{Middle}: Temperature profiles in single band, single species simulations, employing the true molecular mean-opacities. The skin-temperatures scale as $T^4_{\rm skin} \propto F\times \gamma$, where we used $\rm F=3\times10^5\, erg\, cm^{-2}\,s^{-1}$. Note the trend of increasing $T_{\rm skin}$ as the giant planet temperature $T_{\rm irr}$ decreases.  \textbf{Right}: Escape rates resulting from the temperature structure for various species and irradiation temperatures. The red dashed curve is the theoretical toy-model curve from Fig. \ref{fig:varying_gamma_temperatureplot} for $\rm NH_3$.  Species,  irradiation temperatures and opacities are the same as the coloured lines in the middle figure. As expected, realistic setups do not approach the ``energy limit''. }
\label{fig:gamma_vs_mdot}
\end{figure*}


\subsection{\ms{Putting it all together}: Efficient and inefficient escapers}
\label{sec:proofofconcept}

We now characterize the mean opacity ratios of some prominent real gases under the conditions expected for newly formed exo-satellites. Particularly interesting as precursors of modern atmospheres in the outer solar system satellites, proxies we use for our exo-satellites, are species such as $\rm CH_4, \; NH_3,\; H_2O$ \citep{mandt2014}. The temperature structures and escape rates are simulated with \textsc{Aiolos} in single species, single band runs.

We plot the behaviour of the mean opacity ratio of those species as a function of $T_{\rm irr}$ for constant values of $T_{\rm gas}$ in Fig. \ref{fig:gamma_vs_mdot}. As already seen in Fig. \ref{fig:mean_opacity} (Right), a maximum in $\gamma(T_{\rm irr})$ exists for moderate values of $T_{\rm irr}$, which we indeed find for $\rm NH_3$ and $\rm CH_4$, but not for $\rm H_2O$.

The maximum in $\gamma$ leads to the interesting behaviour that at constant irradiation luminosity, a colder black-body can drive higher $T_{\rm skin}$ and hence higher escape rates, as can be seen in Fig. \ref{fig:gamma_vs_mdot} (middle and right). To emphasize this, note the sequence $\dot{M}(400K)>\dot{M}(800K)>\dot{M}(1000K)>\dot{M}(6000K)$ for $NH_3$. 

Again, we emphasize that this counterintuitive result emerges from the relative inability to cool away the absorbed stellar radiation, as the irradiating black-body matches the species' opacity peak. Hence we also loosely refer to species with $\gamma>1$ or $\dot{M}>\dot{M}_{\rm Parker}(T_{\rm eq})$ as \textit{efficient}, even if our subset of species and $T_{\rm irr}$ remains far from the energy-limit, and $\gamma<1$-escaping species as \textit{inefficient}. One can then define an efficiency $\zeta_s$, which measures a species' $s$ mass-loss rate compared to what an isothermal Parker-wind model at the equilibrium temperature and mean molecular mass $\mu_s$ would give:

\begin{align}
    \zeta \equiv \frac{\dot{M}_{\rm simulation}}{\dot{M}_{\rm Parker}(T=T_{\rm eq}, \mu=\mu_s)}
    \label{eq:zetaefficiency}
\end{align}

which we note is different from the efficiency parameter (often given the symbol, $\eta$), used frequently in energy-limited escape theory \citep{baraffe2004, erkaev2007}. 
We particularly note the potential role of $\rm CH_4$ as a strong driver of escape. For $\rm CH_4$ we find $\gamma(T_{\rm irr}$=$800K)\approx 10$, which, assuming a mass reservoir of a Titan-equivalent methane atmosphere ($m_{\rm N_2}\approx 10^{-6}\,m_{\oplus}$ and $m_{\rm CH_4}\approx 10^{-7}\,m_{\oplus}$, about the equivalent of modern Titan's atmospheric mass), corresponds to a lifetime of only 
\ms{
\begin{align}
t_{\rm life} = \frac{M_{\rm CH_4}}{\dot{M}(T_{\rm irr}=800K)} \approx \;2\,\times\,\rm 10^6 yrs
\label{eq:lifetime_estimate}
\end{align}
which can be on the order of, or even shorter than the gas giant's cooling time.
}
This result is significant, as it indicates that under typical bolometric irradiation conditions of Titan-analogues orbiting young, hot gas giants, \textit{the entire atmosphere can be stripped before the gas giant significantly cools}.

$\rm H_2O$, on the other hand, is mainly active as a greenhouse gas and, for that same reason, is also an efficient coolant in the optically thin atmosphere. $\rm H_2O$ remains in the inefficient escape regime with $\gamma<1$ for the expected $T_{\rm irr}$ of exo-satellites. The lower atmospheric temperatures ($r<2 r_{\rm Titan}$) are correspondingly slightly higher, and the upper atmospheric temperatures are lower than those of the other two species.

Other molecules \ms{we investigated} are $\rm HCN$, $\rm CO_2$, $\rm CO$, which are briefly documented in Appendix \ref{sec:appendix_gammacurves}. For the latter two, the strongly peaked opacity functions do not catch the broad-band thermal radiation well, which is why single-band mean opacity calculations of those molecules are inaccurate. Thus, we do not model the effects of these molecules further in this work, and instead focus on mixtures of $\rm CH_4$ and $\rm NH_3$ in multi-band simulations next.

\begin{figure*}
\hspace*{-1.0cm}
\begin{subfigure}{0.40
\textwidth} 
   \centering
   \includegraphics[width=0.9\textwidth]{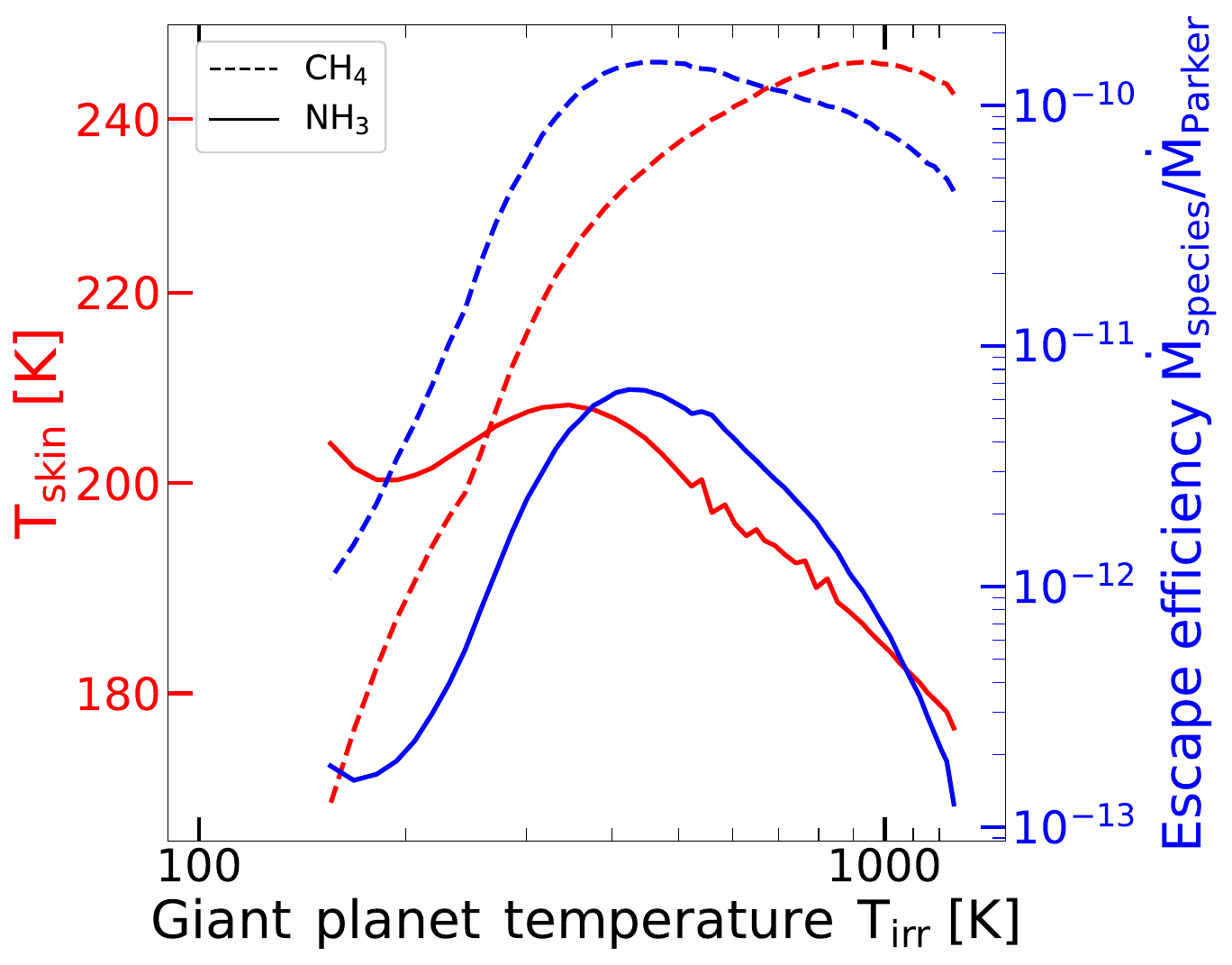}
\end{subfigure}
\begin{subfigure}{0.30
\textwidth} 
   \centering
   \includegraphics[width=1.0\textwidth]{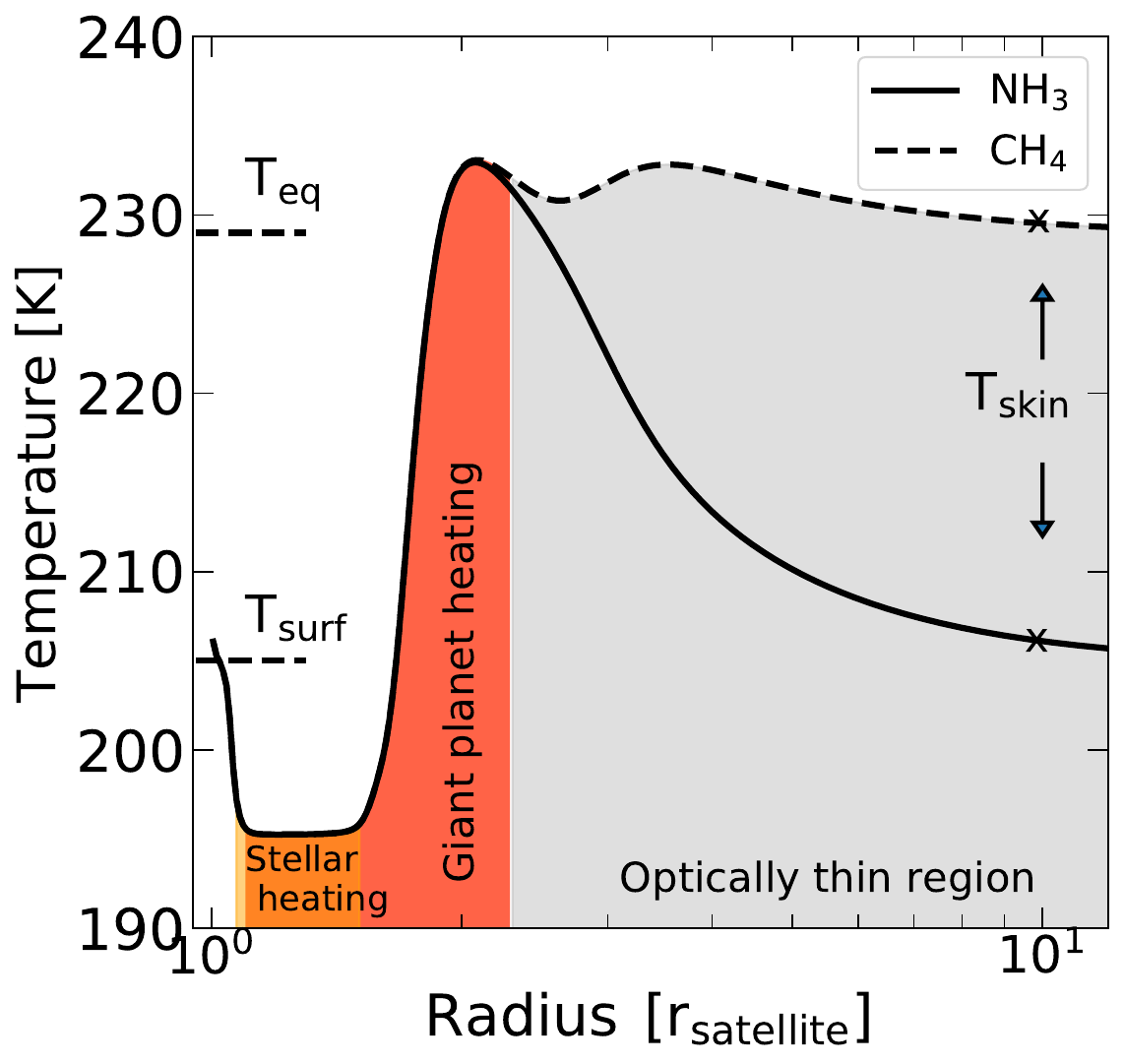}
\end{subfigure}
\begin{subfigure}{0.33
\textwidth} 
   \centering
   \includegraphics[width=1.0\textwidth]{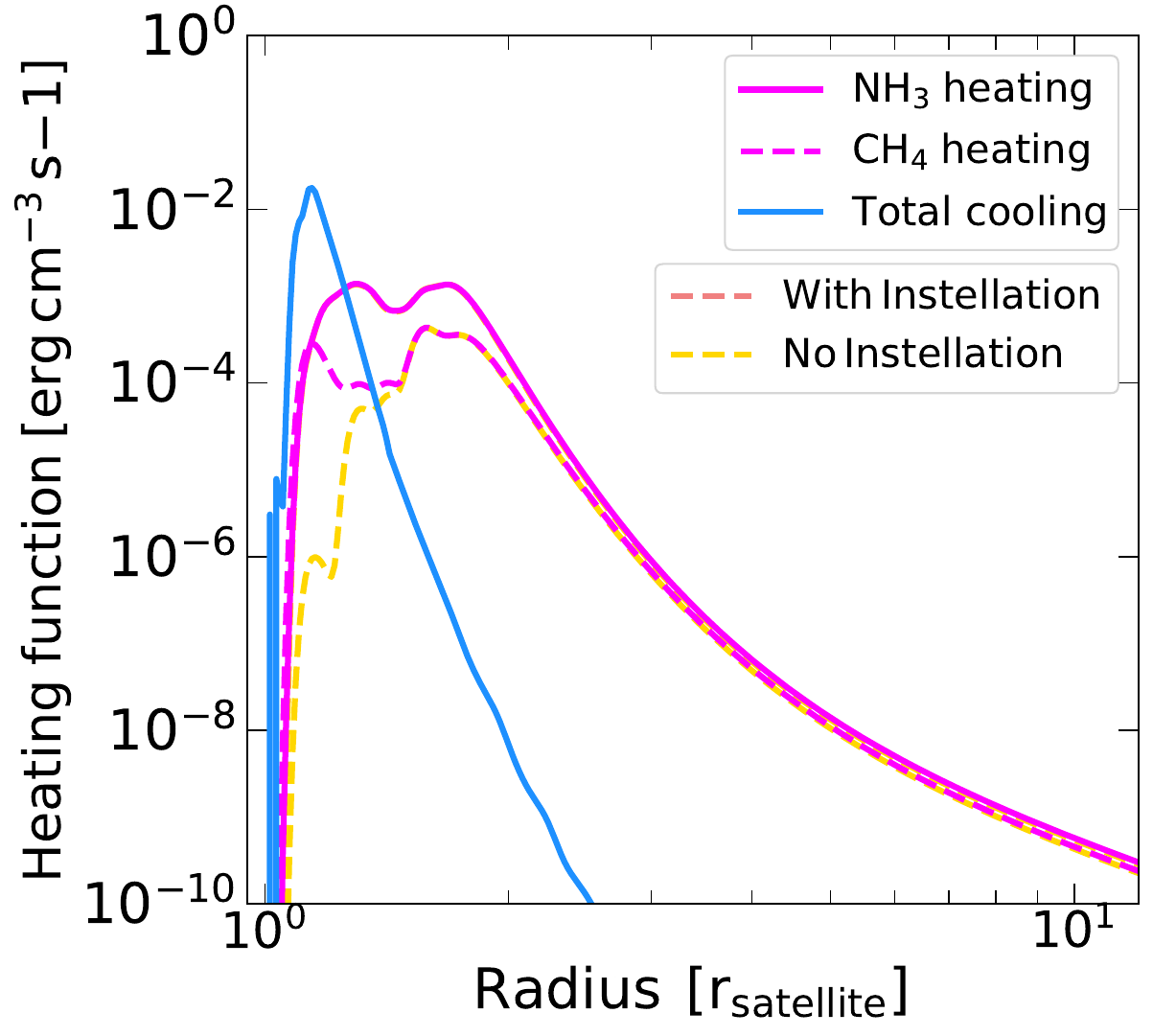}
\end{subfigure}

\caption{ Results from multi-band ($N_b$=80), two-species simulations\ms{, while using a mass mixing ratio of 1$:$1}. \textbf{Left}: By plotting $T_{\rm skin}$ for both species, we emphasize that temperature inversions remain existent. As in Fig. \ref{fig:gamma_vs_mdot}, we varied the giant planet's irradiation temperature while keeping the top-of-the-atmosphere flux constant. As the irradiating black-body matches the opacity function maximum of the atmospheric absorbers, the hottest temperatures and most efficient energy-to-escape rate conversion is reached. {Mass-loss rates are exponentially sensitive to the entire temperature profile, and are lower in the $N_b=80$ models than in the $N_b=1$. To reach similar atmospheric lifetimes, one needs to employ $T_{\rm surf}>300$K, applied in the following sections.} \textbf{Middle} and \textbf{Right}: $T_{\rm skin}$ is any species' optically thin temperature, i.e. $T(r\rightarrow\infty)$, {which can differ between  species due to insufficient collisional heat exchange. } The shown temperature and heating profiles are for $T_{\rm irr}=1200K$. The region heated by the star's black-body is marked orange, T($\rm r=r_{\rm satellite}$) and is slightly above the value set for $\rm T_{\rm surf}=200K$ due to residual radiative diffusion. {We distinguish between the ``stellar heating'' and ``giant planet'' heating regions in order to emphasize their different absorption altitudes, leading to efficiency differences, while the ``optically thin'' region is heated by both.} 
The sonic points in the molecular outflows are marked with crosses. The cooling function is computed from the combined Rosseland-mean opacity of both species. }
\label{fig:efficiency_twospecies}
\end{figure*}

\section{Full models}
\label{sec:fullmodels_section}

While crucial to understanding the physics of temperature inversions, the single-band models from the last section do not predict the correct magnitude of atmospheric escape rates (right panel, Fig. \ref{fig:gamma_vs_mdot}). Furthermore, multiple species can exchange both momentum and internal energy, leading to the acceleration of slower species and the heating of colder species. Thus,  we discuss how the effects of multiple species and multiple radiation bands impact the occurrence of temperature inversions and escape efficiency introduced in the previous section.
We then address whether the cooling of the giant planet can outcompete the mass-loss from lunar bodies by studying the coupled system in long-term evolution simulations.

\subsection{The physics of multi-band and multi-species models}
\label{sec:fullmodels}

So far, we discussed simplified models with double-grey mean opacities (i.e. $N_b$=$1$ \ms{with one in, and one outgoing, radiation band}) and single-species simulations. 
However, multi-species atmospheres will generally feature overlapping opacity bands. This can lead to shadowing effects (see Fig. \ref{fig:appendix_densityvariation}) or simply to an overestimation of the absorption altitude of {incoming} radiation. Hence, typically, $N_b \gg 1$ will be required for more realistic simulations. In our two-species model, we find numerical convergence of the mass-loss rates at $N_b=40$ \ms{with log-equidistant spacing} in the wavelength interval $\lambda \in [0.1:50] \mu$m, but run with $N_b=80$ to ensure convergence (See Fig. \ref{fig:appendix_convergencetests}). The most important conclusion from our tests with spectrally \ms{resolved models is that they give lower temperatures in the optically thin parts of the atmosphere and lower mass-loss rates than those given by a single-band simulation. Thus, while useful for understanding the physics, models with a single value of $\gamma$ do not provide an accurate representation of the mass-loss rates.} 

\ms{In our simulations, the temperature at the satellite's surface ($T_{\rm surf}$) is set to be in thermal equilibrium with the atmosphere. The atmosphere's temperature structure is then relaxed towards the adiabat of the non-condensible gas mixture \citep{graham2021} of $\rm NH_3$ and $\rm CH_4$, which we fully include in the simulation domain. The radiative-convective boundary $r_{\rm rcb}$ is then found self-consistently according to the Schwarzschild criterion. Beyond $r_{\rm rcb}$, the temperature structure follows the radiative solutions of Eqn. \ref{eq:fullequations_energy}. The physics determining the surface temperature is the formation time of the satellite \citep{kuramoto1994}, where this value can be hotter than the atmospheric temperatures at young ages. Given a fixed mass in the atmosphere, this surface temperature controls the density structure of the entire atmospheric density column and, hence, ultimately controls the mass-loss rates.}


While the spectrally resolved simulations are non-isothermal, we find the skin temperature is still a sensible proxy of the mass-loss rates, albeit important differences exist. This result is demonstrated in Fig. \ref{fig:efficiency_twospecies} (Left) where we show the skin temperatures of $\rm NH_3$ and $\rm CH_4$ and their mass-loss rates, normalized to the Parker wind-solution $\dot{M}_{\rm Parker}(T_{eq})$ evaluated at $T_{eq}= (F/(4 \sigma_{SB}))^{1/4}$. 
Normalizing to the Parker mass-loss rate is a more sensible comparison metric for the efficiency of mass-loss rates in multi-band simulations than measuring an effective $\gamma$.
In Fig. \ref{fig:efficiency_twospecies}(Left), only $T_{\rm irr}$ is varied, while the luminosity ($L=4\pi r^2_{\rm Giant}\sigma_{SB} T^4_{\rm irr}$) is kept constant. Essentially, the spectrally integrated top-of-the-atmosphere flux $F=L/(4\pi d^2)$ is kept constant, while the giant planet to satellite semimajor-axis $\rm d=0.00475\;AU=10 r_{Jup}$, is kept constant throughout this experiment. This experiment has $\zeta<1$ although it is $\gamma>1$, because the non-isothermal escape rates now additionally depend on the surface temperature of the satellite $T_{\rm surf}$, and the 'intermediate' temperature, which is close to $T_{\rm eq}/2^{1/4}$ (see the middle panel in Fig. \ref{fig:efficiency_twospecies}).

We notice that the maxima of $\dot{M}/\dot{M}_{\rm Parker}$ are reached for both species at the same value of $T_{\rm irr}\approx 400K$, while the maxima in $T_{\rm skin}$ are reached at different temperatures for both species, around $T_{\rm irr}\approx 350K$ for $\rm NH_3$ and $T_{\rm irr}\approx 900K$ for $\rm CH_4$.
This result arises due to collisional heat exchange in the middle atmosphere that sets an average temperature and a baseline escape rate (see the middle panel in Fig. \ref{fig:efficiency_twospecies}), which is modified by the collisionally decoupled temperatures in the upper atmospheres for both species.
We note from the data in Appendix \ref{fig:appendix_densityvariation}, where the amount of $\rm NH_3$ and $\rm CH_4$ in the atmospheric gas mixture is varied, that $\rm NH_3$ has a net cooling effect on those mid-atmospheric temperatures, which is why the escape rates for both species are lower at a $1:1$ mixing ratio than when $\rm CH_4$ is dominant.

{The cold temperatures relative to $T_{\rm eq}$ can result in extremely low mass-loss rates,  as would follow from Eqn. \ref{eq:parkermdott1.5} which explain the low efficiency. To reach efficiencies of $\approx 1$, one needs to employ $T_{\rm surf}>300$K, which we will be doing in the following sections.}
This is expected in multi-band simulations, \ms{on the one hand, as the coldest temperature $T_{\rm eq}/2^{1/4}$ can always be achieved due to cooling into two equal-sized hemispheres, and on the other hand nonlinear cooling effects of lines, which are often not captured by mean opacities, impact the temperatures further} \citep{parmentier2014, parmentier2015}, and are the key reason for the drops in mass-loss rates. As can be seen from the penetration depths of different black-bodies in Fig. \ref{fig:efficiency_twospecies} (Right) a fraction of the overall heating functions of star and giant planet overlap with the peak of the cooling function, and this radiation is immediately lost. The optically thin region is heated less for the same reason, as the temperature is sensitive to the sum $T^4_{\rm skin} \propto \sum_b F_b \;exp(-\tau_b)$, and the deeper penetrating wavelengths possess low $\tau_b$ in that region.

While thermal decoupling occurs in the upper atmosphere \ms{between different species}, drag \ms{by a hotter species in the lower atmosphere} can aid the escape of cold species. We demonstrate this explicitly using simple, controlled numerical experiments that disentangle those individual effects in Appendix \ref{sec:appendix_densies}.
This effect explains the co-location of the maxima in escape rates in the $\dot{M}/\dot{M}_{\rm Parker}$-curves:
The efficiently escaping $\rm CH_4$, with its mass-loss rate set by its own thermal profile, drags the less efficiently escaping $\rm NH_3$ up and down in mass-loss rates.

Finally, we note that efficient hydrodynamic escape with $ \dot{M}/\dot{M}_{\rm Parker}>1$ remains a possibility \ms{in multi-band models}
if any host giant planet can deliver enough luminosity at $T_{\rm irr}\approx400$K for sufficient time, or if $T_{\rm surf}$ is increased, as we shall see in the next sections. We now move on to quantify the escape rate over time as the irradiation flux from the giant planet decreases.

\begin{figure}
\begin{subfigure}{0.43\textwidth} 
   \centering
   \includegraphics[width=1.0\textwidth]{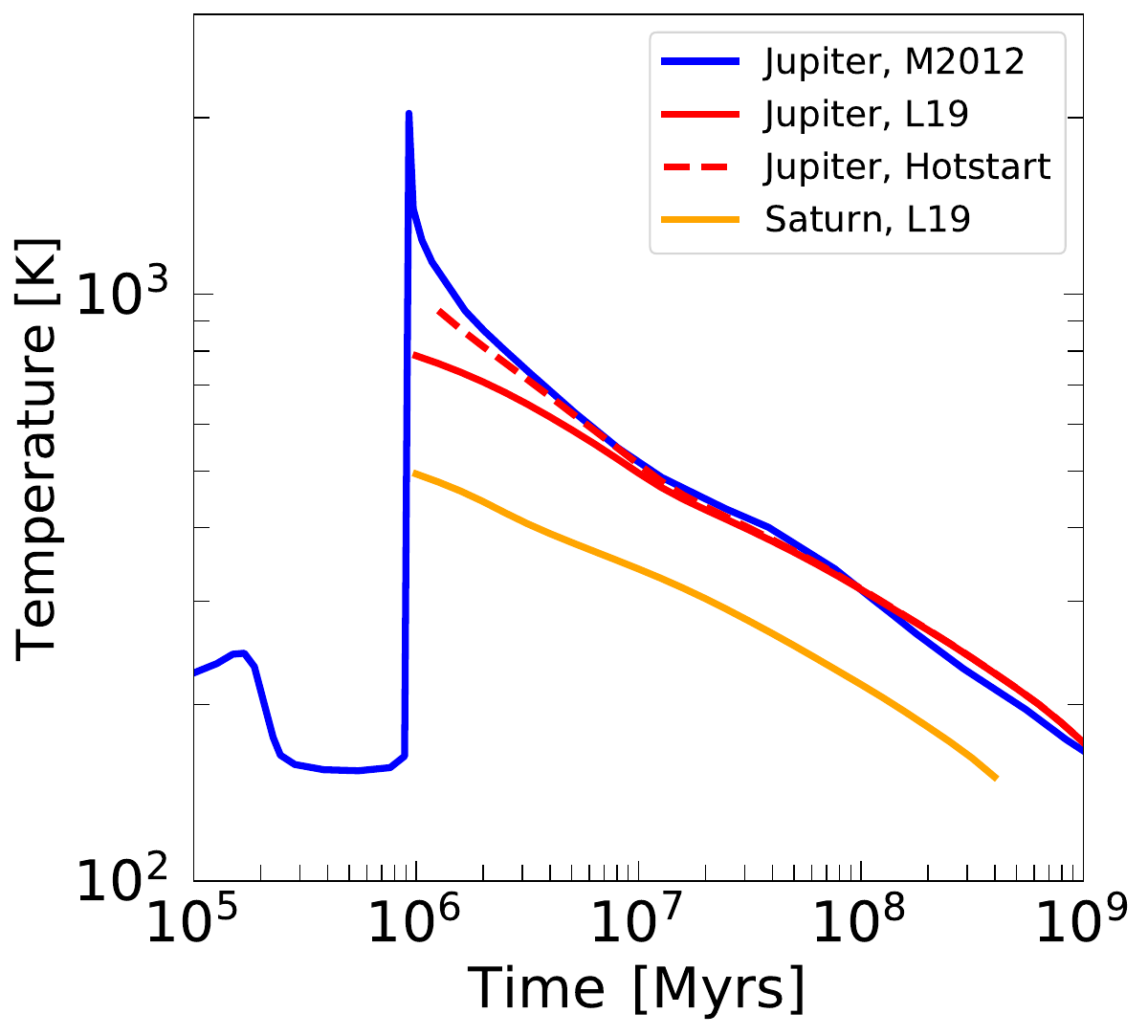}
\end{subfigure}
\caption{Cooling curves for the solar system gas giants taken from the models of \citet{mordasini2012} and \citet{linder2019}. They demonstrate the giant planet's irradiation temperature declines as a function of time. Directly before the occurrence of the runaway accretion shock (peak in the blue curve), the giant planet detaches from its parent disc while simultaneously passing through a phase of rapid contraction. Subsequently, it cools on a Kelvin-Helmholtz timescale and thus irradiates any orbiting satellites.}   
\label{fig:mordasini_curves}
\end{figure}

\subsection{Initial conditions,  \text{composition} and hydrodynamic setup}

It is impossible to follow the evolution of the lunar atmosphere in a fully time-resolved way within a single hydrodynamic simulation because outflows reach steady states on timescales of $10^5$-$10^6$s, much shorter than the thermal evolution times of the satellite body. Thus, we use a ``snapshot'' approach. Each hydrodynamic simulation we perform is a snapshot of the atmospheric mass-loss rate from a satellite, following the cooling curve of its giant planet host in time, see Fig. \ref{fig:mordasini_curves}. We then use the resulting mass-loss rate from this snapshot to evolve to a new evolutionary timepoint and calculate a new snapshot. Every snapshot at timestep $n$, and giant planet age $t_{\rm Giant}$ is fully described by the set of input parameters $(T_{\rm surf}, T_{\rm irr}, L, m_{\rm NH_3}, m_{\rm CH_4})^n$. 
Simulation results are mass-loss rates $(\dot{m}_{\rm NH_3}, \dot{m}_{\rm CH_4})^n$ and the thermal cooling flux emitted from the effective $\tau_R=1$ surface $F_{\rm surf}^n$. Therefore, an evolutionary sequence can be computed as a sequence of appropriately computed hydrodynamic simulations.

In order to compute the mass-loss rates, we set up \textsc{Aiolos} as follows.
We construct a numerically well-balanced hydrostatic atmosphere \citep{kappeli2014} at each \ms{evolutionary} timestep $n$ as an initial condition with the masses $(m_{\rm NH_3}, m_{\rm CH_4})^n$. This requires a guess for the temperature profile upon which the initial hydrostatic density is constructed. Thus, we use $T(r)=T_{\rm surf}$ as an initial guess at all radii in the atmosphere. Then, the external radiation is ramped up over timescales of a few seconds, creating an overpressure that eventually leads to the hydrodynamic escape of the atmosphere. After the mass-loss rates reach a steady state, we \ms{use} those values $(\dot{M}_{\rm NH_3}, \dot{M}_{\rm CH_4})^n$ \ms{to update both species masses in the atmosphere for the next time step.}

The solar system satellites serve to inform plausible scales for the hydrodynamic simulations, i.e. we use $\rm r_{\rm Satellite} =r_{\rm Titan}=2.63\times 10^8 cm$ as lower simulation boundary radius for all simulated exosatellites, $\rm r_{\rm Hill}=a \times \left(m_{\rm Satellite}/3 m_{\rm p} \right)^{1/3}$ {with the semimajor-axis distance from Satellite to Giant planet being $\rm a = 10 r_{\rm Jupiter}$, and $\rm r_{\rm max}=33\; r_{\rm Hill}$ as upper boundary radius of the simulation domain}. We use a satellite mass of $m_{\rm satellite}=0.0225 m_{\oplus}$, and a unit of ``atmospheric mass'' as $m_{A}=10^{-6} m_{\oplus}$, which is approximately the mass of the terrestrial atmosphere in $\rm N_2$ and the mass of Titan's atmosphere converted to $\rm NH_3$.
We initiate our calculations with either $1 m_{A}$ or $10 m_{A}$ mass in both atmospheric components $\rm NH_3$ and $\rm CH_4$.
Those numbers are motivated by estimates for Titan's initial mass content \citep[$\geq 1 m_{A}$][]{lammer2000, erkaev2021} and hence should provide sensible starting conditions for a general population of exosatellites.

We note that $\rm NH_3$ as the main carrier gas of nitrogen atoms has been speculated to be initially accreted on Titan \citep{mandt2014, gilliamlerman2014}. 
\ms{Noble gas isotopes in Titan's atmosphere indicate that it formed within the lifetime of the protosolar disk, and accreted a fraction of nebular gas \citep{glein2017}.} However, no matter the exact molecular form in which $N$ and $C$ atoms were accreted, under warm ($T_{\rm surf}<500 K$) and dense post-formation conditions, the satellites' atmospheric gas should re-equilibrate towards a composition of $\rm CH_4$ and $\rm NH_3$, \citep{heng2016}.
Should the atmospheres be outgassed from the interior instead of accreted, with initially different molecular compositions \citep{melwanidaswani2021}, this argument of re-equilibration at surface conditions should still apply. Therefore, we consider our choice of initial molecules a plausible starting point for studying atmospheric escape from exosatellites. As we have noted before, this choice of composition imparts an effective $\gamma>1$ for both components. Thus, this choice allows us to explore the maximum limits of bolometrically driven escape by giant planets. We do not include $\rm H_2O$ in the atmosphere. This would be required if one were to aim for a more faithful representation of the solar system satellites and should be considered in future work. We furthermore focus on the escape of neutral species only, unimpeded by MHD effects. 
{We also note that we do not model any chemical processes, either in the atmosphere or in an interaction with the surface, in this work. The only way that the composition of the atmosphere in our work evolves is via fractionation, i.e. if the mass-loss rates of the two species are differing relative to each other. }

Lastly, we describe how we choose our initial value of $T_{\rm surf}$ and put timescales into context.
\ms{The formation of a gas giant takes between $10^5$ and $\sim 10^6$yrs (the blue curve in Fig. \ref{fig:mordasini_curves}), mostly determined by the slow hydrostatic gas accretion phase. During the fast runaway collapse of the giant, the Hill-sphere is evacuated, and there is enough space for a circumplanetary disc to grow \citep{ayliffe2009b, tanigawa2012, schulik2020}. In this disc, moons take between $10^3-10^6$yrs to accrete 
\citep{mosqueira2003, canup2006, barr2008}. }In the model of \citet{kuramoto1994}, a formation time of $10^5$yrs corresponds to a surface temperature for a Titan-mass satellite $m_{\rm Satellite}=0.0225 m_{\oplus}$ of $\rm T_{surf}=350 K$, which we take as a nominal value. Hotter initial temperatures of $\rm T_{surf}=450 K$, and $\rm T_{surf}=550 K$ correspond to shorter formation times of $\approx 3\times 10^4$yrs and $10^3$yrs in their work.

For the giant planet cooling models we use the simulation results of \citet{linder2019}, which start just after the detachment and planet contraction phase has begun. Thus, our initial time corresponds to $t_0=t_{\rm detach} + t_{\rm formation}$ in absolute time, but $t_0=t_{\rm formation}$ in the \citet{linder2019} model time, and effectively $t_0\approx t_{\rm detach}$ as the formation time is typically short against the giant planet cooling time. 
Therefore, we assume that $t_{\rm formation}\ll t_{\rm cool}$, which is our reasoning for the choice of $t_0$. Nonetheless, $t_{\rm formation}$ is important as a parameter for choosing the initial surface temperature. This means that our evolution models also represent the maximum irradiation flux possible, and our computed mass-loss rates are to be seen as an upper limit.

\begin{figure*}
\hspace*{-0.5cm}
 \begin{subfigure}{1.05\textwidth} 
   \centering
   \includegraphics[width=1.0\textwidth]{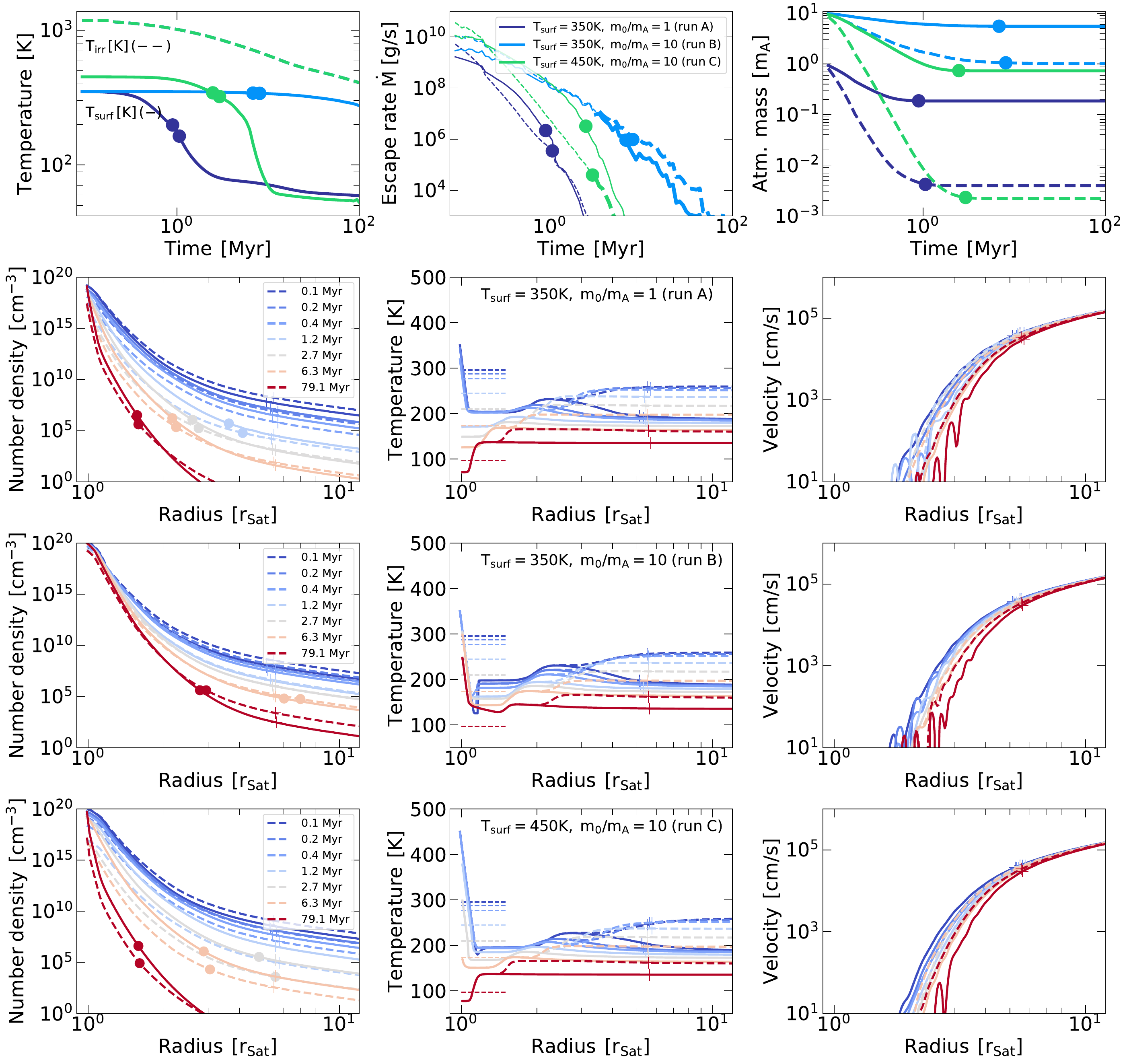}
\end{subfigure}%

\caption{Evolution of an example model for satellites orbiting a $m_{\rm p}$ = $2m_{\rm Jupiter}$ giant. Parameter variations are the initial satellite's surface temperature ($T_{\rm surf}(t=t_0)$) and the initial atmospheric mass, {which is equal in both $\rm CH_4$ and $\rm NH_3$. Dashed curves denote the properties of $\rm CH_4$ and the solid curves denote $\rm NH_3$.
We mark the mass-loss rates according to their efficiency relative to Parker-wind mass-loss rates, Eqn. \ref{eq:zetaefficiency}, where inefficient mass-loss is denoted as thin line, and efficient mass-loss is denoted as thick line. Large dots in the time-dependent plots denote the moment or position when/where the atmosphere becomes non-collisional, according to Eqn. \ref{eq:meanfreepath}. While any mass-loss, independent of efficiency, essentially ceases past this point, we plot the Knudsen number-uncorrected mass-loss rates for reference purposes.
Models with large initial lifetimes, $m/\dot{M}$, (run B, light blue) barely lose any significant mass. Models with short atmospheric lifetimes (run A and C, dark blue and green) models differ in their evolution (see discussion in text). The radial plots show the important hydrodynamic variables, where crosses denote the sonic point, the large dots denote the exobase radius and short horizontal and dashed lines in the temperature plots denote $T_{\rm eq}$ as a reference at the given point in time (indicated by the line colour). The relation of the temperature profile and $T_{\rm eq}$ at any given point explains efficient or inefficient escape.  } } \label{fig:evolution_nominal}
\end{figure*}

\subsection{Evolutionary calculations}
\label{sec:evolution}

After running the hydrodynamic simulations to steady state, we use the simulation outputs to evolve the satellite and its atmospheric mass on\ms{e} evolution step forward and the cycle restarts.

Each species' $s$ atmospheric mass reservoir $m_s$ in the atmosphere is evolved via an implicit evolutionary timestep of length $\Delta t$ according to

\begin{align}
 \frac{    m^{\rm n+1}_s - m^{\rm n}_s}{\Delta t} = - \dot{M}^{n+1}_s \approx- c^n \times  m^{\rm n+1}_s,
 \label{eq:evolution_mass}
\end{align}

where the last step follows from the understanding that Parker-like solutions behave like $\dot{M}\propto m$, which lets us use the atmospheric mass change ratio at timestep $n$, which is is $c^n \equiv  \dot{M}^{n}_s / m^{n}_s$.
\footnote{Therefore, this numerical prescription would be invalid in the regime where $\gamma \gg 1$.}
Thus, the resulting implicit update is
\begin{align}
 m^{\rm n+1}_s = \frac{m^{\rm n}_s}{1+\Delta t \; c^n},
 \label{eq:evolution_mass2}
\end{align}
which is unconditionally stable for large \ms{fractional} mass-loss rates, motivating our choice of this implicit method.
The timesteps are chosen as a small fraction of the giant planet's cooling time, resulting typically in $\approx50-100$ steps over the evolutionary history ($\leq 100 Myr$ in our case).

\ms{The mass-loss rates depend on the entire temperature profile, including  $T_{\rm surf}$, and }
radiative losses of the satellites' thermal energy continue in the post-formation phase. Thus, we also need to compute the temperature evolution of the satellite.
We solve a simple cooling equation for the solid core, assuming its internal thermal profile is isothermal, i.e. $T(r)=T_{\rm surf}$ for all $r<r_{\rm surf}$. The energy equation, integrated over the satellites' volume, then yields:
\begin{align}
    \frac{c_v}{4\pi} m_{\rm satellite} &\frac{T^{n+1}_{\rm surf}-T^{n}_{\rm surf}}{\Delta t} = \nonumber \\
    &-r^2_{\rm surf} \left( F_{\rm surf} \times \frac{1}{1+3\,\tau_{\rm RCB}} - \frac{1}{4}\sum_b S_{\rm b} \exp(-\tau_{\rm b,0}) \right)
    \label{eq:satellite_core_temperature_evolution}
\end{align}
with $F_{\rm surf} = \sigma/\pi (T^{n+1}_{\rm surf})^4$ and $c_v$ the heat capacity at constant volume of the satellite's body. The heat capacity of the lunar material is $c_v= c_{\rm Silicate}$ for rocky, silicate satellites \ms{which we consider in this work}, where we take value provided by \cite{johansen2022a} of $c_{v}=1.2\times 10^7 \rm \;erg/g/K$. 
\ms{The first term on the r.h.s allows for a reduction of the core cooling by atmospheric blanketing, which includes $\tau_{\rm RCB} $. 

Here, $\tau_{\rm RCB}$ is the local Rosseland mean optical depth at the radiative-convective boundary, which \ms{we extract from our hydrodynamic simulations once they have reached steady-state}. To ensure thermal equilibrium between the core and the atmosphere, we set the temperature of the first active simulation cell to $T_{\rm surf}$, and then enforce an adiabatic two-species temperature profile \citep{graham2021} between each subsequent cell until the radiative-convective boundary is found according to the Schwarzschild criterion. The time-dependent radiation transport remains active throughout this process. The second term on the r.h.s includes irradiation, which dominates when the atmosphere becomes optically thin at the satellite surface. This term adds up the giant planet irradiation reaching the satellite's surface, where optical depths $\tau_{b}$ are measured from the top of the atmosphere down to the surface.}

As is clear from Eqn. \ref{eq:parkermdott1.5}, the cooling of the satellite provides a shutdown mechanism for hydrodynamic escape: as the entire atmospheric column is inflated or deflated together with the lower atmosphere at $T=T_{\rm surf}$, satellite cooling can decrease the density at the sonic point drastically, \ms{suppressing mass-loss}.

\ms{The upper atmosphere provides another pathway to shut down mass-loss, which appears when} the gas outflow is not hydrodynamic any more, a situation explored in the works by \cite{johnson2013, volkov2013, volkov2016, volkov2017}.
\ms{For this, the} key parameter to consider is the Knudsen number, encoding how well coupled the individual fluid particles are to the collective behaviour of the gas and whether they feel the pressure gradients from the lower, overheated atmospheric layers. Taking this into account, each species' escape $\dot{m}_s$ rate is corrected by a heuristic prescription using the species' Knudsen number at the sonic point ${\rm Kn}_{cs}$ via 
\begin{align}
    \dot{M}'_s = \dot{M}_s \frac{1}{1+{Kn}_{cs}} 
    \label{eq:knudsen_correction}
\end{align}
which is inspired by \citep{volkov2017}. Therefore, as the ``source'' Knudsen number increases (with decreasing atmospheric mass), the escape rates show a characteristic drop as their upper atmosphere becomes collisionless.
The details of this heuristic correction should not matter greatly, as any reduction of the mass-loss rates proportional to $1/Kn$ (or stronger) will decrease the mass-loss rates rapidly, such that the atmospheric lifetime $m_{\rm atm}/\dot{m}$ increases more than linearly and the escape is effectively shut down on evolutionary timescales.
%
%

The Knudsen number is computed according to
\begin{align}
    Kn(r) = \frac{l_{\rm s}}{H_s}
    \label{eq:knudsennumber}
\end{align}
where $H_s$ is the local gas scale height of a species, and the mean-free path $l_s$ is computed as
\begin{align}
    l_s = \left(\sqrt{2}\,\sum_{s'}^{\rm All\;\, spc} n_{s'} \sigma_{ss'}\right)^{-1}
    \label{eq:meanfreepath}
\end{align}
from all number densities $n_{s'}$ that a particle of species $s$ might encounter, including its own. Finally, the collision cross-sections $\sigma_{ss'}=\pi \left( r_s + r_s' \right)^2$, are obtained from the effective neutral collision radii of the molecules  \citep{schunk1980}, where $r_{\rm CH4}=3.8\times10^{-8}$cm and $r_{\rm NH3}=2.6\times10^{-8}$cm.

\subsection{Nominal cooling models - the impact of surface cooling and atmospheric mass}

In order to highlight the general \ms{features of our evolution models, we first present an analysis of our baseline model results}, shown in Fig. \ref{fig:evolution_nominal}.
For this baseline model, we choose three satellites orbiting a super-Jovian planet ($m_{\rm p} = 600 m_{\oplus}$, a typical distant exoplanetary gas-giant mass, \citealt{Fulton2021}) at two different values of initial surface temperature and initial mass. The upper row shows the time evolution of the atmosphere's bulk quantities, \ms{temperatures, mass-loss rates and masses}, and the lower three rows show the detailed structure evolution of the atmospheres. 

\ms{The dots "$o$" on the curves mark the location of the exobase, and the crosses "$+$" mark the location of the sonic point. The transition from hydrodynamic escape to static atmospheres via Eqn. \ref{eq:knudsen_correction} can then be read-off as a crossing of the exobase inside the $r_s$ in the density plots. Additionally, the dot in the macroscopic quantities marks the point in time when the atmosphere becomes effectively hydrostatic. The reported mass-loss rates beyond this point do not affect the total mass inventory any more.

Generally, all our simulations end their mass loss when the outflows become non-collisional. However, this can be achieved in three different ways: by internal self-regulated shutdown (run A), by giant planet cooling (Run B) or by satellite cooling (run C).

The self-regulated shutdown in run A relies on the disappearance of the adiabatic region in the lower atmosphere, as this region can inflate the atmospheric column. This inflation leads to high densities at the sonic point and high mass-loss rates, provided the lower atmosphere is hot enough. 
The opacity source that sets the radiative-convective boundary $r_{\rm rcb}$ is mostly provided by $\rm NH_3$ in $\kappa_{\rm R}$, where the $\rm CH_4$ contribution is weak due to significant spectral troughs between its bands. 
Hence, after an initial 90$\%$ loss of a significant fraction of $\rm NH_3$, the adiabat disappears, the lower atmosphere becomes isothermal, and the atmospheric column collapses inward after less than $1$ Myr.

}

\begin{figure}
\hspace*{-0.0cm}
 \begin{subfigure}{0.40\textwidth} 
   \centering
   \includegraphics[width=1.0\textwidth]{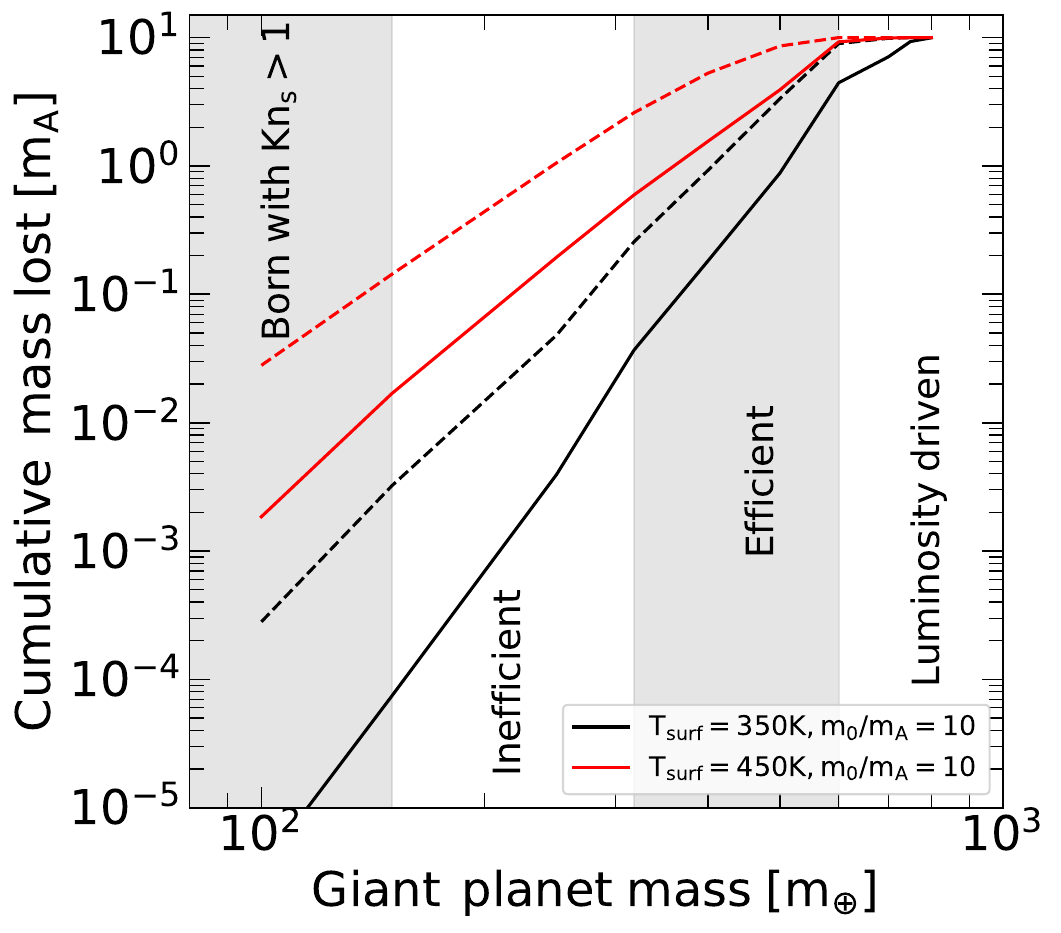}
\end{subfigure}%

\begin{subfigure}{0.40\textwidth} 
   \centering
   \includegraphics[width=1.0\textwidth]{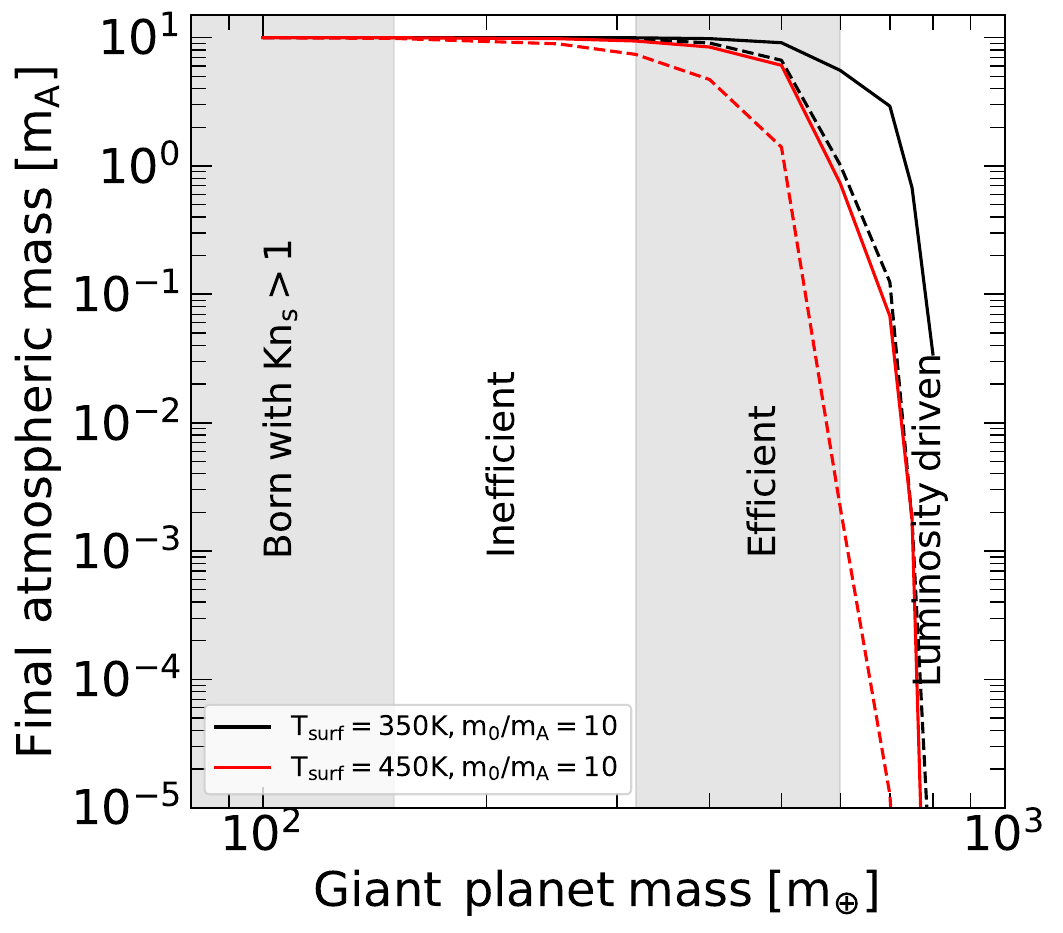}
\end{subfigure}%

\caption{The total mass lost and the remaining atmospheric masses of exo-satellites as a function of the giant planet host mass, once mass-loss ceases (essentially, $Kn_s>1$). Classification of escape regimes is valid for runs with $T_{\rm surf}=350$K, as detailed in the text. The dashed lines indicate results for $\rm CH_4$, and the solid lines indicate results for $\rm NH_3$.
} \label{fig:exosatellites_hostmass}
\end{figure}

\ms {However, the physics of self-regulated shutdown does not happen exclusively in the atmosphere, as two feedback loops play a role in changing the satellite's cooling rate.
As the surface cooling rates are sensitive to $\tau_{\rm rcb}$, loss of optical depth can lead to runaway cooling of the surface, shutting off further escape.
The second loop concerns the functional form of $\kappa_R(T)$. The Rosseland optical depth in the gas mixtures decreases for decreasing temperatures, i.e. $\partial \kappa_R(T)/\partial T > 0$, as line broadening effects weaken for colder temperatures at the same pressure. This also impacts $\tau_{rcb}$ and leads to faster cooling the colder the surface is.
In Run A, however, those effects are slower than self-regulation, as mass-loss continues until $\sim 1$Myr, although the satellite is already cooling. 

It is instructive to compare the radial temperature structures to run B, where the convective region remains present, and the upper atmosphere receives an identical amount of radiation.}
In run B, we increased the starting mass to 10 times larger in both components. This leads to high optical depths at all times, which is high enough to keep the satellite surface temperature nearly constant and the adiabatic region present.
The mass-loss rates are higher compared to run A, but ultimately, it is the cooling of the giant planet that shuts down escape via cold upper atmospheric temperatures.
Inspecting the escape efficiency, we see that regime changes into the high-efficiency regime, i.e. $\dot{M}/\dot{M}_{\rm Parker}>1$, as would have been expected from our previous discussion in Fig. \ref{fig:efficiency_twospecies}.
However, the total mass loss in the high-efficiency regime is lower than that lost at low efficiencies (i.e., between 0 and 8 Myrs) because the giant planet's luminosity has dropped by that time more than the efficiency can compensate for. 
This can change for higher $T_{\rm surf}$, as is shown in Fig. \ref{fig:exosatellites_hostmass} and in run C.

Run C starts hotter ($T_{\rm surf}=450K$) than the previous runs and shows very high mass-loss rates for a longer time than run A. This stems from the absolute amount of $\rm NH_3$, which is higher compared to run A, keeping the satellite hotter for longer, which retains the convective region, keeping the gas mixture collisional for longer. Those combined effects lead to the final mass in $\rm CH_4$ in run C being lower than that for run A. 
A gradual increase in mass-loss efficiency due to giant planet cooling contributes very little to the overall atmospheric mass-loss rates. 

{All mass-loss rates past reaching the $Kn_s=1$ point, when the upper atmosphere at the sonic point becomes collisionless, are corrected to lower values according to Eqn. \ref{eq:knudsen_correction}. The evolution plots only show the non-corrected values, but the mass is evolved with the corrected quantities.. We run the simulations nonetheless to 100Myrs, in order to track the cooling and contraction of the lower atmosphere.}

All runs show higher mass-loss rates in $\rm CH_4$ than for $\rm NH_3$, showing that the relative ordering in mass-loss rates from the previous analysis is maintained, i.e. $\rm CH_4$ is preferentially lost. The relative ordering of mass-loss rates relative to each other and to the Parker-wind rates, \ms{can be understood more in detail} from the middle panels in Fig. \ref{fig:evolution_nominal}, where the short, dashed lines on the left side of the temperature profiles indicate $T_{\rm eq}$. In the spirit of the previous discussion, where we used \ms{Parker wind escape rates} to classify mass loss as \textit{efficient} or \textit{inefficient}, according to Eqn. \ref{eq:zetaefficiency}.

When the atmosphere attains very low mass and becomes transparent, \ms{the escape rates are} not sensitive to the surface temperature any more, resulting in a visible discontinuity between $T_{\rm surf}$ and the radiatively dominated rest of the atmosphere. We also note that due to our choice of $T_{\rm surf}\geq 350K$, the escape efficiencies of the escaping atmosphere are increased over the results shown in Fig. \ref{fig:efficiency_twospecies}, which used $T_{\rm surf}=200 K$.

\section{Population study - Role of the giant planet host mass for the final atmospheric exosatellite mass}
\label{sec:exosatellites}

We now move to investigate whether exo-satellites might still possess significant atmospheres,  depending on the mass of their giant planet host, and we attempt to generalize the understanding of different shut-down mechanisms into this satellite population.

\begin{figure*}
\hspace*{-0.0cm}
 \begin{subfigure}{0.95\textwidth} 
   \centering
   \includegraphics[width=1.0\textwidth]{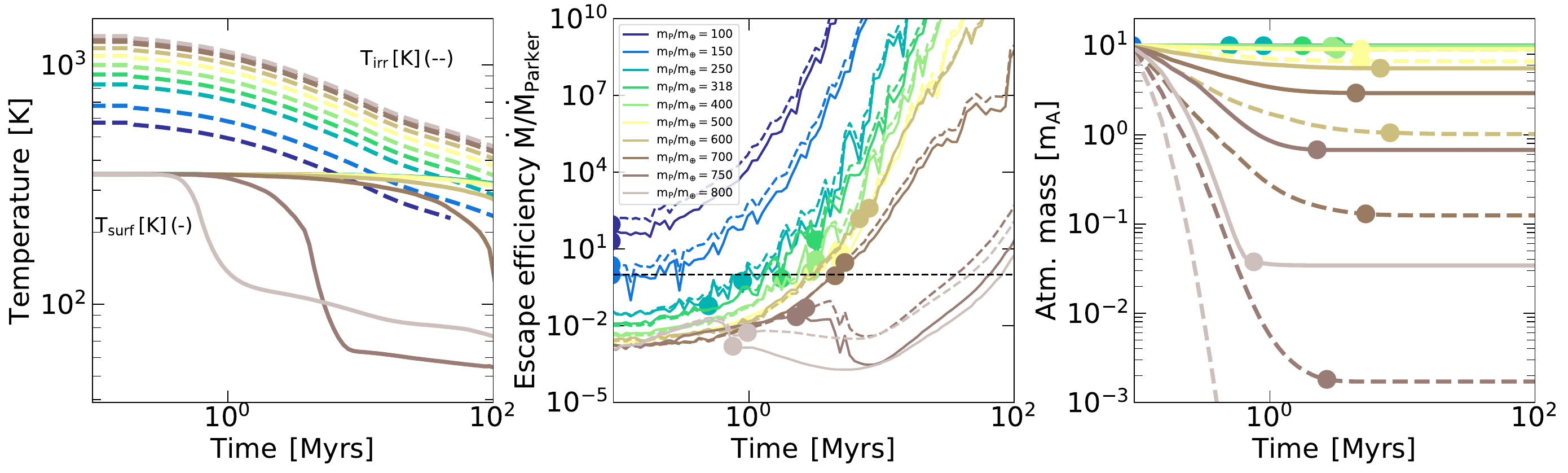}
\end{subfigure}%

\begin{subfigure}{0.95\textwidth} 
   \centering
   \includegraphics[width=1.0\textwidth]{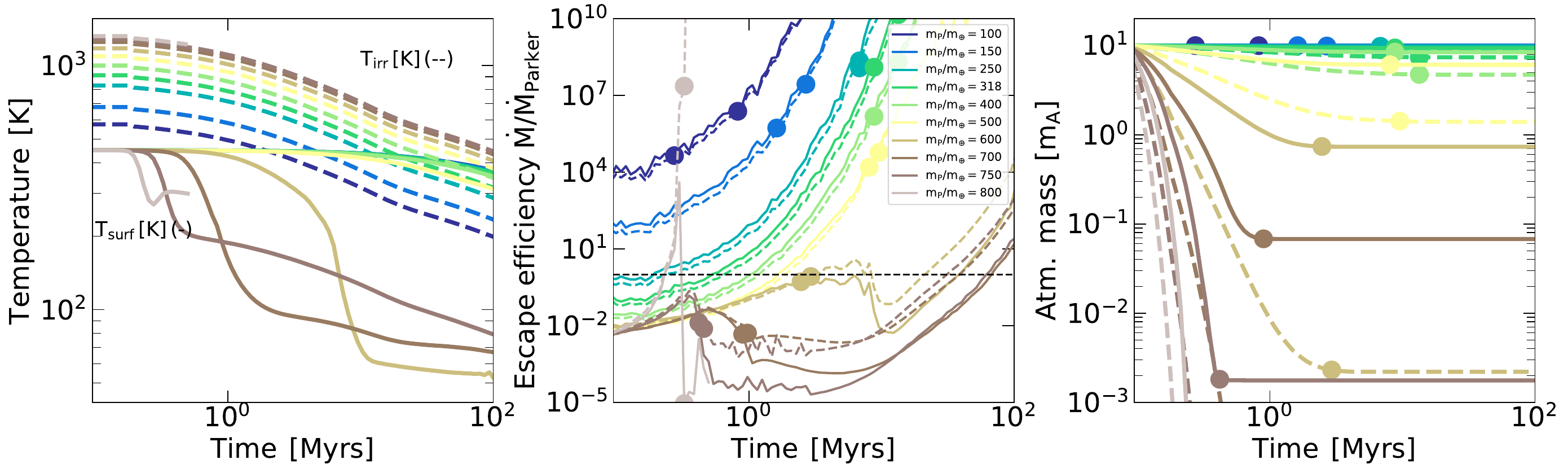}
\end{subfigure}%

\caption{Simulation results for different host giant planet masses (indicated by colour) to study a population of exosatellites. Results are shown for exosatellites starting with atmospheric masses $m_0/m_A=10$ and $T_{\rm surf}=350$K (top) or $T_{\rm surf}=450$K (bottom). In the left panel, dashed lines indicate the evolution of the irradiation temperature, while the solid lines show the evolution of the surface temperature. In the middle and right panels, the dashed lines indicate $\rm CH_4$, and the solid lines indicate $\rm NH_3$. Dots indicate the moment when any species reaches $\rm Kn_s >1$. \ms{ $\rm NH_3$ typically reaches this point earlier than $\rm CH_4$, as its colder temperature outcompetes its higher density  due to lower mass-loss rates. }
} \label{fig:exosatellites_hostmass_detail}
\end{figure*}

\subsection{Host mass and scaling parameters}

In order to determine whether there is an optimum host planet mass for the loss or retention of satellite atmospheres, we fit temperature and luminosity curves from the \citet{linder2019}-models for Saturn, Jupiter and the 2 $m_{\rm Jup}$ planet via 
\begin{align}
    T(m_{\rm p}, t) = T(1\,m_{\rm Jup}, t) \times \left( \frac{m_{\rm p}}{318 m_{\oplus}} \right)^{0.4}
    \label{eq:giant_t_scaling}
\end{align}
and
\begin{align}
    L(m_{\rm p}, t) = L(1\,m_{\rm Jup}, t) \times \left( \frac{m_{\rm p}}  {318 m_{\oplus}} \right)^{1.5}
    \label{eq:giant_l_scaling}
\end{align}

\begin{figure*}
\hspace*{-0.5cm}
 \begin{subfigure}{1.1\textwidth} 
   \centering
   \includegraphics[width=1.0\textwidth]{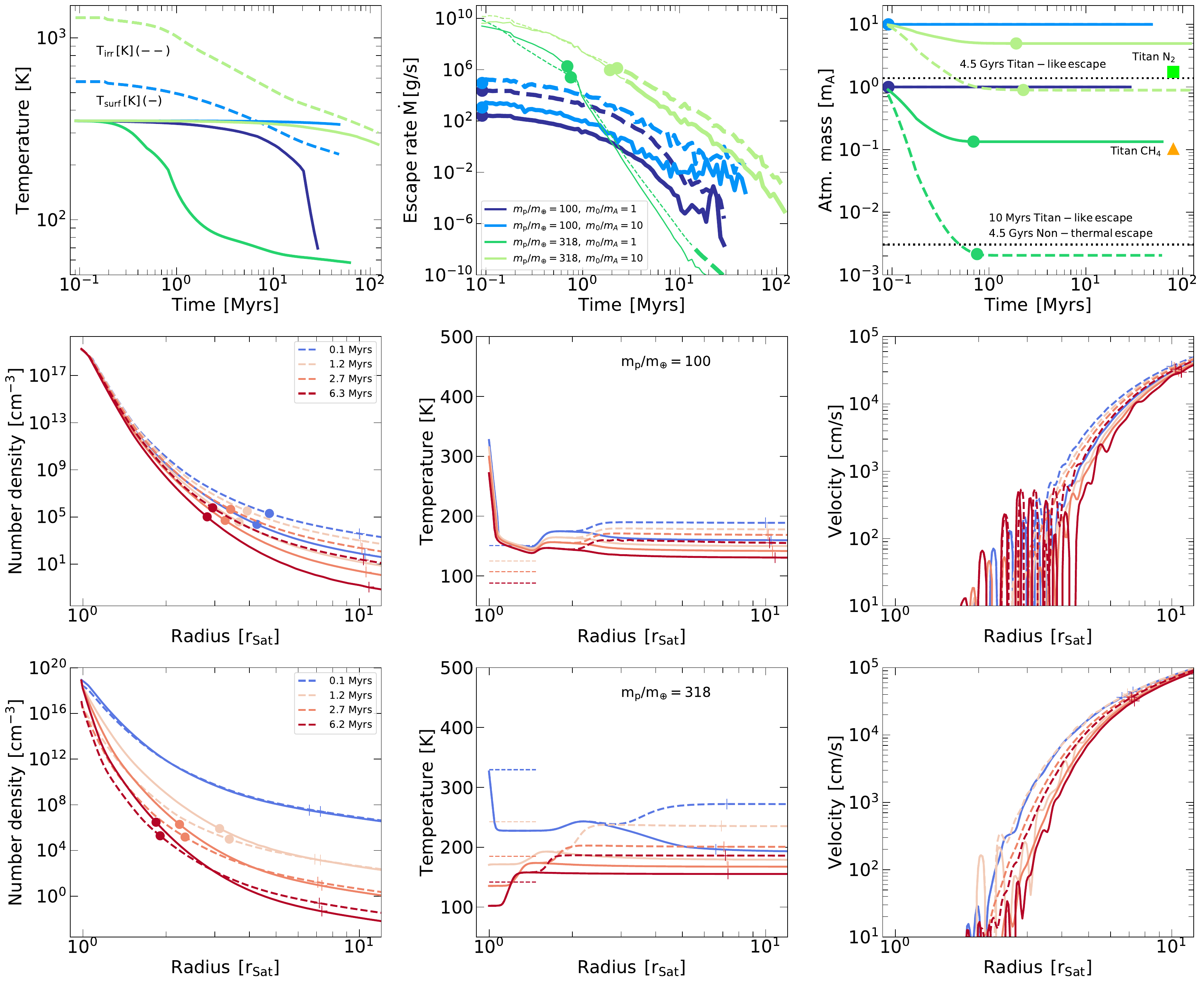}
\end{subfigure}%

\caption{Same as Fig. \ref{fig:evolution_nominal}, but for the evolution of solar system satellite analogues orbiting a hot-start Saturn-mass (blue colours) and a hot-start Jupiter-mass planet (green colours) for two different choices of initial atmospheric mass ($m_0/m_A$=1 \& 10 ) and our nominal choice of $T_{\rm surf}=350$ K. The black dotted lines in the top right panel denotes reference values and an estimate for how much atmospheric loss would occur over $10$Myr/4.5Gyr, based on currently inferred $\rm CH_4$ mass loss on Titan. The modern composition of Titan's atmosphere is marked in that panel with a green square ($\rm N_2$) and orange triangle ($\rm CH_4$) for reference. {The middle and lower rows show the structure evolution of the $m_0/m_A=1$ simulations. }} \label{fig:evolution_ganymede_titan}
\end{figure*}

\subsection{Escape regimes}

In Fig. \ref{fig:exosatellites_hostmass}, we plot the cumulative atmospheric mass lost and the remaining atmospheric mass in both species as a function of the giant planet host mass, for $T_{\rm surf}=350$ K and $T_{\rm surf}=450$ K. We firstly note that those values, apart from their linkage to formation times, \ms{simultaneously} delineate between models that might lead to significant mass-loss around 1$m_{\rm Jup}$ giants. For lower $T_{\rm surf}$, satellites orbiting 1$m_{\rm Jup}$ giants keep their atmospheres.

We define the categories into which we place those satellite atmospheres according to events in their mass-loss histories, which are shown in Fig. \ref{fig:exosatellites_hostmass_detail} (top row).

Discussing first the fates of the $T_{\rm surf}=350$K-satellites, we see that satellite atmospheres born around giants of $m_{\rm p}/m_{\oplus} \leq 150$ would start at high escape efficiency, but due to the low luminosity of these lower-mass giants, the efficiency effect is offset.
These effects result in sufficiently cold upper atmospheres so that the exobase is inside the sonic point, and the escape regime is not hydrodynamic, leading to a non-collisional shutdown.

Giant planets of higher masses $150 \leq m_{\rm p}/m_{\oplus} \leq 320$ emit larger luminosities which are converted into mass-loss at lower efficiencies. See in Fig. \ref{fig:exosatellites_hostmass_detail} how the initial mass-loss efficiencies at $t=0.1$Myrs drop with increasing giant planet mass, demonstrating these atmospheres are in the \textit{inefficient escape regime}.
The total mass lost in those atmospheres is up to 2$\%$ of their initial mass. 

The trend of decreasing initial mass-loss efficiency continues for higher giant masses at $320 \leq m_{\rm p}/m_{\oplus} \leq 600$. But in this mass range, the efficiency pops above 1 at late times before being shut off by the outflow becoming collisionless.
The total mass lost at high efficiencies is, however, lower than at low efficiencies, again due to the large luminosity drop. This is the \textit{efficient escape regime}, where the name is applied regardless of whether the phase of $\dot{M}/\dot{M}_{\rm Parker}>1$ contributes 
significant cumulative mass-loss or not. Both the inefficient and efficient regimes evolve in a manner similar to the nominal run B.

Finally, the giant planets which strip their satellites of atmospheres are situated at $ m_{\rm p}/m_{\oplus} \geq 600$. There, most of the mass is lost at low efficiencies, but due to high luminosities, the absolute mass-loss rates are high enough to strip the entire atmosphere before $\dot{M}/\dot{M}_{\rm Parker}>1$ is reached. This is the \textit{luminosity driven regime}. Those atmospheres evolve similarly to nominal run C, but their evolution transitions closer to being analogous to run A at $ m_{\rm p}/m_{\oplus} = 800$.

\subsection{Hotter starting conditions}

In Fig. \ref{fig:exosatellites_hostmass} (red curves) and Fig. \ref{fig:exosatellites_hostmass_detail} (bottom row) we show the results for $T_{\rm surf}=450$K. The hotter surface temperatures push the escape rates to higher values, according to the expectation from Eqn. \ref{eq:parkermdott1.5}, which leads to the first two regimes from the above discussion disappearing. All atmospheres now exist in the ``efficient'' or ``luminosity-driven'' regimes. 
Now, satellites orbiting $ m_{\rm p}/m_{\oplus} \geq 100$ giants are directly born into the ``efficient'' regime, but no significant mass-loss occurs until a giant mass of $ m_{\rm p}/m_{\oplus} = 400$.
Above $m_{\rm p}/m_{\oplus} = 600$ the satellite atmospheres enter the ``luminosity dominated'' regime, the same value as for lower $T_{\rm surf}$.

\subsection{Luminosity beats efficiency}

We conclude that because the efficiency is a weak function of $T_{\rm irr}$ (see Fig. \ref{fig:efficiency_twospecies}) and $T_{\rm irr}$ is a weak function of $m_{\rm p}$, but $L$ is a much stronger function of $m_{\rm p}$ (see Eqns. \ref{eq:giant_t_scaling} and \ref{eq:giant_l_scaling}), the total luminosity, {rather than the efficiency}, determines the overall retention of a satellites atmosphere.
However, the escape efficiency still differentiates between the evolution of the two species, ultimately determining the overall composition of the remnant atmosphere. We find $\dot{m}_{\rm CH_4}>\dot{m}_{\rm NH_3}$ for all our runs, which is a result of the general radiation transport physics discussed in previous sections, e.g. Fig \ref{fig:gamma_vs_mdot}, {and the temperature separation between the species in the upper atmosphere, which helps their vigorous fractionation}.

\section{Application to solar system exosatellite analogues}
\label{sec:ganymedetitan}

In this section, we aim to investigate whether it is possible to end up 
with a Titan analogue (N-dominated, $\rm CH_4$ subdominant) and a Ganymede analogue (no atmosphere)-like configuration among {extrasatellitary} systems. \ms{It is unknown whether Ganymede possessed an atmosphere and whether Titan's atmosphere is younger than the age of the solar system. However, we initialize both analogues with a Titan-like atmosphere to test whether the differences in luminosity from a $300m_{\oplus}$ and $100m_{\oplus}$ giant planet host might be consistent with their current atmospheric properties.}
We apply our model with identical initial conditions for both satellites, one orbiting a $m_{\rm p}=100m_{\oplus}$ and the other orbiting a $m_{\rm p}=318m_{\oplus}$ giant planet. Both atmospheric component gases start with the same mass.

From previous sections, it is clear that the escape of $\rm CH_4$ is always favoured over that of $\rm NH_3$. 
In \citep{gilliamlerman2014} the authors find a similar result, but while using different physics - in their work Jeans escape rates are employed and the preferential loss of $\rm CH_4$ results from the lower mass compared to $\rm NH_3$. In any case, to evolve into a Titan-like atmosphere, the total initial atmospheric mass for a Titan precursor scenario must be higher than the current atmospheric mass in $\rm CH_4$ on Titan. Also clear from previous results is that {Jupiter-mass hosts, assuming the nominal giant planet cooling curves, are situated on the edge of parameter space for significant atmospheric removal, i.e. do not readily loose all their atmosphere.}   

Therefore, we explore the ``hottest'' possible scenarios for both lower and upper atmospheric temperatures to test those final atmospheres against compatibility with the solar system satellites. 

In practice, this means that we have exchanged the giant planet cooling models employed so far, which are situated in the category of ``cold start'' models, against ``hot start'' models.
The classical idea of a cold start model is that in which the accretion heat of the giant planet is efficiently radiated away during the runaway accretion stage and hence is unavailable at the time when the satellites have formed \citep{marley2007}. Hot start models are the opposite and retain a significant amount of heat during the accretion process.
\cite{marleau2019} indicated that such models are viable options using detailed calculations of the giant planet accretion shock, and \citet{owen2016} showed they could arise through boundary-layer accretion. As the name indicates, those models provide hotter and more luminous initial conditions. They start at $T_{\rm irr}\approx1300$K, relative to the nominal Jovian models at $T_{\rm irr} \approx 910$K, and therefore output a $\sim 4$ times higher irradiation flux onto the satellite atmospheres.

We also explore an upper limit for $T_{\rm surf}$, which is about $T_{\rm surf}=550$ K, corresponding to the shortest possible formation times in \citep{kuramoto1994} of $t_{\rm form}\approx 10^2-10^3$ yrs. In Fig. \ref{fig:evolution_ganymede_titan}, we show the results of these calculations with $T_{\rm surf}=350$ K.

Generally, all satellite atmospheres with $T_{\rm surf} \in [350; 550] K$ orbiting a hot start Saturn-analogue show too little escape to alter their initial compositions or total atmospheric masses, even when the early mass-loss rates are in the hydrodynamic regime. The maximum relative mass lost in the case of $T_{\rm surf} = 550 K$ is about 13 $\%$. Hence, while this result shows that our scenario is consistent with the existence of an atmosphere around Titan-analogues, we would require higher mass-loss rates to \ms{fractionate} $\rm NH_3$ and $\rm CH_4$ into their solar system ratios. The latter issue might, however, be solved by invoking a different birth composition of the atmospheres \citep[e.g.][]{bierson2020}, which we do not explore here. {We note that the latter authors invoke Parker-wind mass-loss rates, which as we have shown in this work, can both over and underestimate the impact of mass-loss in the bolometrically driven regime. This should be addressed by future work exploring wider compositional ranges.}

The Ganymede-analogue readily loses most of its atmosphere when irradiated by a hot-start Jovian host, {even for $T_{\rm surf}=350$K}. Any remaining atmosphere might be removed by kinetic processes over the lifetime of the solar system. {To estimate the latter more quantitatively, we took the published escape rates for $\rm CH_4$ on Titan in (\citealt{mullerwodarg2008, yelle2008}, $\rm \sim 2.3\times10^{27}\,s^{-1}$ or $\rm 6.1\times10^4\,g\,s^{-1}$). The $\rm N_2$ escape rates are poorly constrained, therefore, we do not compare with them. The $\rm CH_4$ escape rates are surely not driven by giant planet bolometric radiation at the current time, but rather by UV-heating \citep{strobel2008a}, where those authors also note that the rates on Titan are $\sim 10^8$ larger than Jeans escape rates, and also $\sim 10^3$ times larger than non-thermal kinetic processes. Those escape rates are not directly measured, but rather inferred via density profile matching in the lower atmosphere, where degeneracies with atmospheric mixing exist (see an extensive discussion in \citealt{mullerwodarg2014titanbook}). Given these difficulties, we nonetheless adopt those numbers as maximum estimate of non-bolometric escape. \\
Next, we extrapolate the total mass-loss over 10 Myrs and over the age of the solar system. The extrapolation over the age of the solar system reveals that under the current mass-loss rate, Titan's atmosphere should not exist. This conundrum might be artificial due to Titan's atmospheric time-variability \citep{Hsu2019}, resupply effects from the geosphere \citep{strobel2012, mandt2014} or even worsened due to photodestruction of $\rm CH_4$ \citep{wilson2004}, but its resolution lies outside of the scope of this current work. Therefore, while it seems implausible that the solar system satellites have experienced the same escape rates over their entire age, we note that 10 Myrs under Titan-like escape would be sufficient to remove the $\rm CH_4$ mass in the atmosphere of a proto-Ganymede.
This is shown as two dotted lines in Fig. \ref{fig:evolution_ganymede_titan} (Top right). If Titan-like escape is not active at all, then non-thermal kinetic escape, which is  about $10^3$ lower \citep{delahaye2007}, would satisfy removal of the residual proto-Ganymede atmosphere over the age of the solar system.  }

{We conclude that a Jovian hot start scenario, whose plausibility is favoured by recent detailed calculations of the hydrodynamic accretion shock in the runaway phase of giant formation \citep{marleau2019} can deliver masses to be broadly consistent with the current solar system atmospheres, but not the compositions, as our Titan-analogue would suffer insufficient escape and fractionation between $NH_3$ and $CH_4$ to match real Titan. This statement stands assuming our starting conditions of initially equal $\rm NH_3$ and $\rm CH_4$ reservoirs, which in reality can be changed by the conditions in the circumplanetary disc \citep{anderson2021}. 
}

{Lastly, we note the prevalence of strong velocity fluctuations in the lower atmosphere of many simulations, particularly visible at low mass-loss rates. We discuss these in Appendix \ref{sec:appendix_velocities} and conclude that the overall mass-loss rate, which has to be constant in the entire atmosphere, is provided by wave action in the lower atomsphere, instead of by large-scale motion. As this mode of mass transport seems to take over mostly in the low mass-loss regime, when the upper atmosphere is collisionless, we do not think this has important implications for the evolutionary state of atmospheres.
}

\section{Summary, discussion and outlook}
\label{sec:discussion}

\subsection{Plausible additional physics}

Our goal was to understand the basic physics of escape from bolometrically driven temperature inversions in general exosatellites, but also to employ enough detail to be able to comment on composition-dependent differences in atmospheric outflows and the evolution of their atmospheres.
Nonetheless, we still employed multiple simplifying assumptions in this work, some of which may impact the computed escape rates, while others might only negligibly affect our results.

The ionisation of gases can impact escape rates \citep{Owen2012, schulik2024} and chemistry \citep{tsai2023}. Ionisation due to stellar X-ray photons and stellar and interstellar cosmic rays might penetrate deep into the satellite's atmosphere \citep{cleeves2013, rodgerslee2021} when the optical depth to UV photons is too high. While strong magnetic fields and high ionization rates might suppress or manipulate MHD-driven outflows \citep{owenadams2014}, the impact of an external (gas giant) magnetic field on partially ionized outflows {can either weaken or strengthen escape \citep{egan2019, presa2024}}, necessitating further work.
While UV radiation penetrates the atmospheres of the modern solar system satellites, this is likely not the case in the direct post-formation era, which is the focus of our study.

Our radiation transport model, while a step forward beyond double-grey models, can be extended to higher accuracies using on-the-fly correlated-k opacities \citep{molliere2015}, see also our discussion in Appendix \ref{sec:appendix_corrk}. Our maximum rule for multi-species Rosseland mean opacities neglects collision-induced absorption, an important ingredient in enhancing the opacitiy in the lower atmosphere \citep{pierrehumbertgaidos2011, heller2015greenhouse}.
Furthermore, the scattering of photons can heat the optically thin atmospheres \citep{mohandas2018}  beyond the temperatures calculated in this work.
It needs to be explored whether the optical or thermal scattering opacities of the molecules considered in our work are sufficient to describe the temperature profiles or whether other strong scatterers, such as aerosols or tholins \citep{mckay1991} are present in young satellite atmospheres.

We neglected any condensation effects \ms{in the atmosphere, both on the 
atmospheric adiabat, as well as on the radiation transport, where clouds and hazes might scatter thermal radiation and induce warmer temperatures than seen in our simulations.}
\ms{Those effects would} materialize as atmospheres cool down and reach the saturation limit of any individual component. A multi-species adiabat allowing for condensation \citep{graham2021} and rainout \citep{Booth2023} (we have used the non-condensible multi-species adiabat) for multiple species should be implemented in a future iteration of this work. Considering that condensed, moist adiabats are shallower than their dry counterparts, \ms{thus, by changing the density at the sonic point, this might have a strong impact on the mass-loss rates.}

The most important species, which we have neglected in our treatment of the lower atmosphere, effectively considering it to be condensed out in the satellite body, is water. Water is a well-known potent greenhouse gas \citep{kuramoto1994}, and for the same reason, it can act as a potent coolant of the upper atmosphere, impacting the outflow rates.
It was argued in the past that devolatilization of accreting solids should happen in the hot, moon-forming circumplanetary disc while the giant planets are luminous \citep{owen2006, alibert2007, bierson2020, anderson2021}; this would form dry moons, however, the solar system satellites might have formed late, when the circumplanetary disc became cold and icy. This scenario would not directly impact our calculations because our model is agnostic to the formation pathway that would create a certain initial atmospheric composition. Nonetheless it is a valid question to ask how strongly our assumed initial conditions depend on the formation physics and vice versa, what we could learn given fixed initial conditions. {We note that also other heat sources such as radiogenic heat \citep{barr2008, lichtenberg2019, johansen2022a} or induction heating of the protosatellite body, similar to that suggested in Io \citep{Piddington1968, goldreich1969}, general satellites \citep{chyba2021} or in exoplanets \citep{kislyakova2017} can impact the thermal state significantly and would allow for our considered range of $T_{\rm surf}$ to be exceeded, releasing constraints on the formation time and prolonging the duration of the hydrodynamic escape phase. Further, induction heating in the atmosphere might influence the local temperatures directly \citep{strugarek2024}. Specifically, \citep{barr2008} note that radiogenic heating can increase the post-formation surface temperature from 200K to 300K for a Callisto-analogue. } 
{The multitude of possible kinetic escape processes which drive atmospheric evolution in the hydrostatic regime, such as classic Jeans escape, sputtering and pickup losses \citep[see e.g.][and references therein]{gunell2018} have been entirely neglected in our work and would need to be introduced separately.}
We discuss the assumed initial composition in the next section.

\subsection{Are exomoons probes of giant planet formation physics?}

\subsubsection{On the composition}

Gas giants accrete most of their mass through the circumplanetary disc, which is simultaneously the source of satellite material. The slow rotation of the solar system giants today \citep{bodenheimer1977, cameron1988} and wide orbit exoplanets \citep{bryan2018} are evidence of the past existence of these discs. 
This final accretion stage occurs mostly at a single location in the circumstellar disc \citep{booth2017, bitsch2019} due to the slowness of type-II migration, especially in low-viscosity discs \citep{mcnally2019, lega2021}. The circumplanetary disc forms at masses similar to Saturn and after the local dust has grown to allow for sufficient cooling \citep{schulik2019, lega2024}. This might be observable \citep{zhu2015, taylor2024, choksi2025}, particularly as circumplanetary discs form under thermodynamic conditions which also renders them observable \citep{schulik2025}.


Hydrogen-rich gas in the circumstellar disc will form $\rm CH_4$ and $\rm NH_3$ as the dominant $\rm C,\,N$ carriers in chemical equilibrium below the temperatures of 500K \citep{heng2016}, consistent with moderate satellite formation times \citep{kuramoto1994} and giant planet formation at large distances \citep{Chiang1997}. Therefore, our initial composition is consistent with delivery by, but rapid escape of circumstellar $\rm H_2$ gas {e.g. via boil-off \citep{owenwu2016, ginzburg2016, Rogers2024} and fractionation of heavier species.}

However, there is no guarantee that gas in this temperature range has reached thermochemical equilibrium after a few Myrs. \cite{liggins2022} estimate the time to reach equilibrium for various plausible atmospheric compositions of rocky bodies. Their results indicate that for a gas temperature of $T= 500$~K at 1 bar, the equilibration timescales for $\rm NH_3$ and $\rm CH_4$ range between $10^6$-$10^9$yrs. 
As the giant planet system cools at a comparable timescale or even faster, the chemistry of either the accreted circumplanetary or circumstellar material would be frozen into the exosatellite's atmospheres. 
The chemical timescales are further set by the actual temperatures and ionization rates through which gas is processed in the circumplanetary disc \citep{fujii2017, schulik2020, shibaike2024}, but those complexities are beyond the scope of this work.

In the absence of important catalysts, satellite migration or planet-planet scattering events \citep{Frelikh2019}, chemical inheritance from the circumstellar disc could be used to learn directly about the composition of the giant planet. This information would then complement the idea of inverting formation pathways of giant planets via the measurement of $[C/O]$ ratios \citep{oberg2011}, and even extending it, as it encodes its information at the level of individual molecules. 

The 'inverting formation pathways' approach invokes species phase jumps at iceline locations in the protoplanetary disc, and traces a giant planet's migration and accretion history through those jumps. It then attempts to identify the formation locations of giant planets by relating a planet's $[C/O]$ value to a local disc $[C/O]$ value.
This idea has sparked considerable interest in recent years \citep[e.g.][]{kirk2024, penzlin2024}. However, the idea has been shown to suffer from considerable challenges due to model degeneracies \citep{molliere2022}. 

Those problems necessitate independent tests to show whether giant planet composition and disc compositions actually match.
While the latter can be provided by ongoing surveys with ALMA \citep{miotello2023} and JWST \citep{vandishoeck2023faraday, tabone2023, keyte2024}, the former might be most reliably provided by information on giant planet satellites, given their compositional information is not lost due to re-equilibration. Particularly cold Jupiters, such as the PDS 70 planets \citep{Keppler2018, isella2019}, should be the targets for those searches, as Hot Jupiters are improbable to host satellites \citep[e.g.][]{Wei2024}. This is where satellites might provide valuable insight - detecting their atmospheric components would directly inform about the composition of the majority of the accreted giant planet mass, if the satellites never equilibrated.

This could be compared to the accreted and reduced chemistry of their host giant planets, a test that might be withing reach with the upcoming ELTs \citep[e.g.][]{vanWoerkom2024}.

\subsubsection{On cooling curves}

Giant planet cooling curves are of great importance  in planet formation theory \citep[e.g.][]{burrows1997}, as they are often the only possibility to constrain wide-orbit giant planet's masses \citep[e.g.][]{marois2008, haffert2019}. However, uncertainties in these cooling curves remain, often giving freedom to invoke the two extremes of the 'hot' and 'cold' start scenarios \citep{marley2007}. Considerable progress has been made in recent years \citep{berardo2017, marleau2017, marleau2019} in distinguishing between those extremes, but as our usage of the \cite{linder2019} model showed, some important modeling uncertainties remain.

Under the assumption of no or negligible satellite migration after circumplanetary disc dissipation, for which there is some evidence in the Galilean moons \citep{dekleer2024}, the masses of satellite atmospheres as a function of their giant planet host masses would inform about the cooling curves of giant planets. Particularly the sharp cliff in final atmospheric satellite masses seen in Fig. \ref{fig:exosatellites_hostmass} can distinguish between hot and cold start models \ms{of giant planet formation}. The migration of giant planets would also impact satellite atmospheres once they reach a certain proximity to the star: the stellar bolometric radiation will start being important in supporting atmospheric escape inside a giant-planet-to-star semimajor axis of $\sim$1 AU for a Sun-like star. Large pre-main-sequence luminosities \citep[e.g.][]{burrows1997,Baraffe1997} would exacerbate this effect. Therefore, Grand-tack-like excursions to the inner system \citep{walsh2011}, with a return to a cold region in the disk, might  leave imprints through the lack of satellite atmospheres.

\subsection{Summary}
In this paper, we have explored the dependency of atmospheric escape models for exosatellites driven by bolometric radiation and parameterized its impact by the opacity ratio parameter $\gamma$. We computed realistic values of $\gamma$ \ms{for general atmospheres first} and found that values of $\gamma>1$ are conducive to powerful escape. In those cases, regions of the atmosphere heat up above the equilibrium temperature. Such a scenario occurs when matching molecular opacities of cold gas temperatures ($T_{\rm gas}\approx 200$K) with the temperature of the radiation field around $400-800$K. This exact situation corresponds to the upper atmospheres of warm, post-formation satellites orbiting hot, luminous young gas giants. 

Our nominal scenario predicts that for reasonable assumptions of the formation times, corresponding to a range of surface starting temperatures, satellites orbiting $\geq 2m_{\rm Jup}$ gas giants at a distance of $\sim 10 R_{\rm Jup}$ distances should be stripped of their atmospheres due to the early bolometric irradiation from their gas giant host. Satellites orbiting lower mass giant planets inherit the atmospheres given by the post-formation initial conditions (or any outgassing) with only minor alterations of their composition due to weak hydrodynamic escape.  For gas giant masses $\leq 150 m_{\rm \oplus}$, and 
 surface temperatures $\leq350$ K, escape is so weak that the state of the atmosphere directly jumps into the hydrostatic and Jeans escape regime after formation, completely freezing the formation compositions.

While the condition for shutting off escape is always for the outflow to become non-collisional, the typical reasons for this can differ in three main ways: Cooling of the upper atmosphere for atmospheres with long lifetime, cooling of the surface, or self-termination after the atmosphere becomes relatively optically thin, and the hot adiabatic region of the atmosphere disappears.
A surprising result for the modelled population of satellites is that while our scenario of ``efficient'' escape for which $\dot{M}/\dot{M}_{\rm  Parker}>1$ is found over a range of the giant planet mass-space, the majority of the satellite atmospheric mass is lost in the inefficient regime due to the stronger evolution of the giant planet's luminosity over atmospheric temperature.

Our final application of hot-start cooling curves for giant planets shows that the solar system satellite atmospheres of Titan and Ganymede can be consistent with Saturn's inability to strip Titan's atmosphere. In contrast, Jupiter can strip Ganymede's atmosphere down to a mass at which kinetic escape can remove its remainder during the lifetime of the solar system.

\section*{Acknowledgements}
We thank the anonymous referee for helping to improve the quality of this manuscript. MS wants to thank William Misener, Hilke Schlichting and James Rogers for discussions about internal cooling and contraction of envelopes. This publication makes use of The Data \& Analysis Center for Exoplanets (DACE), which is a facility based at the University of Geneva (CH) dedicated to extrasolar planets data visualisation, exchange and analysis. DACE is a platform of the Swiss National Centre of Competence in Research (NCCR) PlanetS, federating the Swiss expertise in Exoplanet research. The DACE platform is available at https://dace.unige.ch. JEO is supported by a Royal Society University Research Fellowship. This project has received funding from the European Research Council (ERC) under the European Union’s Horizon 2020 research and innovation programme (Grant agreement No. 853022).

\section*{Data Availability}

The {\sc Aiolos} code is publically available at \url{https://github.com/Schulik/aiolos}.



\bibliographystyle{mnras}
\bibliography{bibs_merged} 




\appendix

\begin{figure}
   \centering
   \includegraphics[width=0.40\textwidth]{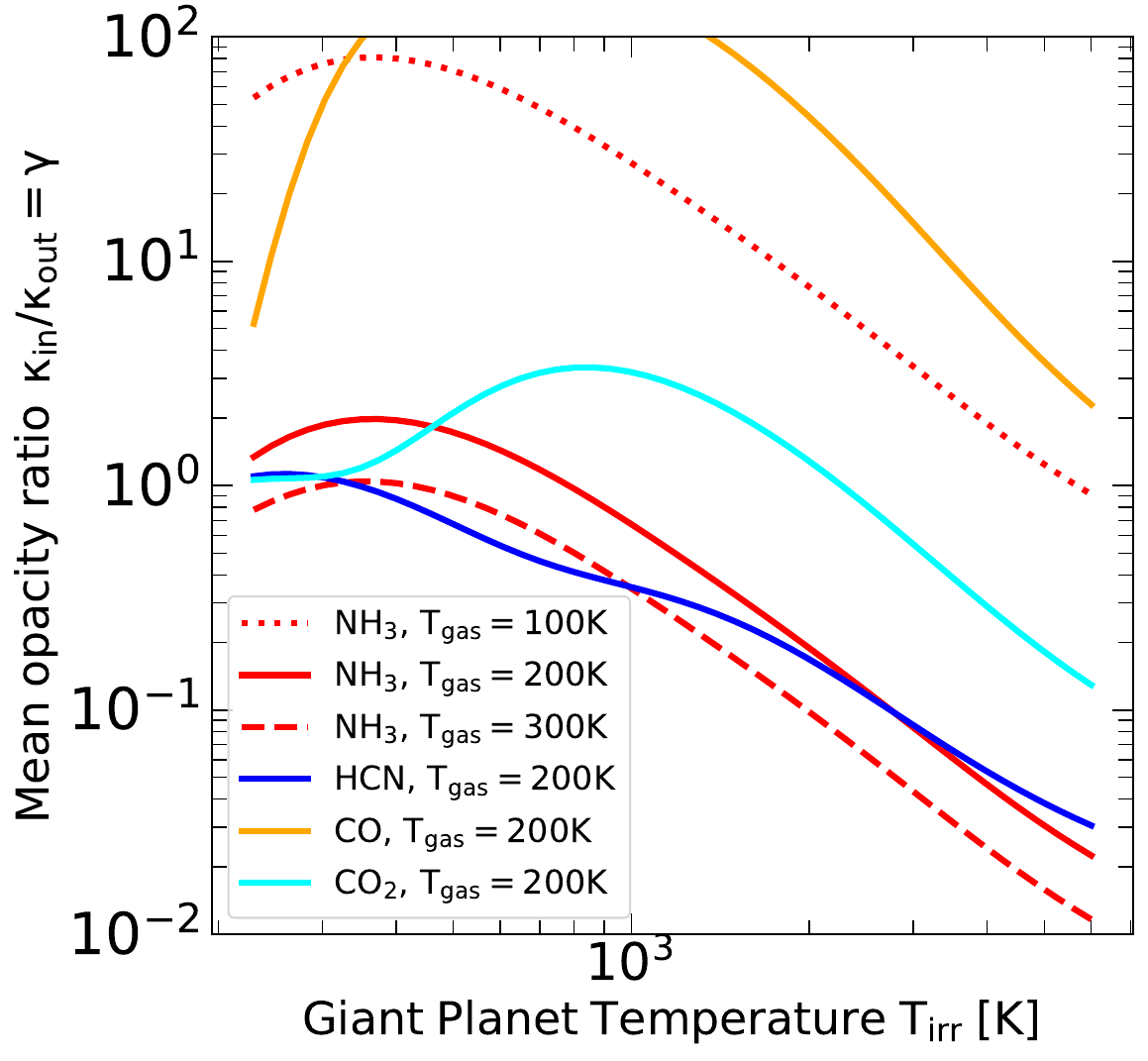}

     \includegraphics[width=0.40\textwidth]{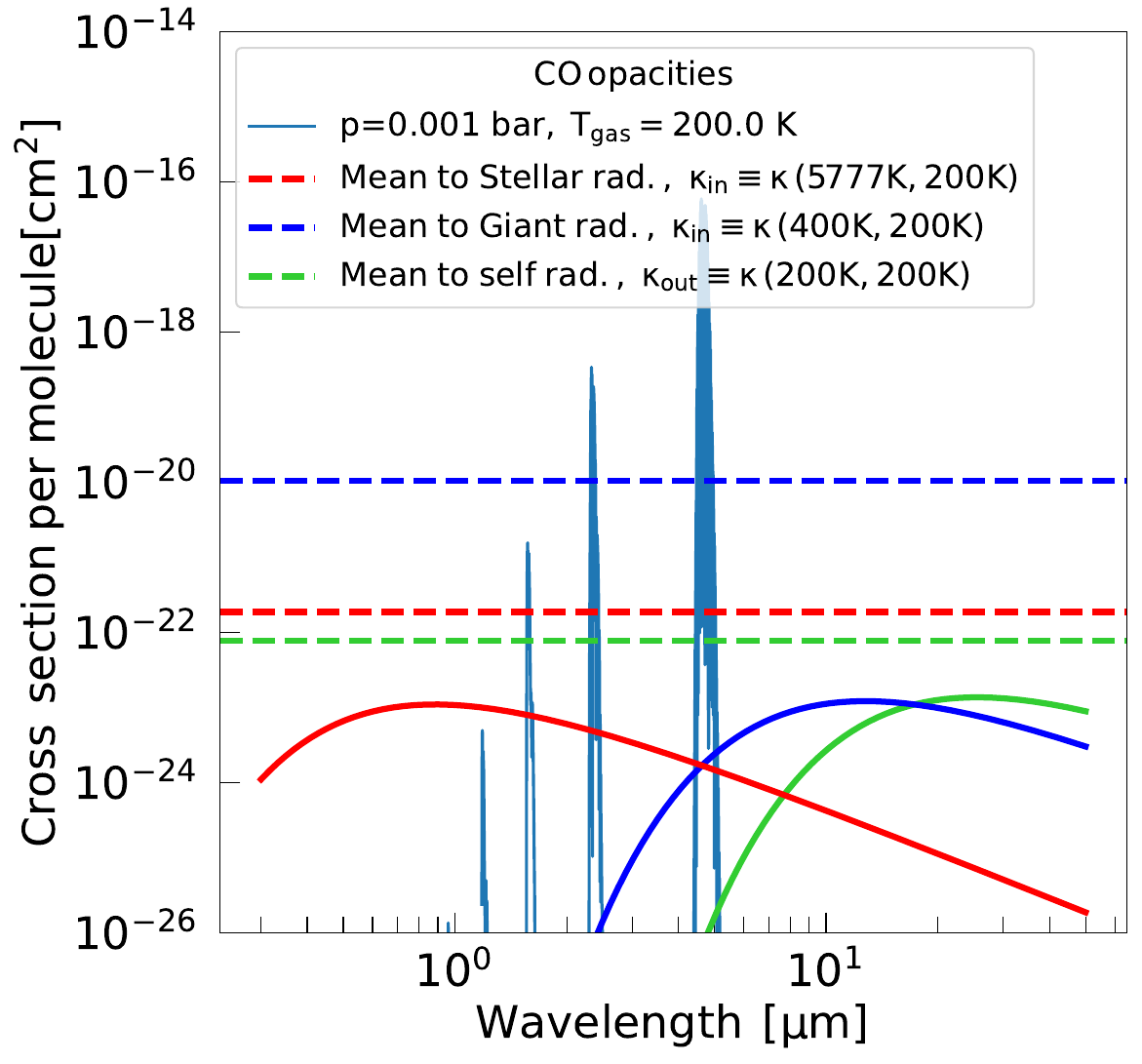}

\caption{An extension of Fig. \ref{fig:gamma_vs_mdot} (Left) to indicate that large values of $\gamma$ alone need not be sufficient for significant heating, see the top panel for CO - $\gamma$ needs to be representative of continuum opacities, which CO does not provide, see the lower panel. }
\label{fig:appendix_gammacurves}
\end{figure}

\appendix
\ms{
\section{When can $\gamma$-curves be trusted?}
\label{sec:appendix_gammacurves}
{One might be interested to find more molecules with the property $\gamma>1$ at some $T_{\rm irr}$ of interest. We present and explain an important pitfall that might occur when performing such a search.}
In Fig. \ref{fig:gamma_vs_mdot} we showed how the mean opacity ratio behaves for different values of the irradiating blackbody's temperature. For the sake of completeness, we document those ratios for other molecules which might be present in primitive, cold atmospheres in Fig. \ref{fig:appendix_gammacurves}. We want to caution the reader, as inspection of this data might lead to the conclusion that CO should be a strong driver of bolometrically-driven atmospheric escape. However, in this case, just the application of mean opacity ratios can be strongly misleading: the opacity function for CO at low pressure and temperature is dominated by just a few sharp peaks. Thus, the mean opacity ratio does not represent the continuum, and its use would yield a poor representation of the actual temperature structure. $\rm CO_2$ suffers from a similar property, albeit in a less extreme manner. Thus, these molecules emphasize the need for multi-band treatments, as we have adopted in our more detailed calculations.
}

\section{Constant $\kappa_s$-spectrum - justification via corr-k comparison}
\label{sec:appendix_corrk}
\begin{figure}
     \includegraphics[width=0.40\textwidth]{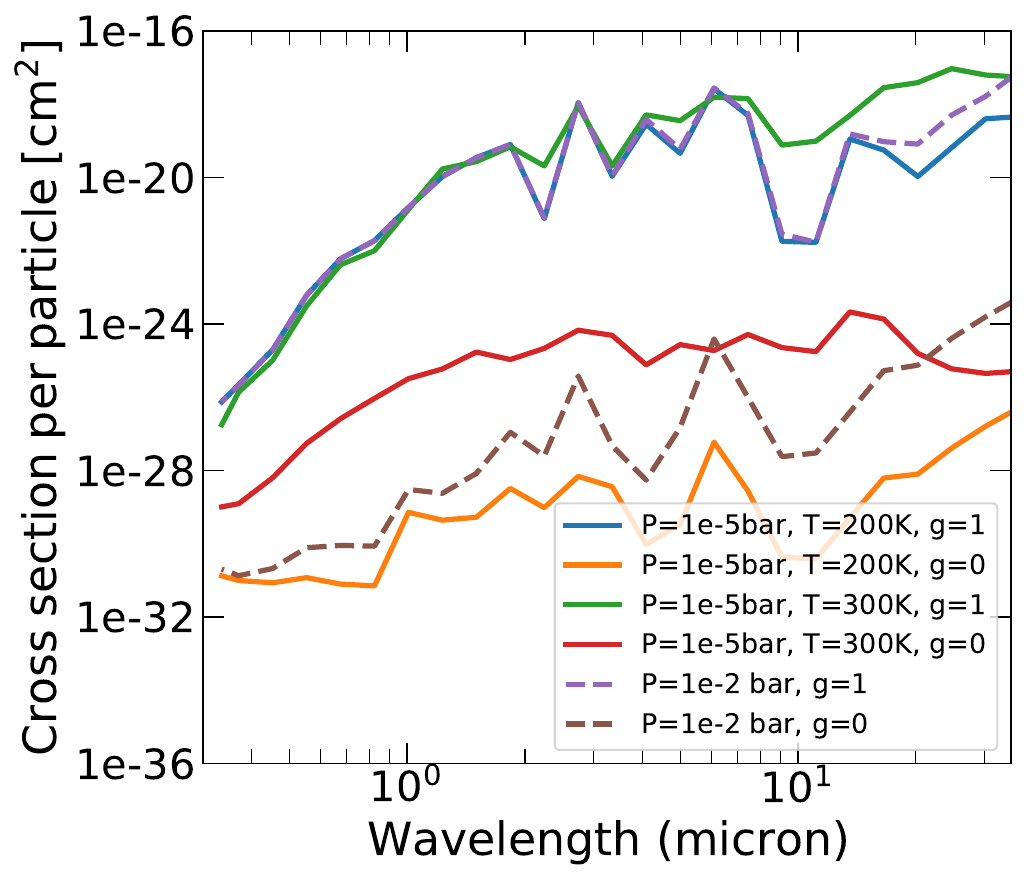}
     \includegraphics[width=0.40\textwidth]{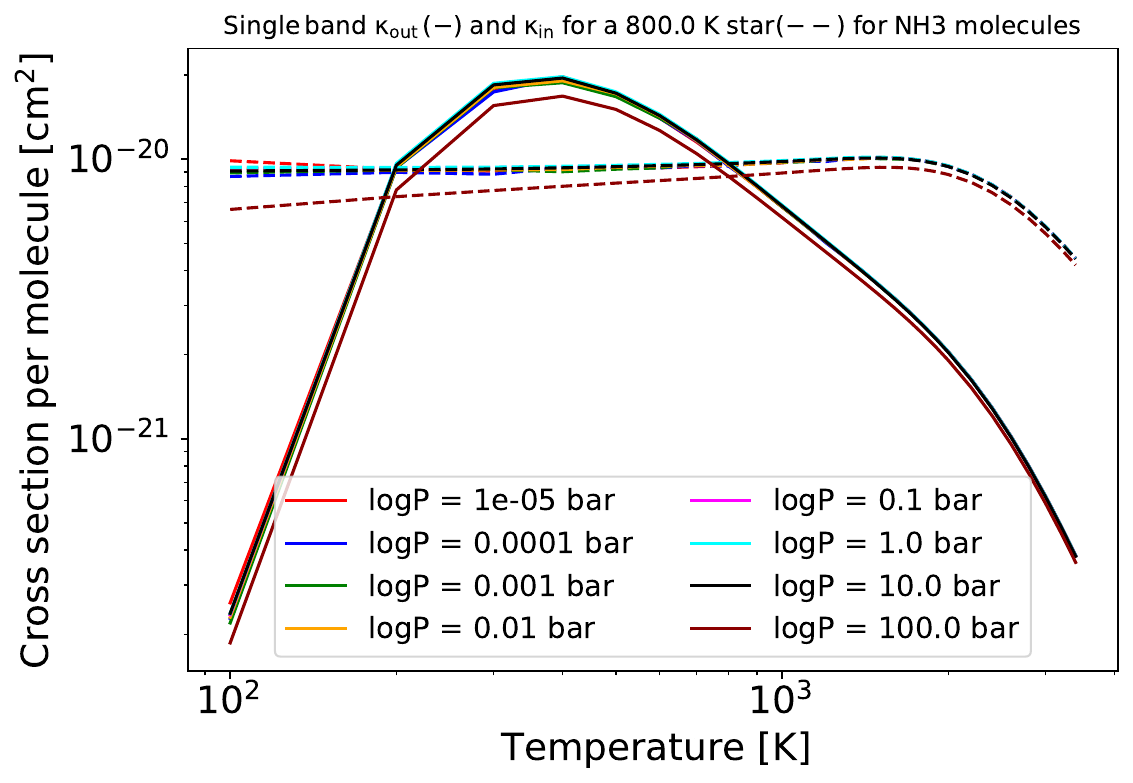}

\caption{ Downgraded exo-k line distribution (left) and resulting mean single-band $\kappa_{\rm in}$. }
\label{fig:appendix_exok}
\end{figure}

{
Here we document why it is justified to use either a single-band $\kappa_{\rm in}$ or a multi-band $\kappa^b_{\rm in}$ as constant function of $P,T$.
The downgraded line distributions are plotted in Fig. \ref{fig:appendix_exok} for $\rm NH_3$ show only small variations over the P/T range relevant for absorption. The ‘top opacities’ of the correlated-k distribution, encoded by 'g$=$1', are barely reacting to P and T in our chosen range.
It is exactly those opacities that our mean opacities encode. However as the overall multi-band variation remains small, this is also true for the wavelength resolved $\kappa^b_{\rm in}$, which we therefore do \textit{not} treat as a function of $P,T$. Particularly at the irradiation temperatures around $T=800K$, corresponding to ~1 $\mu$m, the impact on $\kappa_{\rm in}$ is negligible, see the lower plot and the set of dashed lines for this.
The outgoing opacities $\kappa_{\rm out}$, determined by the outgoing blackbody at $\geq 10 \mu$m do react strongly to T and P broadening, but this effect on $\kappa_{\rm out}$ is included in our calculations, as T,P variations in $\kappa_{\rm out}$ and $\kappa_{\rm Rosseland}$.
We note that the lower end of the opacity distribution 'g$=$0' reacts strongly to both T and P, which should provide for interesting solution behaviour in full correlated-k radiation transport simulations, which are currently prohibitevly expensive in a hydrodynamic setting, as this would add three more computational dimensions $(P,T,g)$ in order to compute the heating function. }

\section{Numerical convergence of the heating function and mass-loss rates in multi-band simulations}
\label{sec:appendix_resolution}

\begin{figure*}
\hspace*{-0.5cm}
 \begin{subfigure}{0.34\textwidth} 
   \centering
   \includegraphics[width=1.0\textwidth]{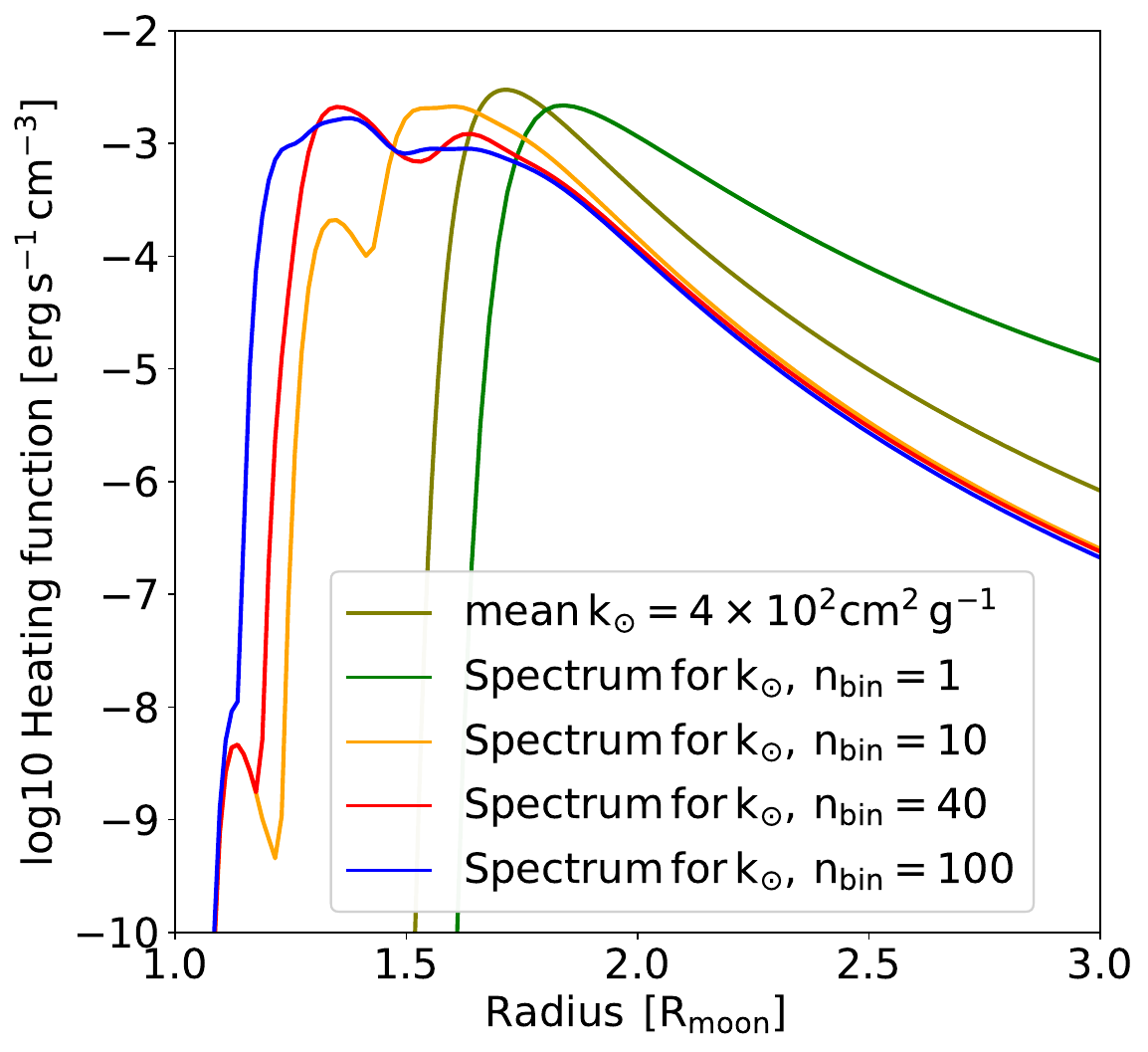}
\end{subfigure}%
\begin{subfigure}{0.34\textwidth} 
   \centering
   \includegraphics[width=1.0\textwidth]{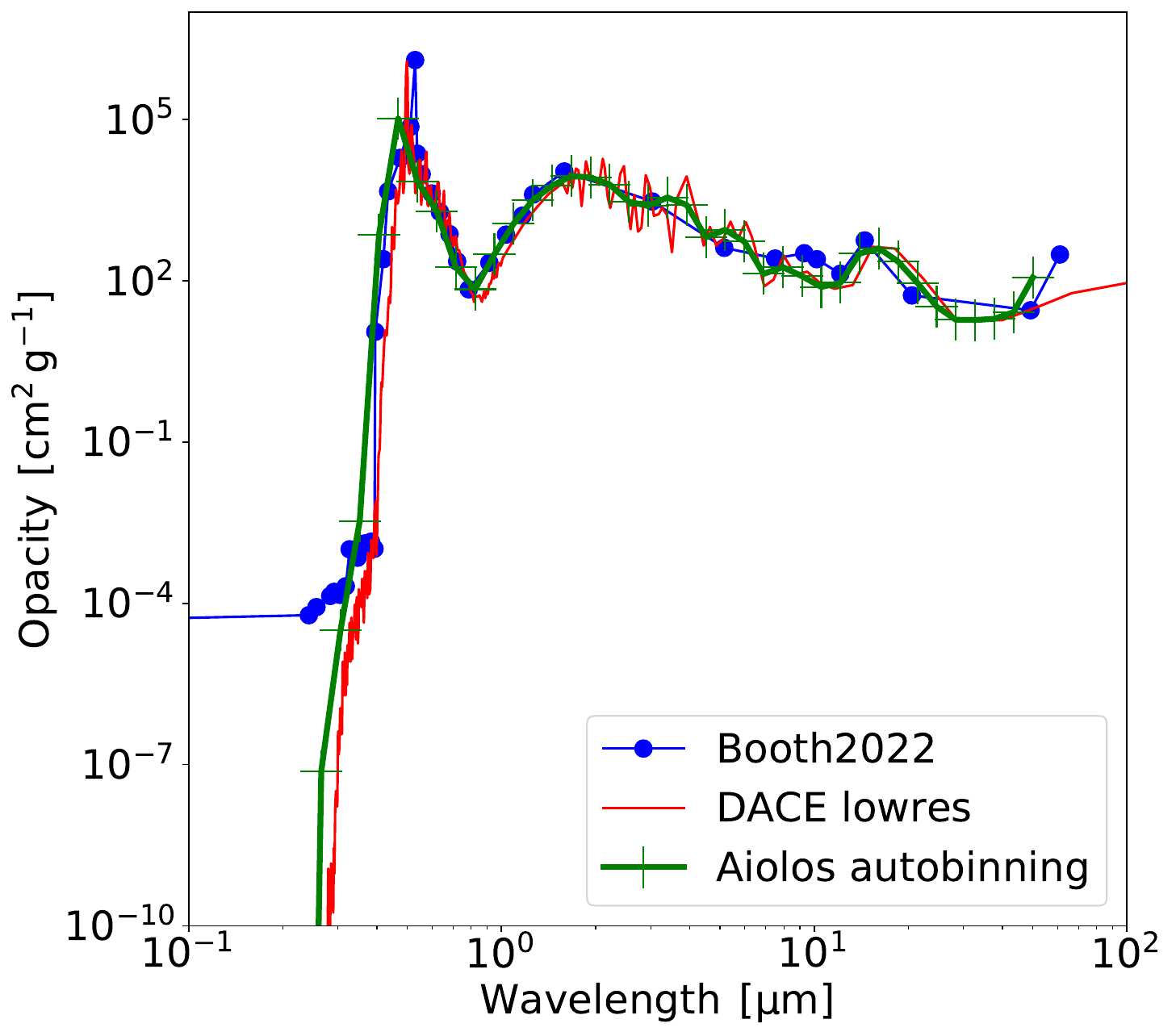}
\end{subfigure} 

\hspace*{-0.5cm}
 \begin{subfigure}{0.34\textwidth} 
   \centering
   \includegraphics[width=1.0\textwidth]{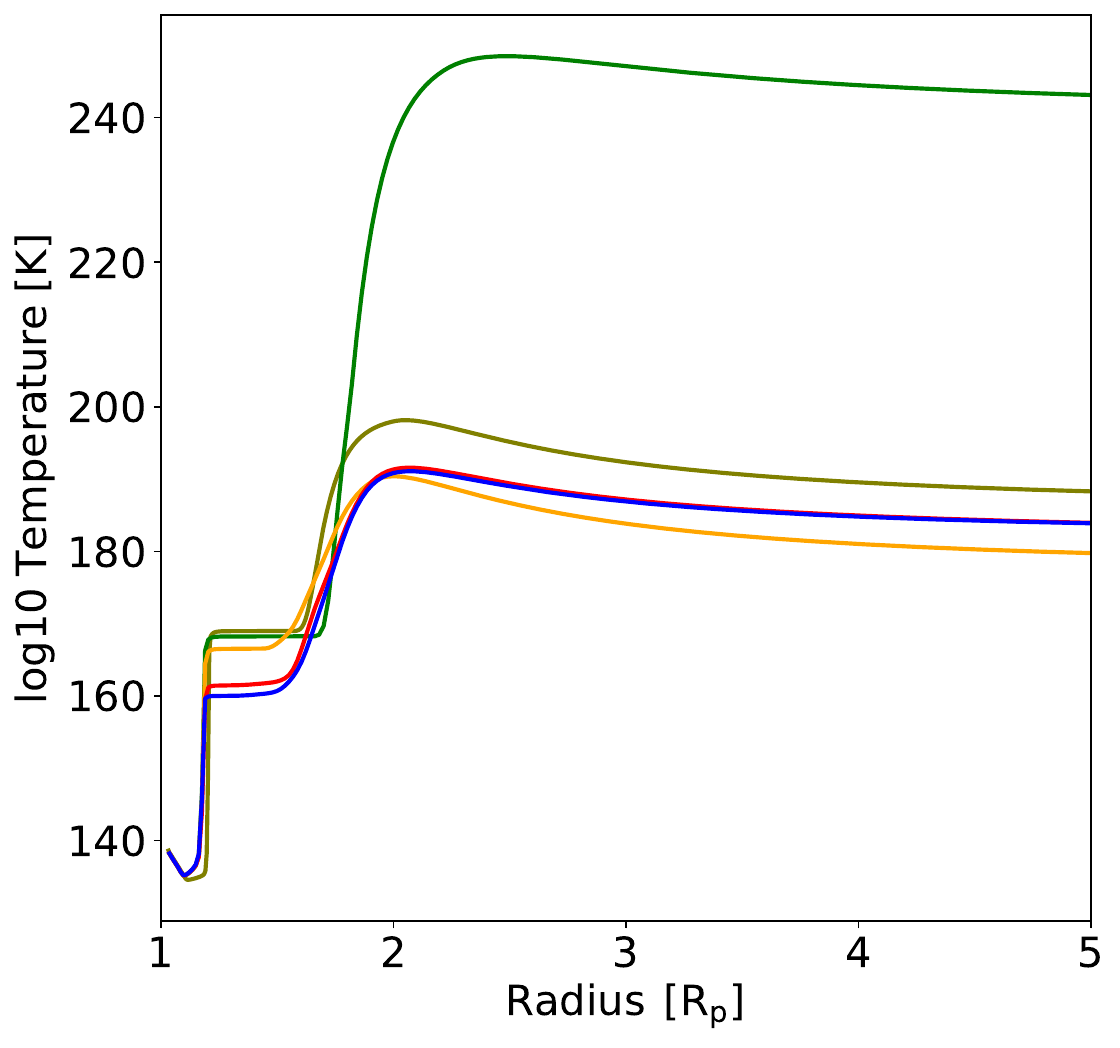}
\end{subfigure}%
 \begin{subfigure}{0.34\textwidth} 
   \centering
   \includegraphics[width=1.0\textwidth]{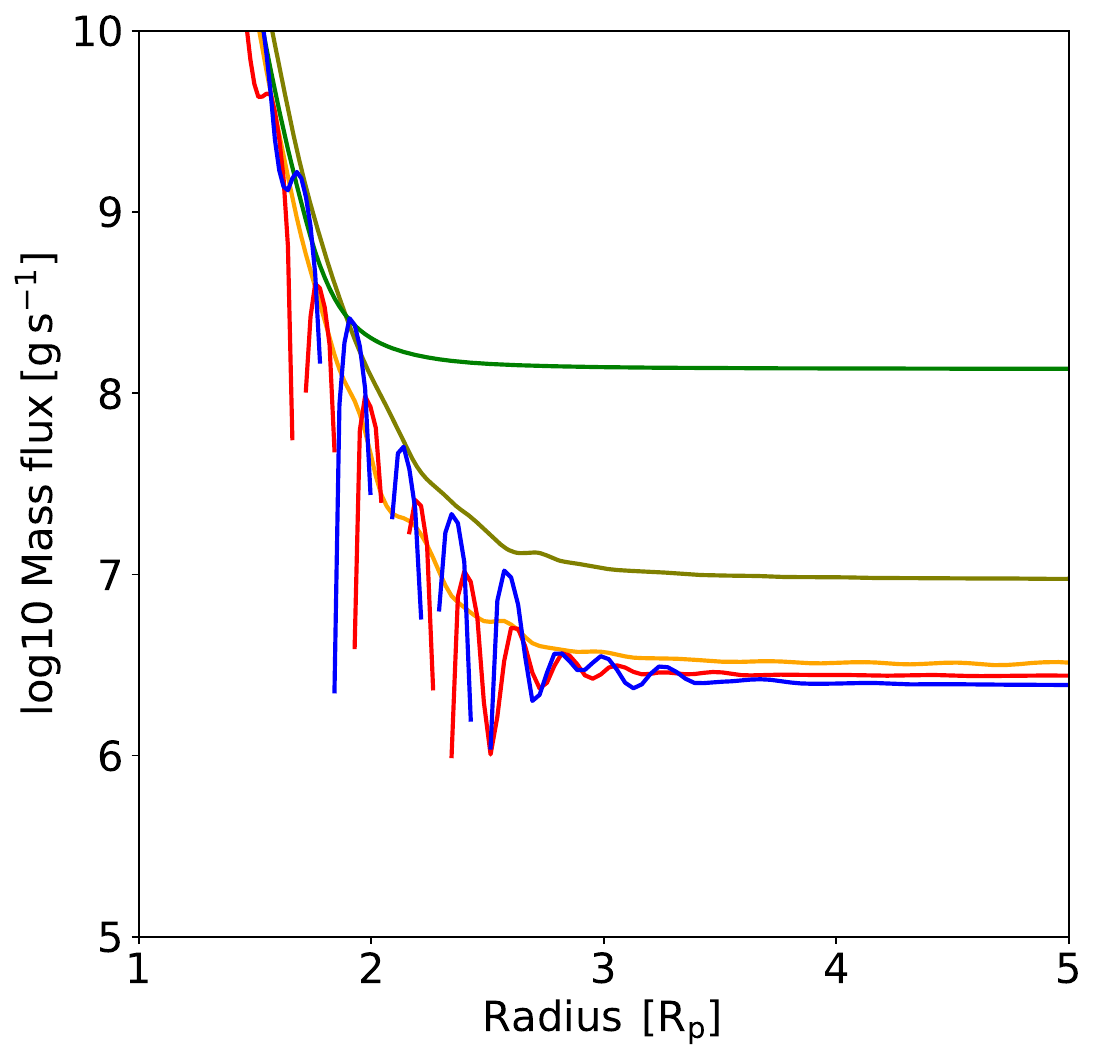}
\end{subfigure}%

\caption{Convergence tests of the heating function (\textbf{Left}) of $\rm NH_3$ with varying number of bins and the opacity of the mineral $\rm MgO$ at high spectral resolution with different binning techniques (\textbf{Right}). Spectral resolution power is key to obtain a sensible representation of the heating function in the simulation domain. The heating and hence temperatures in the upper atmosphere ($r>1.7 R_{\rm satellite}$) converge rapidly at a low spectral resolution of $N_b=10$, but the deeper atmospheric layers require more resolution, depending on the exact amount of substructure found in a species' opacity function. On the right, the opacity function of an example species is shown.
The blue data points from \citep{Booth2023} took two weeks to compute from line-lists on a single-core machine and a sophisticated binning technique, whereas the green data points are generated using a low-resolution opacity spectrum from the DACE database, which is binned on-the-fly by our AIOLOS autobinning function with $N_b=40$. The results in opacity structure between the very expensive and very cheap method are quite agreeable. We then continue to show the resulting temperature structure and mass loss rates (\textbf{lower panels}) }
\label{fig:appendix_convergencetests}
\end{figure*}

In Fig. \ref{fig:appendix_convergencetests}, we show convergence tests as a function of resolution in wavelength space. This shows that from a band number of $b=40$ or higher, the escape rates converge to the same value. Thus, our choice of $b=80$ is appropriate. 

\section{Mass loss rates of primary and secondary species - effects of drag and thermalisation}
\label{sec:appendix_densies}
\begin{figure*}
\hspace*{-0.5cm}
 \begin{subfigure}{0.34\textwidth} 
   \centering
   \includegraphics[width=1.0\textwidth]{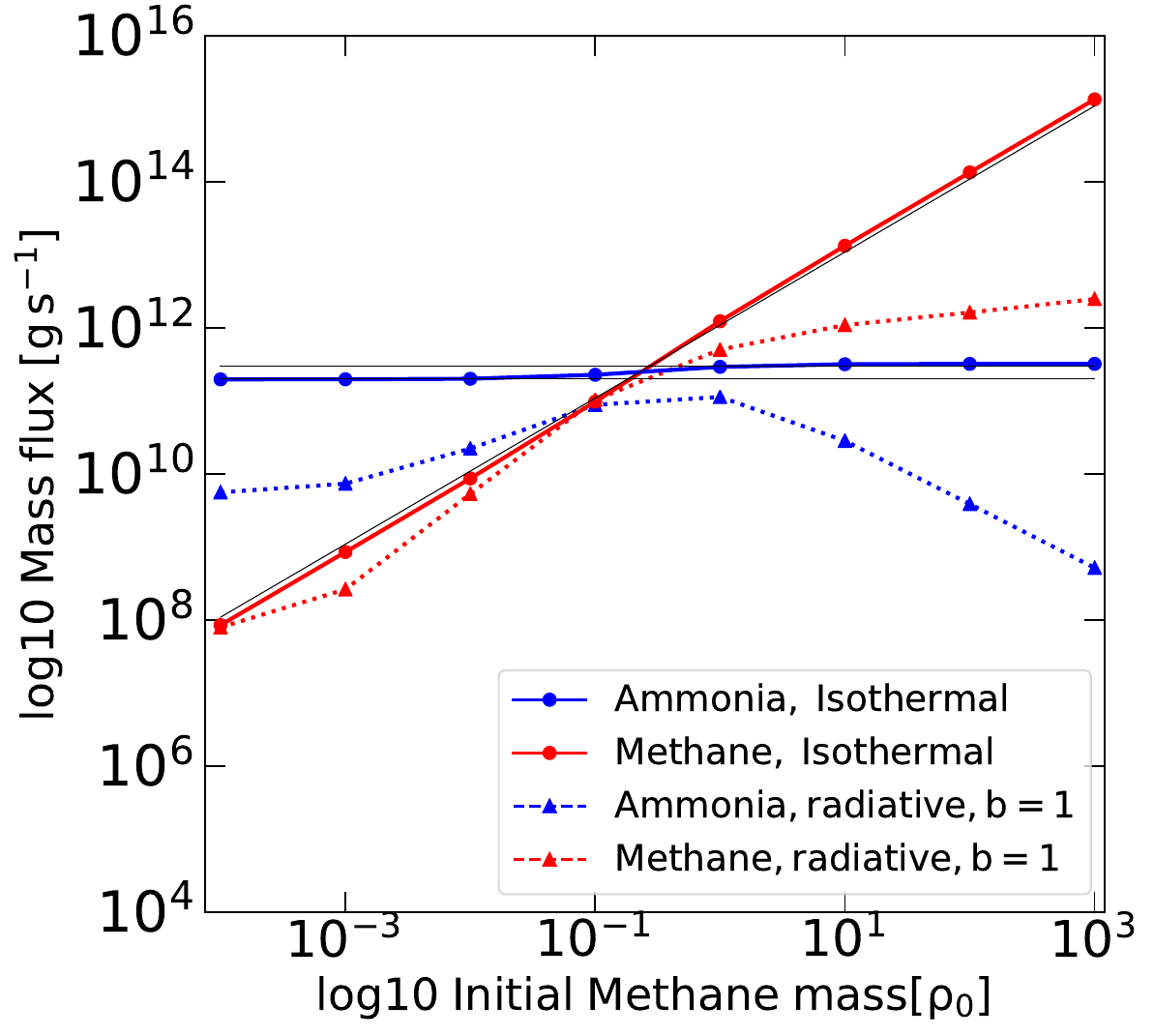}
\end{subfigure}%
\begin{subfigure}{0.34\textwidth} 
   \centering
   \includegraphics[width=1.0\textwidth]{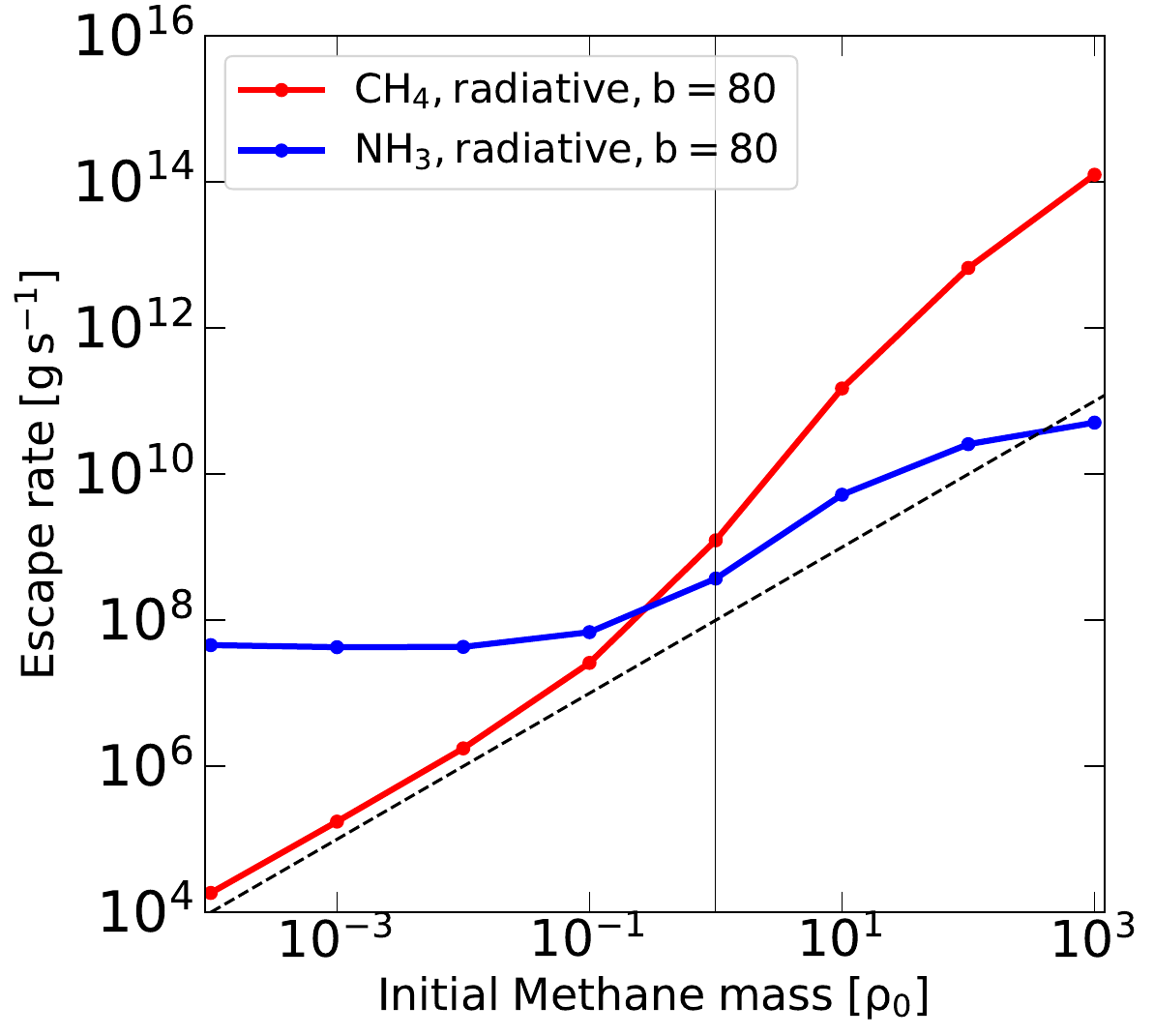}
\end{subfigure} 
\begin{subfigure}{0.34\textwidth} 
   \centering
   \includegraphics[width=1.0\textwidth]{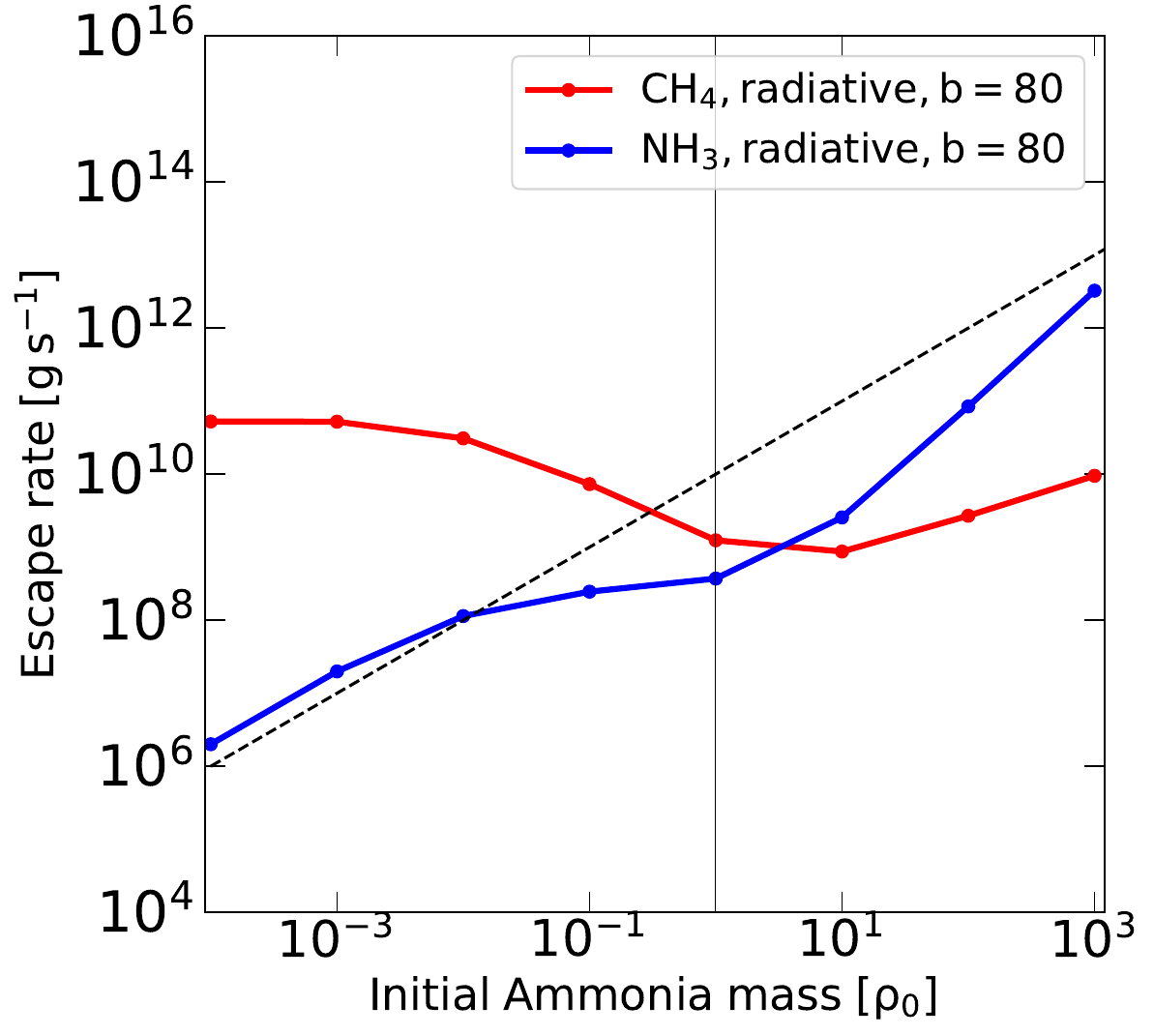}
\end{subfigure} 

\caption{Effects of treatment of the energy equation, drag and temperature domination for two-species solutions in various approximations. \textbf{Left}: Isothermal simulations at $T=200K$ and radiative $N_b=1$ simulations are compared to each other. The solid black horizontal lines denote the uncoupled and coupled limits according to \citep{zahnlekasting1986}, considering $\rm CH_4$ as the dominant escaping species and $\rm NH_3$ as the trace species. The difference between uncoupled and coupled regime for $\rm NH_3$ is minuscule due to the similar masses of both molecules. For the radiative case, with increasing $\rm CH_4$, the $\rm NH_3$ is shielded and less coupled into the outflow as the absorption altitude of bolometric radiation shifts to larger values. \textbf{Middle} and \textbf{Right}: Highly resolved $N_b=80$ simulations for which the density of one component is varied, while the other is left constant. A $\rm NH_3$-dominated atmosphere is colder than a $\rm CH_4$ dominated one, at high altitudes. This causes the escape rates to increase, as $\rm CH_4$ increases, evident as $\dot{m}_{CH_4} = c \times m_{CH_4}$ changes to a different constant $c$. Correspondingly, the opposite effect occurs for increasingly $NH_3$ dominated atmospheres.}
\label{fig:appendix_densityvariation}
\end{figure*}

In our simulations, two effects can contribute to species interacting with each other: drag (collisional momentum exchange) and collisional heat exchange. We investigate their relative importance in Fig. \ref{fig:appendix_densityvariation} by varying the base densities of both escaping components.

 Isothermal simulations at $T=200K$ show that decreasing the species density only has a negligible effect on the secondary species. This behaviour is indicative of weak drag coupling, fully expected at our low temperatures and neutral components. The single-band radiative simulations ($b=1$) show that thermal effects, in this case, shadowing of the single opacity band, are much more important in driving escape. This effect is emphasized by the fact that as $\rho_{\rm CH_4}$ increases, $\dot{m}_{\rm NH_3}$ decreases, opposite to the expectations in a drag-dominated regime.
At low $\rho_{\rm CH_4}$, the mass-loss rates roughly follow $\dot{m}_{\rm CH_4} \propto\rho_{\rm CH_4}$, as one would expect from the Parker-like escape regime.

When moving to highly resolved simulations, it becomes clear that: (i) the escape rates drop overall by a factor of $\sim10^2$; (ii) the decrease in $\rm NH_3$ escape rates in the $\rm CH_4$ dominated regime is an artefact of the low spectral resolution simulations, arising from shadowing and (iii) the species common temperatures are set by whichever component has the highest cooling/heating capability per mass. As we vary the methane content, it is obvious how the proportionality constant in $\dot{m}_{\rm CH_4} \propto  \;\rho_{\rm CH_4}$ changes from a smaller to a higher value. Since with high $\rm CH_4$ content, the thermal profiles approach that dictated by $\rm CH_4$, and vice versa.

We conclude that in our simulations, the most important influence of species on one another is collisional heat exchange. In principle, the species could also interact only via the commonly shared thermal photon field $J$, even when the collisional exchange terms are neglected. However, one would expect the thermal profiles to decouple and both species to ``go their separate ways'', which is more akin to what is seen in the isothermal simulation.

\begin{figure*}
\hspace*{-0.5cm}
 \begin{subfigure}{1.0\textwidth} 
   \centering
   \includegraphics[width=1.0\textwidth]{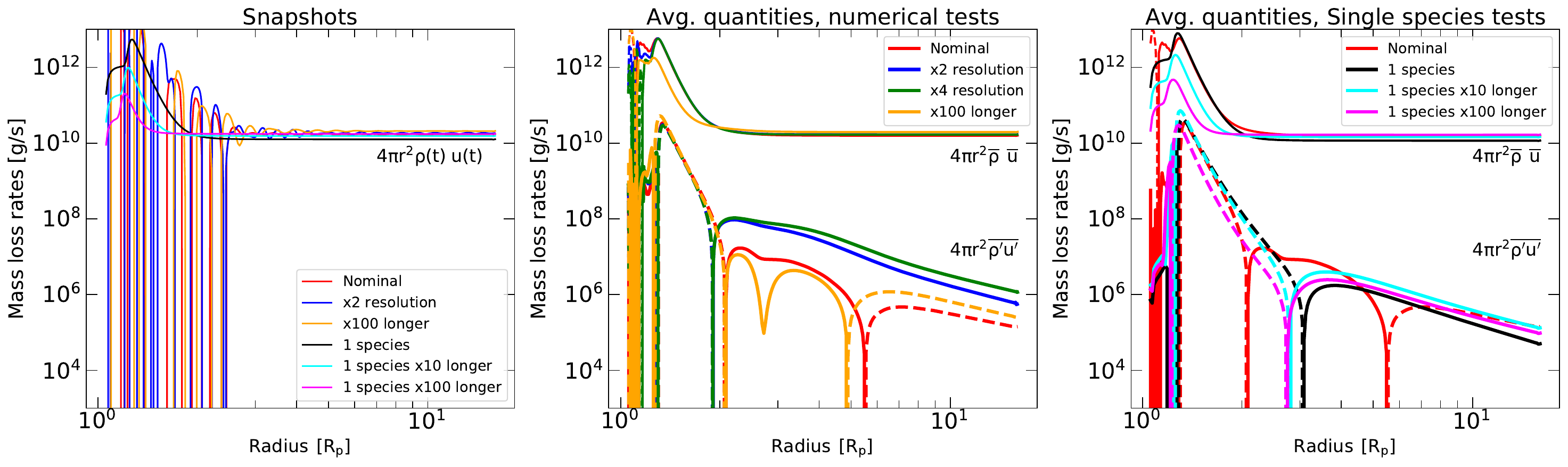}
\end{subfigure}%
\caption{Analysis of wave fluxes in the deep atmosphere. 
We compare instantaneous snapshots of the mass-loss rates $\rm 4\pi r^2 \rho(t) u(t)$ in the left panel with their time-averaged mean and fluctuating counterparts in the middle and right panels. The wave-driven mass-transport $\rm 4\pi r^2 \overline{\rho' u'}$ is the lower set of curves in those panels. Solid lines indicate positive values and dashed lines indicate negative values.}
\label{fig:appendix_fluctuations_new}
\end{figure*}

\section{Fluctuation-driven mass transport and the role of apparent waves}
\label{sec:appendix_velocities}

{
Here we work to understand the role of the velocity fluctuations in the lower atmosphere, such as seen in Fig. \ref{fig:evolution_ganymede_titan}, which seem to consistently appear in many simulations. These fluctuations seem to be robust against running numerical convergence and long-term simulation tests, and hence might indicate a physically real phenomenon. Furthermore, our application of the gravitational well-balancing scheme by \citep{Kappeli2016}, indicates that those fluctuations should pose real oscillations around a stable hydrostatic equilibrium. In order to understand the action of the fluctuations, we compute their average mass-transport properties. 
We ran a temperature-inverted simulation to convergence, which is reached after the time of $10^7$s, quantified by a non-changing mass loss rate $4\pi r^2 \overline{\rho u}$.
The mean and fluctuating parts of the average mass-loss rate can be decomposed as $\overline{\rho u} = 
\overline{\rho} \; \overline{u}+ \overline{\rho' u'}$. The terms $\overline{\rho u}$ and $\overline{\rho} \; \overline{u}$ can be calculated as running means of the simulation state variables $\rho(t)$, $u(t)$ directly.
From this one can obtain the fluctuation-driven mass flux $\overline{\rho' u'}$ is obtained via subtracting the two.}

{
In investigating the source of the oscillations in $\rho(t)u(t)$ in the lower atmosphere, we note their persistence after running for a longer time and with a higher spatial resolution, as presented in Fig. \ref{fig:appendix_fluctuations_new}. We also note the absence of the fluctuations in single-species simulations, which leads us to attribute their existence to frictional coupling in the hydrostatic atmosphere.
The middle panel shows the separation in magnitudes between the mean-flow escape rates and the fluctuation-driven ones. The fluctuation-driven values are orders of magnitude weaker than the instantaneous fluctuations, which average to match the directed escape rates through the sonic point. We note that for longer simulation times for both single- and two-species simulations, the average escape rate profiles decrease in the hydrostatic atmosphere to match those in the hydrodynamic atmosphere, as expected. We conclude that the instantaneous fluctuations are possibly an artefact of our multi-fluid treatment in the lower-hydrostatic atmosphere. These instantaneous oscillations do not correspond to dominant fluctuation-driven mass-transport. The averaged fluctuation-driven flux which in fact becomes negative for both species in the hydrostatic atmosphere, can be delivered by physical wave action, but it is not entirely excluded that numerical effects remain important in this term.}


\bsp	
\label{lastpage}
\end{document}